\documentclass[prd,aps,10pt,nofootinbib,twocolumn,superscriptaddress,preprintnumbers,balancelastpage,longbibliography]{revtex4-1}

\usepackage{amsmath,amssymb}	
\usepackage{mathtools}
\usepackage{fontawesome}
\usepackage[dvipsnames]{xcolor}
\usepackage{hyperref}
\usepackage{xspace}
\usepackage{fancyhdr}
\usepackage{braket}
\usepackage{graphicx}
\usepackage{siunitx}
\usepackage{blindtext}
\usepackage{nicefrac}

\usepackage{afterpage}
\newcolumntype{L}[1]{>{\raggedright\let\newline\\\arraybackslash\hspace{0pt}}m{#1}}
\newcolumntype{C}[1]{>{\centering\let\newline\\\arraybackslash\hspace{0pt}}m{#1}}
\newcolumntype{R}[1]{>{\raggedleft\let\newline\\\arraybackslash\hspace{0pt}}m{#1}}
\usepackage{longtable}
\newcommand{\nbicon}{{\color{linkcolor}\faFileCodeO}\xspace}
\newcommand{\nblink}[1]{\href{https://github.com/smsharma/dark-photons-perturbations/blob/apr-2020/notebooks/#1.ipynb}{\nbicon}}
\newcommand{\githubmaster}{\href{https://github.com/smsharma/dark-photons-perturbations/}{\faGithub}\xspace}

\newcommand{\dd}{\mathrm{d}}
\newcommand{\mAp}{m_{A^\prime}}

\definecolor{deepgreen}{rgb}{0.2,0.8,0.2}

\colorlet{linkcolor}{BrickRed}

\hypersetup{colorlinks=true,
linkcolor=linkcolor,
citecolor=linkcolor,
urlcolor=linkcolor,
,linktocpage=true
,pdfproducer=medialab}

\defcitealias{Caputo:2020bdy}{Paper~I}

\DeclareSIUnit \h {\ensuremath{\mathit{h}}}
\DeclareSIUnit\electronvolt{e\kern-.05em V}
\DeclareSIUnit\parsec{pc}

\begin{document}

\title{Modeling Dark Photon Oscillations in Our Inhomogeneous Universe}

\author{Andrea Caputo}
\email{andrea.caputo@uv.es}
\thanks{ORCID: \href{https://orcid.org/0000-0003-1122-6606}{0000-0003-1122-6606}}
\affiliation{Instituto de F\'{i}sica Corpuscular, CSIC-Universitat de Valencia, Apartado de Correos 22085, E-46071, Spain}

\author{Hongwan Liu}
\email{hongwanl@princeton.edu}
\thanks{ORCID: \href{https://orcid.org/0000-0003-2486-0681}{0000-0003-2486-0681}}
\affiliation{Center for Cosmology and Particle Physics, Department of Physics, New York University, New York, NY 10003, USA}
\affiliation{Department of Physics, Princeton University, Princeton, NJ 08544, USA}

\author{Siddharth Mishra-Sharma}
\email{sm8383@nyu.edu}
\thanks{ORCID: \href{https://orcid.org/0000-0001-9088-7845}{0000-0001-9088-7845}}
\affiliation{Center for Cosmology and Particle Physics, Department of Physics, New York University, New York, NY 10003, USA}

\author{Joshua T. Ruderman}
\email{ruderman@nyu.edu}
\thanks{ORCID: \href{https://orcid.org/0000-0001-6051-9216}{0000-0001-6051-9216}}
\affiliation{Center for Cosmology and Particle Physics, Department of Physics, New York University, New York, NY 10003, USA}

\date{\today}

\begin{abstract}
A dark photon may kinetically mix with the Standard Model photon, leading to observable cosmological signatures. The mixing is resonantly enhanced when the dark photon mass matches the primordial plasma frequency, which depends sensitively on the underlying spatial distribution of electrons. Crucially, inhomogeneities in this distribution can have a significant impact on the nature of resonant conversions. We develop and describe, for the first time, a general analytic formalism to treat resonant oscillations in the presence of inhomogeneities. Our formalism follows from the theory of level crossings of random fields and only requires knowledge of the one-point probability distribution function (PDF) of the underlying electron number density fluctuations. We validate our formalism using simulations and illustrate the photon-to-dark photon conversion probability for several different choices of PDFs that are used to characterize the low-redshift Universe. \githubmaster
\end{abstract}

\maketitle

\section{Introduction}

A dark photon $A'$ which kinetically mixes with the Standard Model (SM) photon, $\gamma$, is one of the simplest extensions of the SM~\cite{Holdom:1985ag}. 
The range of possible $A'$ masses $m_{A'}$ spans many orders of magnitude, and an intense theoretical and experimental program is ongoing to constrain and test dark photon models. 
At low masses ($m_{A'} \lesssim \SI{e-9}{\eV}$), the Compton wavelength of $m_{A'}$ starts to exceed the size of typical experiments, and terrestrial probes start to become increasingly insensitive to the presence of $A'$, motivating probes on larger length scales.
Light dark photons in this mass range are also a well-motivated candidate for dark matter~\cite{Redondo:2008ec,Nelson:2011sf,Arias:2012az,Fradette:2014sza,An:2014twa,Graham:2015rva,Agrawal:2018vin,Dror:2018pdh,Co:2018lka,Bastero-Gil:2018uel,Long:2019lwl}, while relativistic $A'$ produced by decaying dark matter which then resonantly convert into $\gamma$ has also been proposed as a new-physics explanation~\cite{Pospelov:2018kdh,Choi:2019jwx} and can be detected by 21-cm observations. Probes of the dark photon over cosmological scales are therefore critical to constraining its properties.

Existing experimental measurements are sensitive to oscillations of cosmic microwave background (CMB) photons into dark photons, $\gamma \to A'$, or to oscillations of dark photon dark matter into low-energy photons, $A' \to \gamma$. 
The probability of these conversions at a particular redshift $z$ and position in space $\vec{x}$ depends on the photon plasma mass at that point, $m_\gamma(z,\vec{x})$, and becomes resonantly enhanced whenever it becomes equal to $m_{A'}$.  
$\gamma \to A'$ conversions can leave a distortion in the energy spectrum of the CMB due to a disappearance of photons from the spectrum, while $A' \to \gamma$ conversions for low mass dark photons produce SM photons that are readily absorbed by baryons and electrons, resulting in an increase of the intergalactic medium (IGM) temperature.

Under the assumption of a completely homogeneous Universe, constraints on the kinetic mixing parameter $\epsilon$ for the case of $\gamma \to A'$ were obtained using the COBE/FIRAS~\cite{Fixsen:1996nj} measurement of the CMB energy spectrum, which shows no significant evidence of distortion from a pure blackbody spectrum~\cite{Mirizzi:2009iz,Kunze:2015noa}. More recently, Ref.~\cite{McDermott:2019lch} presented new homogeneous constraints for $A' \to \gamma$ in the case of dark photon dark matter, finding strong limits on $\epsilon$ using IGM temperature measurements during HeII reionization, among other novel cosmological constraints.    

This paper is part of a pair of companion papers with the overarching goal of establishing a new formalism for understanding both the physics and the experimental consequences of $\gamma \to A'$ and $A' \to \gamma$ oscillations in our inhomogeneous Universe. In Ref.~\cite{Caputo:2020bdy}, hereafter referred to as~\citetalias{Caputo:2020bdy}, we briefly introduce our formalism and present \emph{(i)} the $\gamma \to A'$ CMB spectral distortion and \emph{(ii)}~$A' \to \gamma$ dark photon dark matter IGM temperature constraints on the kinetic mixing parameter $\epsilon$. We find that limits derived under the assumption of a homogeneous photon plasma were not conservative, and that including inhomogeneities allows for constraints to be set over a much broader mass range of $A'$. In this paper, we provide a detailed description of the formalism and its mathematical derivation, as well as an elaboration on the cosmological inputs that go into the $A'$ limits obtained in~\citetalias{Caputo:2020bdy}. 

This paper is organized as follows. 
We begin Sec.~\ref{sec:oscillations} with a quantum mechanical derivation of the oscillation probability of $\gamma \leftrightarrow A'$ for both relativistic and nonrelativistic $A'$ with multiple resonance crossings. 
We then introduce our analytic formalism for computing the expected probability of conversion for both $\gamma \to A'$ and $A' \to \gamma$, taking as input the one-point probability density function (PDF) of baryon inhomogeneities in our Universe, described in Sec.~\ref{Sec:Formalism}. 
In Sec.~\ref{sec:linear_regime}, we explore our formalism in the regime where fluctuations are Gaussian to gain some analytic understanding. 
We then move on to describe the two main cosmological inputs that are needed to apply our results to our Universe: the one-point probability density functions of baryon fluctuations in Sec.~\ref{sec:PDFs}, and the variance of fluctuations (characterized by power spectra for the number density fluctuation of baryons and free electrons), in Sec.~\ref{sec:variance_of_fluctuations}. 
We validate our formalism against several simulations of baryon fluctuations and $\gamma \leftrightarrow A'$ conversions, which we describe in Sec.~\ref{sec:simulations}. 
Some results for the $\gamma \leftrightarrow A'$ conversion probability obtained from our analytic formalism for various cosmological inputs are presented in Sec.~\ref{sec:results}. 
We finally conclude in Sec.~\ref{sec:conclusion}. 
In our appendices, we provide a comparison between our work and several recent papers treating inhomogeneities~\cite{Bondarenko:2020moh,Garcia:2020qrp,Witte:2020rvb}, along with other details of the formalism.

Throughout this work, we use natural units with $\hbar = c = k_\text{B} = 1$, as well as the \emph{Planck} 2018 cosmology~\cite{Aghanim:2019ame}. In the spirit of reproducibility, we provide links in the figure captions (\nbicon) pointing to the Jupyter notebooks used to generate them. 

\section{Oscillations}
\label{sec:oscillations}

$\gamma \leftrightarrow A'$ oscillations are described by the same formalism as neutrino flavor oscillations, which have been studied extensively in the literature. 
In this section, we follow the neutrino discussion of Ref.~\cite{Dasgupta:2005wn} closely, first reviewing $\gamma \to A'$ oscillations and highlighting any differences between $\gamma \leftrightarrow A'$ and neutrino oscillations whenever they arise. 
$A' \to \gamma$ oscillations are similar, and are discussed at the end of this section.

\begin{figure}
    \centering
    \includegraphics[width=0.45\textwidth]{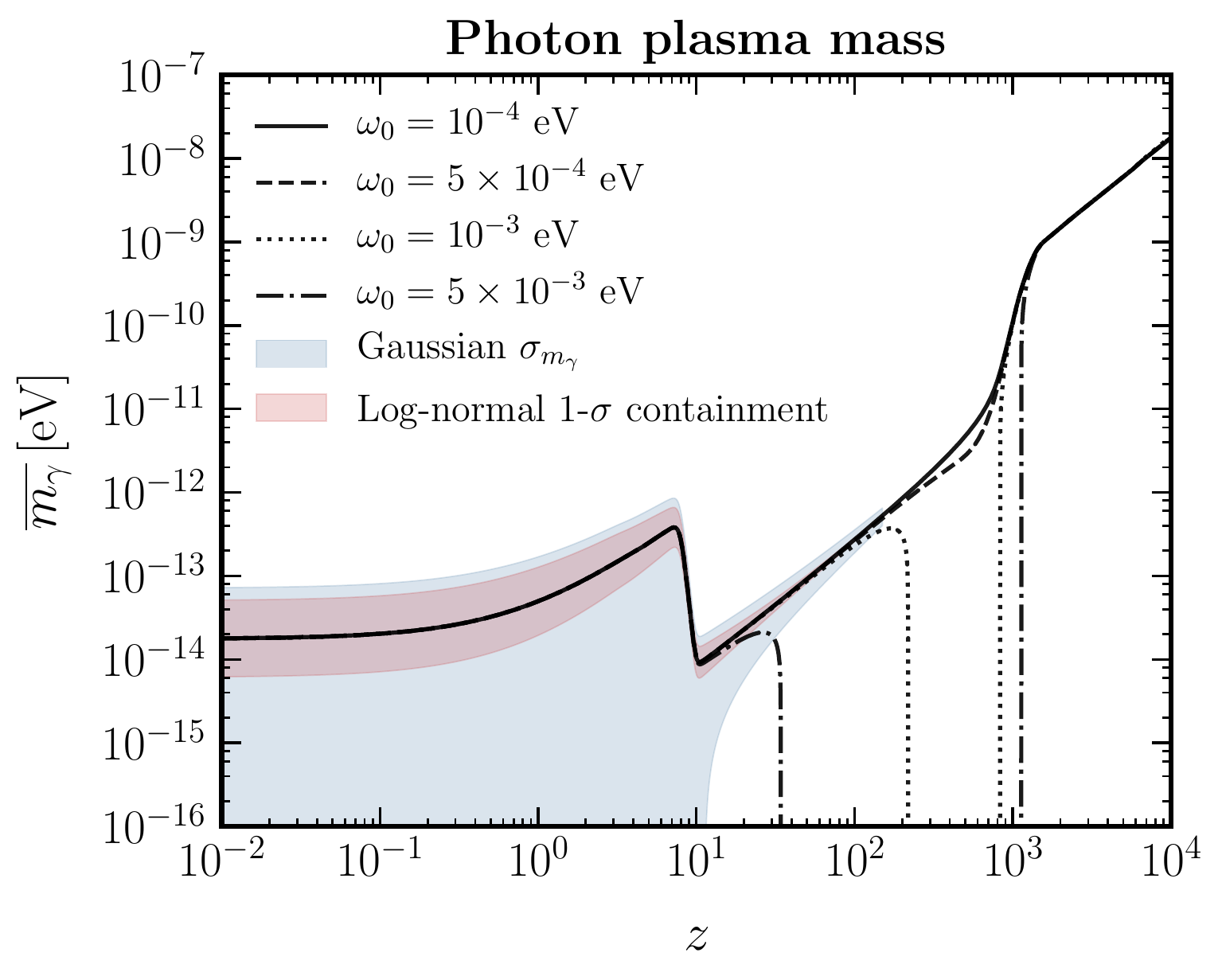}
    \caption{The photon plasma mass as a function of redshift for several values of the present-day photon energy $\omega_0$. The Gaussian standard deviation of plasma mass fluctuations $\sigma_{m_\gamma}$, informed by the linear baryon power spectrum for illustration, is shown as the blue band. The equivalent middle-68\% containment of fluctuations assuming a log-normal description of the PDF is shown as the red band.~\nblink{18_plasma_mass_plot_formalism}}
    \label{fig:m_Ap_single_LN}
\end{figure}

Consider a single photon passing through some worldline from the early Universe to us.
Along this path, parametrized by $t$, there are variations in the number densities of free electrons and neutral atoms, leading to variations in the plasma properties, giving rise to a plasma mass $m_\gamma(t)$~\cite{Mirizzi:2009iz}:
\begin{alignat}{2}
    m_\gamma^2(t) &\simeq&& \,\, \frac{4 \pi \alpha_\text{EM} n_\text{e}(t)}{m_\text{e}} - 2 \omega^2(t) \left(\mathfrak{n}_\text{HI}(t) - 1 \right) \nonumber \\
    &\simeq&& \,\, \SI{1.4e-21}{\eV\squared} \left(\frac{n_\mathrm{e}(t)}{\SI{}{\per\centi\meter\cubed}}\right)  \nonumber \\
    & &&- \SI{8.4e-24}{\eV\squared} \left( \frac{\omega(t)}{\SI{}{\eV}} \right)^2 \left(\frac{n_{\mathrm{HI}}(t)}{\SI{}{\per\centi\meter\cubed}}\right)  \,.
    \label{eq:m_gamma_sq}
\end{alignat}
Here, $\alpha_\text{EM}$ is the electromagnetic fine structure constant, $m_\text{e}$ is the electron mass, $\mathfrak{n}_\text{HI}$ is the refractive index of monatomic hydrogen~\cite{pauling1985introduction}, $\omega(t)$ is the photon energy, and $n_\text{e}(t)$ and $n_\text{HI}(t)$ are the local free electron and neutral hydrogen densities along the path.\footnote{Our expression clarifies the actual species densities that enter the plasma mass expression in Ref.~\cite{Mirizzi:2009iz}, and corrects earlier expressions for the photon mass, which mistakenly used the refractive index of diatomic hydrogen gas and not monatomic hydrogen.}
We similarly define $\overline{m_\gamma^2}$ as the homogeneous value of $m_\gamma^2$, evaluated with the mean cosmological values of $n_\text{e}$ and $n_\text{HI}$. 
We neglect helium, which makes up only 8\% by number density and has a smaller index of refraction. 
If fluctuations in free electron density $x_\text{e}$ are small, \emph{i.e.},\ $x_\text{e}$ has essentially the same value everywhere in space at each point in time, then fluctuations in $n_\text{e}$ and $n_\text{HI}$ track fluctuations in the number density of baryons, $n_\text{b}$,
\begin{alignat}{1}
    \frac{m_\gamma^2(t)}{\overline{m_\gamma^2}(t)} = \frac{n_\text{b}}{\overline{n}_\text{b}} \,.
    \label{eq:m_gamma_sq_propto_n_b}
\end{alignat}
Further discussion of this proportionality and the effect of fluctuations in $x_\text{e}$ can be found in Sec.~\ref{sec:variance_of_fluctuations}. Figure~\ref{fig:m_Ap_single_LN} illustrates the variation of the photon plasma mass as a function of redshift for several representative values of the present-day photon frequency $\omega_0$.

The kinetic mixing between $\gamma$ and $A'$ induces oscillations between these two interaction eigenstates, described by the Schr\"{o}dinger equation
\begin{alignat}{1}
    i \frac{\dd}{\dd t} \begin{pmatrix}
        \gamma \\ A'
    \end{pmatrix} = \mathsf{H} \begin{pmatrix}
        \gamma \\ A'
    \end{pmatrix} \,,
\end{alignat}
where $\mathsf{H}$ is the Hamiltonian (assuming all particles are relativistic)~\cite{Dasgupta:2005wn}
\begin{alignat}{1}
    \mathsf{H} = \frac{1}{4\omega(t)} \begin{pmatrix}
        m_\gamma^2(t) - \mAp^2 & 2\epsilon \mAp^2 \\
        2 \epsilon \mAp^2 & - m_\gamma^2(t) + \mAp^2
    \end{pmatrix} \,.
    \label{eq:hamiltonian}
\end{alignat}
The plasma mass is an in-medium effect similar to the Mikheyev-Smirnov-Wolfenstein (MSW) effect in the case of neutrino oscillations~\cite{Wolfenstein:1977ue,Mikheev:1986gs}, leading in our case to the familiar correction of $m_\gamma^2/2\omega$ relative to the propagation phase of a massless particle~\cite{Hook:2018iia,Battye:2019aco}.
$\mathsf{H}$ can be conveniently written in terms of Pauli matrices,
\begin{alignat}{1}
    \mathsf{H} = \phi(t) \sigma_3 + \eta(t) \sigma_1 \,,
\end{alignat}
where
\begin{alignat}{1}
    \phi(t) \equiv \frac{m_\gamma^2(t) - m_{A'}^2}{4 \omega(t)}\,, \quad \eta(t) = \frac{\epsilon m_{A'}^2}{2\omega(t)} \,.
\end{alignat}
$\phi(t)$ has the intuitive interpretation of being half the relative phase between $\gamma$ and $A'$. 

Starting with an initial state of $\gamma$, the Schr\"{o}dinger equation can be solved perturbatively in $\epsilon$, with $\eta(t)\sigma_1$ as an interaction Hamiltonian. 
To first order in $\epsilon$, we obtain
\begin{alignat}{1}
    A'(t) = -i e^{i\alpha(t)} \int_0^t \dd \xi \, \eta(\xi) e^{-2i \alpha(\xi)} + \mathcal{O}(\epsilon^2) \,,
\end{alignat}
where we have defined
\begin{alignat}{1}
    \alpha(s) \equiv \int_0^s \dd \xi \, \phi(\xi) \,,
\end{alignat}
the accumulated phase between 0 and $s$.
This leads to the probability of disappearance at $t$, given by $|A'(t)|^2$, or explicitly,
\begin{alignat}{1}
    P_{\gamma \to A'}(t) = \left| \int_0^t \dd \xi \, \eta(\xi) e^{-2i \alpha(\xi)} \right|^2 + \mathcal{O}(\epsilon^3)\,.
    \label{eq:raw_conv_prob}
\end{alignat}

Away from regions of space where $m_\gamma^2(t) \sim m_{A'}^2$, $\phi(t)$ is given parametrically by
\begin{alignat}{1}
    \phi(t) \sim \frac{\SI{200}{\per\kilo\parsec}}{1+z(t)} \left(\frac{\big|\overline{m_\gamma^2}(t) - m_{A'}^2 \big|}{\SI{e-26}{\eV^2}}\right) \left(\frac{10}{\omega_0/T_{\text{CMB},0}}\right) \,,
    \label{eq:phi_parametric}
\end{alignat}
where $T_{\text{CMB},0}$ is the temperature of the CMB today, and $\omega(t) = \omega_0(1+z(t))$, with $z(t)$ being the cosmological redshift at $t$. 
The FIRAS experiment detects photons in the range $1.2 \lesssim \omega_0 / T_{\text{CMB},0} \lesssim 11.3$~\cite{Fixsen:1996nj}.
Over cosmological distances, the integral of $\phi(t)$ therefore oscillates rapidly with $t$, except when $m_\gamma^2(t) \sim m_{A'}^2$; we can therefore evaluate the integral in Eq.~\eqref{eq:raw_conv_prob} over the entire worldline of the photon using the stationary phase approximation, giving
\begin{alignat}{1}
    P_{\gamma \to A'} = \pi \left|\sum_i \frac{\eta(t_i)}{\sqrt{|\phi'(t_i)|}} e^{-2i \alpha(t_i)} e^{i\beta_i} \right|^2 + \mathcal{O}(\epsilon^3) \,,
\end{alignat}
where $i$ indexes positions $t_i$ where $m_\gamma^2(t_i) = m_{A'}^2$, $\phi'$ is the derivative of $\phi$, and $\beta_i = \pm \pi / 4$, with the sign given by the sign of $\phi'(t_i)$.

If there is only one point $t_r$ where $m_\gamma^2(\xi) = m_{A'}^2$, then oscillations from $\gamma$ to $A'$ occur resonantly at $t_r$, giving
\begin{alignat}{1}
    P_{\gamma \to A'} \simeq \frac{\pi \eta(t_r)^2}{|\phi'(t_r)|} = \frac{\pi \epsilon^2 m_{A'}^2}{\omega(t_r)} \left| \frac{\dd \ln m_\gamma^2(t)}{\dd t} \right|^{-1}_{t = t_r} \!\!\! .
\end{alignat}
This is the same expression derived using the Landau-Zener approximation for non-adiabatic transitions of $\gamma \to A'$ in Ref.~\cite{Mirizzi:2009iz}; it is also similar to expressions for the $A'$ production rate in stars under the narrow width approximation~\cite{Hardy:2016kme,Redondo:2013lna}. 
The stationary phase approximation has also been used to calculate appearance and disappearance probabilities with resonant oscillations in the context of neutrino oscillations~\cite{Dasgupta:2005wn} and in axion-photon conversions in magnetic fields~\cite{Raffelt:1987im,Hook:2018iia,Battye:2019aco}. 

When multiple resonances exist, the probability becomes
\begin{alignat}{1}
    P_{\gamma \to A'} \simeq \sum_{i} \frac{\pi \eta(t_i)^2}{|\phi'(t_i)|} + \sum_{i < j} \frac{2 \pi \eta(t_i) \eta(t_j) \cos \theta(t_i,t_j)}{\sqrt{|\phi'(t_i)|}\sqrt{|\phi'(t_j)|}} \,,
    \label{eq:prob_with_cross_terms}
\end{alignat}
where $\theta(t_i, t_j) \equiv2\alpha(t_j) - 2\alpha(t_i) + \beta_i - \beta_j$. 
The second summation in Eq.~\eqref{eq:prob_with_cross_terms} corresponds to quantum interference from resonance conversion of $\gamma \to A'$ between two resonance points~\cite{Dasgupta:2005wn}.\footnote{Corrections to the probability due to multiple conversions, \emph{e.g.},\ $\gamma \to A' \to \gamma$, which is treated classically in Ref.~\cite{Mirizzi:2009iz}, only appear at higher order in $\epsilon$.} 
Interference from conversions along a trajectory due to stochastic matter fluctuations can be important in understanding neutrino oscillations within supernovae~\cite{Dasgupta:2005wn,Fogli:2006xy,Friedland:2006ta}.
To assess the importance of this for cosmological $\gamma \to A'$ oscillations, we note two things. First, the minimum comoving size of baryonic fluctuations is given by the Jeans length, which we discuss in more detail in Sec.~\ref{sec:linear_regime}. 
We can estimate this minimum expected value by setting $T_\text{b}$ to its expected value without reionization effects at $z = 20$, giving us $R_\text{J,min} \sim \SI{10}{\kilo\parsec}$. 
Second, the size of fluctuations in $m_\gamma^2$ and hence $\phi$ is determined by the standard deviation of baryon density fluctuations $\sigma_\text{b}$ (see Fig.~\ref{fig:sigma_lin_baryon} for some typical values of these fluctuations); in other words, we expect that between resonances when $m_\gamma^2 = m_{A'}^2$, fluctuations in $m_\gamma^2$ can typically reach values of around $(1 \pm \sigma_\text{b})m_{A'}^2$.
These two estimates and Eq.~\eqref{eq:phi_parametric} show that the phase difference between consecutive resonances is roughly
\begin{multline}
    \theta(t_i, t_{i+1}) \gtrsim \frac{4 \times 10^3}{(1 + z_\text{h})^2} \left(\frac{R_\text{J}}{\SI{10}{\kilo\parsec}}\right) \min \left[ \sigma_\text{b}(z_\text{h}), 1 \right]  \\
    \times \left(\frac{m_{A'}}{\SI{e-13}{\eV}}\right)^2 \left(\frac{10}{\omega_0/T_{\text{CMB},0}}\right) \,,
\end{multline}
where $z_\text{h}$ is the lowest redshift at which $\overline{m_\gamma^2} = m_{A'}^2$, since all of the resonances occur in a redshift window centered at redshifts where $\overline{m_\gamma^2} = m_{A'}^2$, and transitions are more adiabatic at lower redshifts. 
The relative phase between $A'$s produced at any two resonance points is thus many times larger than $2\pi$ throughout the history of the Universe, so that $\cos \theta(t_i,t_j)$ is expected to be uncorrelated with the location of the resonances. 
We will ultimately be interested in the mean value of $P_{\gamma \to A'}$ across all possible photon worldlines, such that uncorrelated interference effects average out.
We therefore do not expect the second summation in Eq.~\eqref{eq:prob_with_cross_terms} to contribute to the overall probability of conversion, obtained by averaging over all worldlines, each with a different distribution of resonance points.
The total probability of oscillations is thus obtained by summing up the conversion probability of each resonance, each given by the Landau-Zener expression:
\begin{alignat}{1}
    P_{\gamma \to A'} \simeq \sum_i \frac{\pi \epsilon^2 m_{A'}^2}{\omega(t_i)} \left| \frac{\dd \ln m_\gamma^2(t)}{\dd t} \right|^{-1}_{t = t_i} \!\!\! . 
    \label{eq:prob_gamma_to_Ap}
\end{alignat}
We note that the Landau-Zener approximation holds for any crossing encountered by the photon. For $\epsilon \ll 1$, since the resonance timescale, $\tau_\text{res} \sim \epsilon |\dd \ln m_\gamma^2 / \dd t|^{-1}$, is much smaller than the timescale over which $m_\gamma^2$ changes, $|\dd \ln m_\gamma^2 / \dd t|^{-1}$, at any crossing, allowing the use of the Landau-Zener approximation of taking the density profile over the resonance to be linear. The suitability of the Landau-Zener approximation in the context of neutrino oscillations, starting from a similar Hamiltonian to Eq.~\eqref{eq:hamiltonian}, is derived in Ref.~\cite{Petcov:1987xd}. 

Following a similar derivation, we can show that relativistic dark photons undergoing $A' \to \gamma$ conversions will also have a conversion probability that is identical to Eq.~\eqref{eq:prob_gamma_to_Ap}, with $\omega$ now specifying the $A'$ energy. 
If $A'$ is the dark matter, however, the assumption of relativistic particles assumed in Eq.~\eqref{eq:hamiltonian} breaks down. 
Nevertheless, there are several ways to see that the conversion probability $P_{A'\to \gamma}$ is identical to $P_{\gamma \to A'}$ with $\omega(t_i) \to m_{A'}$. 
First, it can be derived in thermal field theory~\cite{Arias:2012az} by applying a narrow-width approximation (see App.~\ref{app:thermal}). 
Second, the probability of conversion $P_{A' \to \gamma}$, shown on the right-hand side of Eq.~\eqref{eq:prob_gamma_to_Ap}, is Lorentz invariant, as all transition probabilities should be. 
Evaluating the probability in the rest frame of the dark matter $A'$ gives
\begin{alignat}{1}
    P_{A' \to \gamma} \simeq \sum_i \pi \epsilon^2 \mAp \left| \frac{\dd \ln m_\gamma^2(t)}{\dd t} \right|^{-1}_{t=t_i} \!\!\!,
    \label{eq:prob_Ap_to_gamma}
\end{alignat}
consistent with the result in Ref.~\cite{Arias:2012az}. 
Under standard cosmology scenarios where the magnitude of $\delta_\text{b}$ grows monotonically with redshift, each value of $\mAp$ has at most one resonance transition point; our formalism, however, does not rely on this assumption. 

The results in Eqs.~\eqref{eq:prob_gamma_to_Ap} and~\eqref{eq:prob_Ap_to_gamma} form the starting point for understanding $\gamma \leftrightarrow A'$ conversions along a single worldline, as well as for all of the results presented in~\citetalias{Caputo:2020bdy}.

\section{Formalism}
\label{Sec:Formalism}

In the presence of inhomogeneities, the resonance condition can be met many times along a path, even at times when the homogeneous plasma mass $\overline{m}_\gamma$ is far from $\mAp$ and no resonance is present in the homogeneous limit. 
Each worldline passes through a different series of perturbations, leading to conversions that vary significantly in number and in distance from the observer. 

\subsection{\texorpdfstring{$\gamma \to A'$ oscillations}{Photon-dark photon oscillations}}

We will now discuss how to determine the expected probability of $\gamma \to A'$ conversion, $\langle P_{\gamma \to A'} \rangle$. The derivation of our results is closely related to the derivation of the mean number of times a stationary process crosses a fixed level per unit time~\cite{rice1944mathematical,lindgren2013stationary}. 

To average over all worldlines, we first begin by rewriting the probability of conversion along a worldline as
\begin{alignat}{1}
    \frac{ \dd P_{\gamma \to A'}}{\dd t} = \frac{\pi \mAp^2 \epsilon^2}{\omega(t)} \delta_\text{D}(m_\gamma^2(t) - \mAp^2) \, m_\gamma^2(t) \,,
\end{alignat}
where $\delta_\text{D}$ is the Dirac delta function.
We can check that Eq.~\eqref{eq:prob_gamma_to_Ap} is recovered by performing the substitution
\begin{alignat}{1}
    \dd t = \left| \frac{\dd \ln m_\gamma^2}{\dd t} \right|^{-1} \!\! \frac{\dd m_\gamma^2}{m_\gamma^2}
    \label{eq:t_to_m_gamma_sq_sub}
\end{alignat}
and integrating the delta function over the entire worldline. 
The mean value of $P_{\gamma \to A'}$ is then obtained by integrating over all possible values of $m_\gamma^2$ at each point along the path, weighted by the probability density function (PDF) $f(m_\gamma^2;t)$ of $m_\gamma^2$:
\begin{multline}
    \frac{\dd \langle P_{\gamma \to A'} \rangle}{\dd z} = \frac{\pi \mAp^2 \epsilon^2}{\omega(t)} \left| \frac{\dd t}{\dd z} \right| \\
    \times \int \dd m_\gamma^2 \, f(m_\gamma^2;t) \, \delta_\text{D}(m_\gamma^2 - m_{A'}^2) \, m_\gamma^2 \,.
    \label{eq:dP_dz}
\end{multline}
Note that the PDF evolves with time since $m_\gamma^2$ tracks the baryon density (in the limit of small fluctuations in the free electron fraction), as shown in Eq.~\eqref{eq:m_gamma_sq_propto_n_b}. 
We can now perform the integral to give
\begin{alignat}{1}
    \frac{\dd \langle P_{\gamma \to A'} \rangle}{\dd z} = \frac{\pi \mAp^4 \epsilon^2}{\omega(t)} \left| \frac{\dd t}{\dd z} \right| f(m_\gamma^2 = \mAp^2; t) \,.
    \label{eq:dP_dz_simplified}
\end{alignat}
The problem of determining the averaged probability therefore reduces to finding the PDF of $m_\gamma^2$, which we discuss in detail in subsequent sections. Note that Eqs.~\eqref{eq:dP_dz} and~\eqref{eq:dP_dz_simplified} both apply equally to relativistic $A' \to \gamma$ oscillations as well. 

As an example, let us consider the homogeneous limit where $m_\gamma^2 = \overline{m_\gamma^2}$ everywhere; in this case, the PDF is trivially given by
\begin{alignat}{1}
    f_\text{h}(m_\gamma^2;t) = \delta_\text{D}(m_\gamma^2 - \overline{m_\gamma^2}(t)) \,.
    \label{eq:f_homo_limit}
\end{alignat}
We therefore see that the mean homogeneous conversion probability is
\begin{alignat}{1}
    \langle P_{\gamma \to A'} \rangle_\text{h} &= \int \dd t \, \frac{\pi m_{A'}^4 \epsilon^2}{\omega(t)} \delta_\text{D}(m_\gamma^2 - \overline{m_\gamma^2}(t)) \nonumber \\
    &= \sum_i \frac{\pi m_{A'}^2 \epsilon^2}{\omega(t_i)} \left| \frac{\dd \ln \overline{m_\gamma^2}(t)}{\dd t} \right|^{-1}_{t=t_i} \!\!\!,
\end{alignat}
where $i$ indexes times $t_i$ when $\overline{m_\gamma^2}(t_i) = \mAp^2$, and we have again made use of the substitution shown in Eq.~\eqref{eq:t_to_m_gamma_sq_sub}. This recovers the homogeneous limit expressions found in Eq.~\eqref{eq:prob_gamma_to_Ap} and Ref.~\cite{Mirizzi:2009iz}.

\subsection{\texorpdfstring{$A' \to \gamma$ oscillations}{Dark photon-photon oscillations}}

For $A' \to \gamma$ conversions with $A'$ dark matter, in the range of $\mAp$ of interest, the converted photons are absorbed quickly by electrons in the IGM via free-free absorption~\cite{McDermott:2019lch}, heating the IGM\@. 
The quantity of interest is therefore the average energy injected into the plasma per baryon, $\langle E_{A' \to \gamma} \rangle$. The derivation of $\langle E_{A' \to \gamma}\rangle$ proceeds in a similar fashion, except that the energy injected per volume along the worldline is given by $P_{A' \to \gamma}(t) \rho_{A'}(t)$, where $\rho_{A'}(t)$ is the mass density of $A'$ dark matter at the point on the worldline $t$. 
The rate of energy injected per baryon along the worldline of the massive dark photon is therefore
\begin{alignat}{1}
    \frac{\dd E_{A' \to \gamma}}{\dd t} = \pi m_{A'} \epsilon^2 \frac{\overline{\rho}_{A'}}{\overline{n}_\text{b}} \frac{\rho_{A'}(t)}{\overline{\rho}_{A'}(t)} \, \delta_\text{D}(m_\gamma^2(t) - \mAp^2)\, m_\gamma^2(t) \,,
\end{alignat}
where $\overline{n}_\text{b}$ is the homogeneous baryon number density, with $\overline{\rho}_{A'}/\overline{n}_\text{b}$ being a time-independent quantity. 

To obtain the mean value, we technically need to perform an integral over the joint distribution of both $m_\gamma^2$ and $\rho_{A'}$. However, two points make this unnecessary.
First, if fluctuations in the free electron fraction are small, then as we argued in Eq.~\eqref{eq:m_gamma_sq_propto_n_b}, $m_\gamma^2 \propto n_\text{b}$. 
This assumption is true during the period of HeII reionization, the regime we study in~\citetalias{Caputo:2020bdy} to obtain limits on $\epsilon$ in the case of $A'$ dark matter, since the Universe is almost completely ionized at this time except for HeII, while fluctuations in baryon density are large compared to the mean. 
Second, we adopt the standard assumption that baryon density fluctuations track matter density fluctuations $\rho_\text{m}$ with a bias $b \sim \mathcal{O}(1)$. 
This means that
\begin{alignat}{1}
    \frac{\rho_\text{m}}{\overline{\rho}_\text{m}(t)} = \frac{1}{b} \frac{n_\text{b}}{\overline{n}_\text{b}} = \frac{1}{b} \frac{m_\gamma^2}{\overline{m_\gamma^2}(t)} \,,
    \label{eq:rho_n_m_gamma_relation}
\end{alignat}
where in the case of $A'$ dark matter, $\rho_\text{m} \simeq \rho_{A'}$. Note that in~\citetalias{Caputo:2020bdy}, we assumed $b = 1$ for simplicity, although including a small bias consistent with values reported in Ref.~\cite{Hurtado-Gil:2017dbm} does not change the result significantly. With this relation, we find
\begin{multline}
    \frac{\dd \langle E_{A' \to \gamma} \rangle}{\dd z} = \pi \mAp \epsilon^2 \frac{\overline{\rho}_{A'}}{b\,\overline{n}_\text{b}} \left| \frac{\dd t}{\dd z} \right| \\
    \times \int \dd m_\gamma^2 \, \frac{m_\gamma^2}{\overline{m_\gamma^2}(t)} f(m_\gamma^2;t) \, \delta_\text{D}(m_\gamma^2 - \mAp^2)\, m_\gamma^2 \,,
    \label{eq:dE_dz_general}
\end{multline}
and as before we can perform the integral to obtain
\begin{alignat}{1}
    \frac{\dd \langle E_{A' \to \gamma} \rangle}{\dd z} = \frac{\pi \mAp^5 \epsilon^2}{\overline{m_\gamma^2}(t)} \frac{\overline{\rho}_{A'}}{b\,\overline{n}_\text{b}} \left| \frac{\dd t}{\dd z} \right| f(m_\gamma^2 = \mAp^2; t) \,.
\end{alignat}
This treatment implicitly assumes that the conversion probability of $A' \to \gamma$ is small, which is required if $A'$ is all of the dark matter. A more general treatment is possible by allowing $b(z)$ to vary as a function of the total conversion up to $z$. 

In deriving Eq.~\eqref{eq:dE_dz_general}, we have assumed that the energy deposited by the conversion is distributed uniformly across all baryons, enabling us to characterize the entire plasma with a single temperature. 
This is in contrast to the assumption made in Ref.~\cite{Witte:2020rvb}, where energy deposition is local. 
The corresponding expression under this assumption can be obtained by replacing $\overline{n}_\text{b} \to n_\text{b}$ inside the integral,
\begin{multline}
    \frac{\dd \langle E_{A' \to \gamma} \rangle_\text{local}}{\dd z} = \pi \mAp \epsilon^2 \frac{\overline{\rho}_{A'}}{b\,\overline{n}_\text{b}} \left| \frac{\dd t}{\dd z} \right| \\
    \times \int \dd m_\gamma^2 \, f(m_\gamma^2;t) \, \delta_\text{D}(m_\gamma^2 - \mAp^2)\, m_\gamma^2 \,,
    \label{eq:dE_dz_local}
\end{multline}
which we can integrate to obtain
\begin{alignat}{1}
    \frac{\dd \langle E_{A' \to \gamma} \rangle_\text{local}}{\dd z} = \pi m_{A'}^3 \epsilon^2 \frac{\overline{\rho}_{A'}}{b\,\overline{n}_\text{b}} \left| \frac{\dd t}{\dd z} \right| f(m_\gamma^2 = \mAp^2; t) \,.
    \label{eq:dE_dz_local_simple}
\end{alignat}
These results agree with the analogous expression in Ref.~\cite{Witte:2020rvb}. We leave a detailed comparison of our results to App.~\ref{app:comparison_with_previous_work}.

Eqs.~\eqref{eq:dP_dz} and~\eqref{eq:dE_dz_general} were presented in~\citetalias{Caputo:2020bdy}, and with several different choices of the PDF, $f(m_\gamma^2;t)$, were used to derive all of the relevant bounds on the existence on $A'$. 
The rest of the paper will now focus on determining the analytic form of $f(m_\gamma^2;t)$, and checking these results with simulation. 

\section{Understanding the Formalism}
\label{sec:linear_regime}

We are now in a position to evaluate Eqs.~\eqref{eq:dP_dz} and~\eqref{eq:dE_dz_general} numerically. 
To gain some intuition regarding our formalism and highlight some important physics, we begin our discussion assuming Gaussian fluctuations, a valid assumption at redshifts $z \gg 20$, where density perturbations are well described by linear perturbation theory. 
In this limit, $\langle P_{\gamma \to A'} \rangle$ and $\langle E_{A' \to \gamma} \rangle$ have analytic solutions, which serve as a useful pedagogical example for our full treatment.
We will first discuss the various inputs that determine $f(m_\gamma^2;t)$, before discussing the analytics of the result in the Gaussian regime. 

\subsection{PDF, variance of fluctuations and power spectrum}

We begin by taking the limit where we neglect fluctuations in the free electron fraction, as in Eq.~\eqref{eq:m_gamma_sq_propto_n_b}. The baryon density fluctuation $\delta_\text{b}(\vec{x})$ at each point in space is defined as
\begin{alignat}{1}
    \delta_\text{b}(\vec{x}) \equiv \frac{\rho_\text{b}(\vec{x}) - \overline{\rho}_\text{b}}{\overline{\rho}_\text{b}} \,,
\end{alignat}
where $\rho_\text{b}(\vec{x})$ is the baryon mass density at $\vec{x}$ and $\overline{\rho}_\text{b}$ is the mean, homogeneous baryon mass density. In the linear regime, the fluctuations follow a Gaussian distribution, given by the one-point PDF of baryon density fluctuations,
\begin{alignat}{1}
    \mathcal{P}_\text{G}(\delta_\text{b};z) = \frac{1}{\sqrt{2\pi \sigma_\text{b}^2(z)}} \exp \left(- \frac{\delta_\text{b}^2}{2 \sigma_\text{b}^2(z)}\right) \,,
    \label{eq:gaussian_pdf}
\end{alignat}
with the variance of the distribution $\sigma_\text{b}^2$ directly related to the baryon (auto) power spectrum, $P_\text{bb}(k,z)$ through
\begin{alignat}{1}
    \sigma_\text{b}^2(z) = \int \frac{\dd^3 \vec{k}}{(2\pi)^3} P_\text{bb}(k,z) \,.
    \label{eq:sigma_sq_def}
\end{alignat}
 In linear perturbation theory, $P_\text{bb}$ is the \textit{linear} baryon power spectrum, $P_\text{bb,L}(k,z)$. Fig.~\ref{fig:sigma_lin_baryon} shows $\sigma_\text{b}(z)$, computed using the value of $P_\text{bb,L}(k,z)$  produced by CLASS~\cite{Blas:2011rf}. With this function, we have fully specified the one-point PDF:
\begin{alignat}{1}
    f(m_\gamma^2;t) = \frac{\dd \delta_\text{b}}{\dd m_\gamma^2} \mathcal{P}(\delta_\text{b};t) = \frac{\mathcal{P}(\delta_\text{b}(m_\gamma^2); t)}{\overline{m_\gamma^2}(t)} \,,
    \label{eq:f_from_one_pt_PDF}
\end{alignat}
directly relating the PDF for $m_\gamma^2$ to a cosmological observable. We discuss the issue of perturbations in $x_\text{e}$ in Sec.~\ref{sec:variance_of_fluctuations}. The blue band in Fig.~\ref{fig:m_Ap_single_LN} shows the standard deviation of plasma mass fluctuations induced by baryon Gaussian fluctuations, for illustration.

\subsection{Jeans scale and sensitivity to small scales}
\label{sec:jeans_scale}

In linear perturbation theory, the linear \textit{matter} power spectrum $P_\text{mm,L}(k,z)$ scales as $k^{-3}$ at large $k$, so that the variance in \textit{matter} fluctuations, calculated using Eq.~\eqref{eq:sigma_sq_def} with $P_\text{mm,L}(k,z)$, theoretically exhibits a $\log k$ ultraviolet divergence. 
This divergence is regulated by the fact that measurements and simulations of matter density are always averaged over some smoothing scale $R$; $P_\text{mm,L}(k,z)$ needs to be convolved with a windowing function (\emph{e.g.},\ a top-hat function) with characteristic size $R$, giving a variance as a function of $R$. 
For \textit{baryons} in the linear regime, baryonic structures have the Jeans length as a physical cut-off scale: the formation of structures with comoving size less than $R_\text{J}$ is suppressed due to gas pressure counteracting the gravitational collapse, defined by
\begin{alignat}{1}
     R_\text{J}(z) = \frac{2 \sqrt{2} \pi}{\sqrt{3}} \frac{(1+z)}{H(z)} \sqrt{\frac{\gamma T_\text{b}(z)}{\mu m_\text{p}}} \,,
     \label{eq:jeans_length}
\end{alignat}
where $\gamma = 5/3$ is the adiabatic index for an ideal monatomic gas, $\mu = 1.22$ is the mean molecular weight of the neutral IGM, $m_\text{p}$ is the proton mass, $T_\text{b}$ is the baryon temperature, $c_s(z)$ is the baryon sound speed, and $H(z)$ is the Hubble parameter. 
Numerically, this is
\begin{alignat}{1}
    R_\text{J}(z) \sim \SI{1.4}{\mega\parsec} \left(\frac{1.0}{1+z}\right)^{1/2} \left(\frac{T_\text{b}}{\SI{e4}{\kelvin}}\right)^{1/2} \,,
    \label{eq:jeans_length_numeric}
\end{alignat}
with a minimum value of $R_\text{J,min} \sim \SI{e-2}{\mega\parsec}$ at $z \sim 20$ with $T_\text{b} \sim \SI{10}{\kelvin}$, before reionization heats baryons significantly. 
In terms of wavenumber, the Jeans length ensures that $P_\text{bb,L}(k,z)$ is suppressed above $k_\text{J} \sim 2 \pi / R_\text{J}$, which lies between $10^2$ and \SI{e3}{\per\mega\parsec} for $z \gtrsim 20$. 

\begin{figure}
    \centering
    \includegraphics[width=0.45\textwidth]{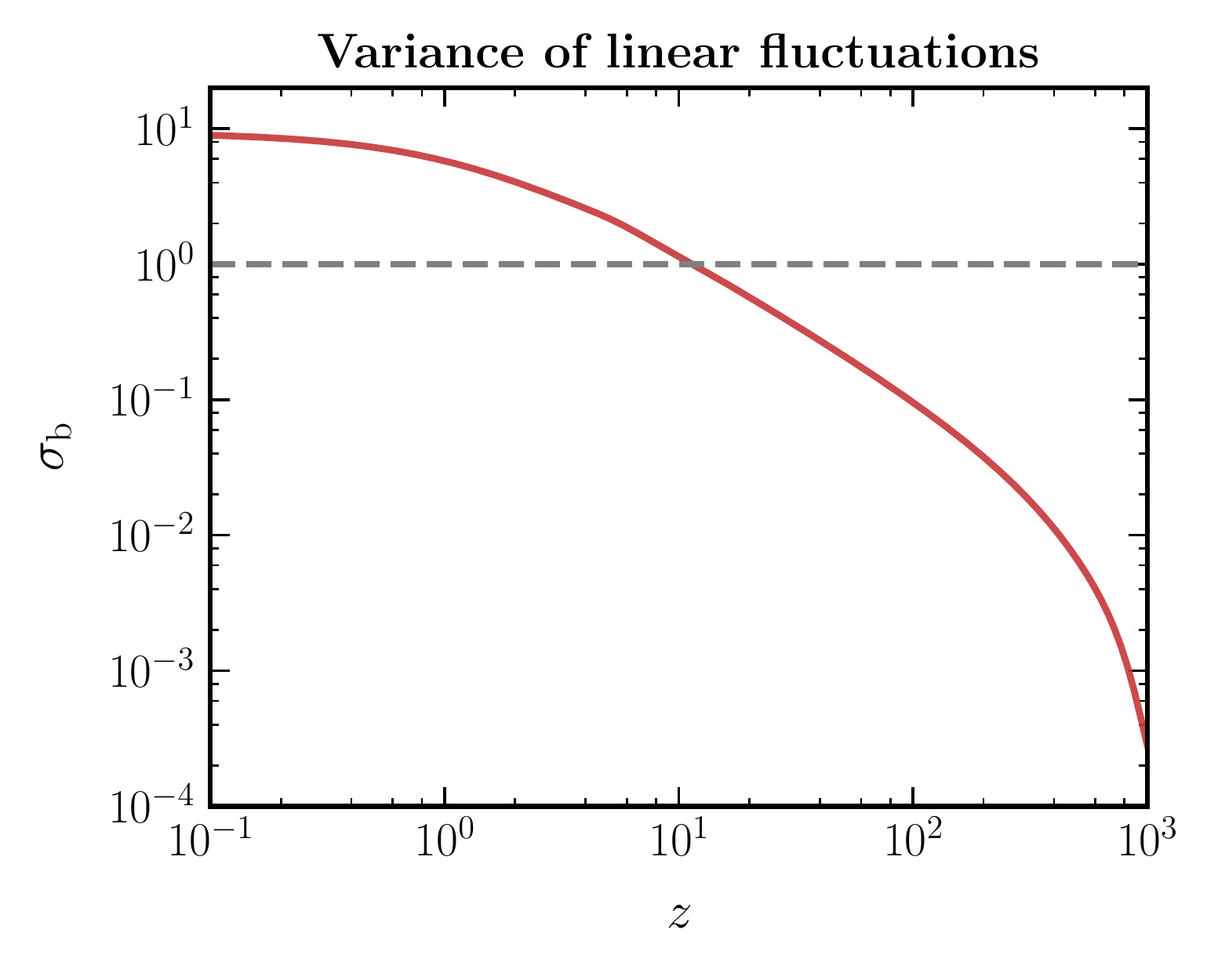}
    \caption{Standard deviation of baryon fluctuations $\sigma_\text{b}$ in linear perturbation theory (red). The dashed line indicates where the typical size of fluctuations becomes comparable to the mean density.~\nblink{12_formalism_plots_linear}}
    \label{fig:sigma_lin_baryon}
\end{figure}

Once reionization begins, Eq.~\eqref{eq:jeans_length_numeric} shows that $k_\text{J}$ decreases rapidly due to the increase in baryon temperature.  
Fluctuations also become increasingly nonlinear during this epoch.
On the other hand, Boltzmann codes like CLASS~\cite{Blas:2011rf} and CAMB~\cite{Lewis:2007zh} compute the linear baryon power spectrum $P_\text{bb,L}(k,z)$ with a suppression at $k_\text{J}$ \textit{without} reionization sources included when computing $T_\text{b}$, leading to a suppression scale of $k_\text{J} \sim \SI{700}{\h\per\mega\parsec}$, instead of $k_\text{J} \sim \SI{10}{\h\per\mega\parsec}$ as estimated from Eq.~\eqref{eq:jeans_length_numeric}.     
However, power above $k_\text{J} \sim \SI{10}{\h\per\mega\parsec}$ is actually unsuppressed due to the increasingly nonlinear behavior of baryons at late times; this lack of suppression is confirmed by baryon power spectra extracted from high-resolution hydrodynamic $N$-body simulations with baryonic physics included~\cite{vanDaalen:2019pst}. 
In light of this, we continue to adopt the linear power spectrum computed by CLASS for $P_\text{bb,L}$ with power suppressed above roughly $\SI{700}{\h\per\mega\parsec}$, and defer a complete discussion of this to Sec.~\ref{sec:variance_of_fluctuations}. 
We will also refer to the Jeans scale and corresponding Jeans length as the value of $k$ at which the linear power spectrum of CLASS shows a suppression of power relative to the matter power spectrum, instead of Eq.~\eqref{eq:jeans_length}. 

Since the baryon power spectrum $P_\text{bb,L}$  like $P_\text{mm,L}$ also scales as approximately $k^{-3}$ at large $k$ up to $k_\text{J}$, and nonlinear effects usually lead to the baryon power spectrum $P_\text{bb}$ exceeding $P_\text{bb,L}$ at large $k$, the value of $\sigma_\text{b}^2$ and hence the probability of conversion is sensitive to the smallest unsuppressed length scales in $P_\text{bb}(z)$. 
This exhibits one of the key peculiarities of dark photon oscillations in the presence of inhomogeneities: \textit{the resulting physics is sensitive to small-scale perturbations, depending on the details of the baryon power spectrum at scales as small as \SI{e3}{\per\mega\parsec}}, providing a rare example of a cosmological phenomenon that is ultraviolet-sensitive to perturbations. 
We will discuss our treatment of the baryon power spectrum beyond the linear regime in significant detail in Sec.~\ref{sec:variance_of_fluctuations}. 

Finally, although the Gaussian distribution is well-motivated at high redshifts when fluctuations are small, the Gaussian PDF shown in Eq.~\eqref{eq:gaussian_pdf} breaks down once $\sigma_\text{b} \sim 1$, since large negative fluctuations which lead to an overall negative density is assigned a sizable probability. Fig.~\ref{fig:sigma_lin_baryon} shows that the applicability of the Gaussian PDF starts becoming questionable once $z \lesssim 20$. 

\subsection{Analytics}

Substituting the expression for $f(m_\gamma^2;t)$ in Eq.~\eqref{eq:f_from_one_pt_PDF} into Eq.~\eqref{eq:prob_Ap_to_gamma} gives
\begin{multline}
    \frac{\dd \langle P_{\gamma \to A'} \rangle_\text{G}}{\dd z} = \frac{\pi m_{A'}^4 \epsilon^2}{\overline{m_\gamma^2}(z) \omega(z)} \left| \frac{\dd t}{\dd z} \right|\\
    \times \frac{1}{\sqrt{2\pi \sigma_\text{b}^2(z)}} \exp \left[- \frac{(\mAp^2/\overline{m_\gamma^2}(z) - 1)^2}{2 \sigma_\text{b}^2(z)}\right] \,,
    \label{eq:dP_dz_gaussian}
\end{multline}
where the subscript `G' stands for Gaussian. The corresponding energy deposited per baryon is
\begin{multline}
    \frac{\dd \langle E_{A' \to \gamma} \rangle_\text{G}}{\dd z} = \frac{\pi \mAp^3 \epsilon^2}{\overline{m_\gamma^2}(z)} \frac{\mAp^2}{\overline{m_\gamma^2}(z)} \frac{\overline{\rho}_{A'}}{b \overline{n}_\text{b}} \left| \frac{\dd t}{\dd z} \right| \\
    \times \frac{1}{\sqrt{2\pi \sigma_\text{b}^2(z)}} \exp \left[- \frac{(\mAp^2 / \overline{m_\gamma^2}(z) - 1)^2}{2 \sigma_\text{b}^2(z)}\right] \,.
\end{multline}
Given $\sigma_\text{b}(z)$ and $\overline{m_\gamma^2}(z)$ from Eq.~\eqref{eq:m_gamma_sq}, these compact results can now be integrated numerically to obtain $\langle P_{\gamma \to A'} \rangle$. 

In the $\sigma_\text{b}^2 \to 0$ limit, the Gaussian narrows, and can eventually be approximated by a Dirac-delta function; this expression then converges to the homogeneous result, as a corollary of the discussion around Eq.~\eqref{eq:f_homo_limit}. On the other hand, for some finite value of $\sigma_\text{b}^2$, the characteristic redshift width $\Delta z$ over which transitions occur is given by
\begin{alignat}{1}
    \Delta z \sim \sigma_\text{b} \left| \frac{\dd \ln \overline{m_\gamma^2}}{\dd t} \frac{\dd t}{\dd z} \right|^{-1} \,,
    \label{eq:z_range_of_conversion_less_approx}
\end{alignat}
which during periods when $x_\text{e}$ does not change significantly (\emph{e.g.},\ before recombination, during the dark ages and after reionization is complete) is approximately
\begin{alignat}{1}
    \Delta z \sim 3.3 \left(\frac{1+z_\text{h}}{100}\right) \left( \frac{\sigma_\text{b}(z_\text{h})}{0.1} \right) \,,
    \label{eq:z_range_of_conversion}
\end{alignat}
where $z_\text{h}$ is the redshift at which $\overline{m_\gamma^2} = m_{A'}^2$. 
In the linear regime, fluctuations grow linearly with the scale factor, and thus $\sigma_\text{b} \propto 1/(1+z)$; this implies that $\Delta z$ stays relatively constant throughout the dark ages.
We can see that the range of redshifts over which conversions can happen can be very large, with $\Delta z \gtrsim z$ at low redshifts. 

Similarly, a range of $\mAp^2$ can now convert with significant probability at any given redshift $z$. At a particular value of $z_h$, this range is roughly
\begin{alignat}{1}
    \Delta \mAp^2 \sim \pm \sigma_\text{b} \overline{m_\gamma^2}(z_\text{h}) \,.
\end{alignat}
We note that when $\sigma_\text{b}$ exceeds one at $z \lesssim 20$, this range of $\mAp$ includes negative values, highlighting the fact that the Gaussian PDF becomes unphysical in this range, as we discussed above. 
However, the lesson here is clear: the presence of under- and overdensities allows conversions well above and below the homogeneous value $\overline{m_\gamma^2}(z_\text{h})$, allowing \emph{(i)} conversions with $\mAp \lesssim \SI{e-14}{\eV}$, \emph{i.e.},\ below the homogeneous plasma mass at any point in the history of the Universe, and \emph{(ii)} lower redshift conversions for $\SI{e-13}{\eV} \lesssim \mAp \lesssim \SI{e-12}{\eV}$, which have a higher probability of conversion. 

In the Gaussian limit, we can derive the ratio of the probability calculated under the homogeneous assumption to the probability given a Gaussian PDF analytically. 
We begin by defining the variable $\Delta \equiv \mAp^2 / \overline{m_\gamma^2} - 1$, and rewrite the conversion probability with the Gaussian PDF shown in Eq.~\eqref{eq:dP_dz_gaussian} as
\begin{alignat}{1}
    \langle P_{\gamma \to A'} \rangle_\text{G} = \int_{-1}^{\Delta_0} \dd \Delta \, \frac{g(\Delta)}{\sqrt{2\pi \sigma_\text{b}^2}} \exp \left(-\frac{\Delta^2}{2\sigma_\text{b}^2} \right) \,,
\end{alignat}
where we have defined
\begin{alignat}{1}
    g(\Delta) \equiv \frac{\pi \mAp^2(\Delta + 1) \epsilon^2}{\omega(\Delta)} \frac{\dd t}{\dd \Delta} \,,
\end{alignat}
and $\Delta_0 = \mAp^2 / \overline{m_\gamma^2}(z = 0) - 1$, with $\omega$ now being a function of $\Delta$. 
Observe that $g(0) = \langle P_{\gamma \to A'} \rangle_\text{h}$ provided $\Delta_0 \geq 0$, where $\langle P_{\gamma \to A'} \rangle_\text{h}$ is the homogeneous conversion probability. 
Since the contribution to the integral is centered at $\Delta = 0$, we can set $g(\Delta) \approx g(0) + g'(0)\Delta$ and take $\sigma_\text{b}$ to be constant, giving
\begin{multline}
    \frac{\langle P_{\gamma \to A'} \rangle_\text{G}}{\langle P_{\gamma \to A'} \rangle_\text{h}} \simeq \frac{1}{2} \left[\text{erf} \left(\frac{1}{\sqrt{2 \sigma_\text{b}^2}}\right) + \text{erf} \left(\frac{\Delta_0}{\sqrt{2 \sigma_\text{b}^2}}\right)\right] \\
    + \frac{g'(0)}{g(0)} \frac{ \sigma_\text{b}}{ \sqrt{2\pi}} \left[\exp \left(-\frac{1}{2\sigma_\text{b}^2}\right) - \exp \left(- \frac{\Delta_0^2}{2\sigma_\text{b}^2}\right)\right]
    \label{eq:homo_vs_gauss_error}
\end{multline}
for $\Delta_0 > 0$. 
In the limit of constant $x_\text{e}$ and a matter dominated Universe, $g'(0)/g(0) = 5/6$.
The ratio of probabilities would be greater than one if the homogeneous assumption is conservative with respect to the Gaussian result. 
Moreover, in the limit when $\sigma_\text{b} \to 0$, this expression tends to 1, as expected. 

\begin{figure}
    \centering
    \includegraphics[width=0.45\textwidth]{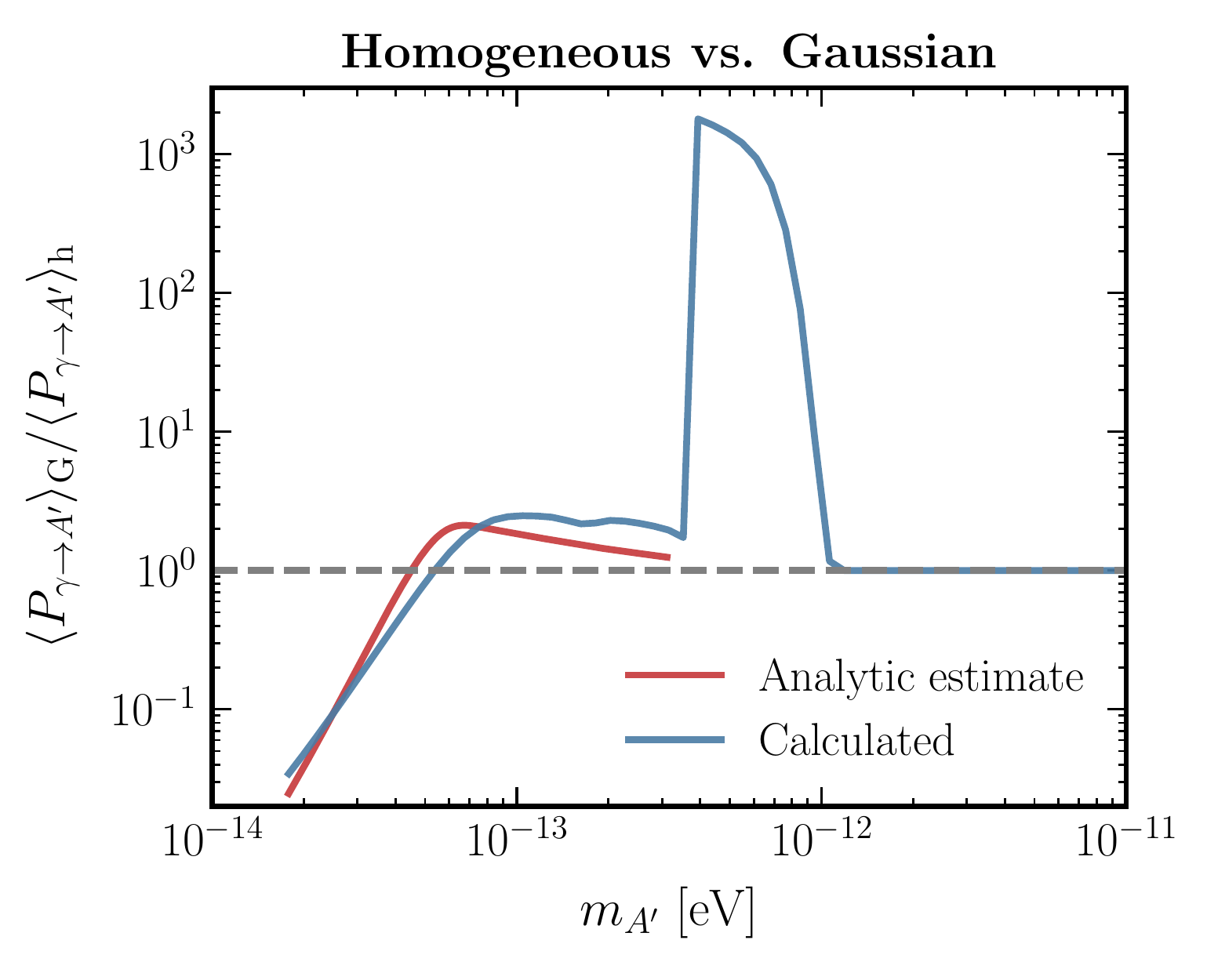}
    \caption{An analytic estimate for the ratio of the probability of conversion with the Gaussian PDF v.s.\ that of the homogeneous assumption (red) for conversions that happen at $z < 6$, with agreement between the two approaches corresponding to a conversion probability ratio of one (gray). The full numerical result is shown for comparison (blue). There are no conversions in the homogeneous limit for $\mAp \lesssim \SI{2e-14}{\eV}$.~\nblink{12_formalism_plots_linear}}
    \label{fig:homo_vs_gauss_error}
\end{figure}

In Fig.~\ref{fig:homo_vs_gauss_error}, we plot the conversion probabilities ratio as a function of $\mAp$. 
The analytic estimate in Eq.~\eqref{eq:homo_vs_gauss_error} is evaluated with $\sigma_\text{b}$ at the homogeneous resonance redshift $z_\text{h}$, and is shown for homogeneous conversions that occur at $z < 6$.  
We also include the exact probability ratio computed numerically.
Large values of the ratio of Gaussian to homogeneous conversion probabilities occur for values of $\mAp$ where the homogeneous limit resonance is deep in the dark ages, but overdensities allow for significant conversions at $z \sim 6$ with the Gaussian PDF (see Fig.~\ref{fig:m_Ap_single_LN}).
At large values of $\mAp$, the Gaussian and homogeneous conversion probabilities rapidly converge as the variance of fluctuations decreases.  

This ratio is significantly less than one for later conversions, \emph{i.e.}, lighter $\mAp$.
Qualitatively, the Gaussian PDF spreads out the probability of conversion over a range $\Delta z$ given in Eq.~\eqref{eq:z_range_of_conversion} compared to the homogeneous assumption; for small $z_\text{h}$, this can mean that most of the probability of conversion lies in the future, even though $z_\text{h} > 0$. 
For sufficiently large $z_\text{h}$, however, the probability of conversion in the future is negligible while the total conversion probability in the Gaussian limit is larger, since conversions happening below $z_\text{h}$ have higher values of $\dd P/\dd z$, increasing the overall integrated probability. 

\subsection{Main takeaways}

Having gone through the example of a Gaussian PDF, we are now ready to understand how to arrive at a numerical result for $\langle P_{\gamma \to A'} \rangle$ and $\langle E_{A' \to \gamma} \rangle$ in general. We need two inputs, both of which need to be evaluated correctly in the nonlinear regime:
\begin{enumerate}
    \item \textbf{Functional form for baryon one-point PDF}\@. In the linear regime, the PDF has a Gaussian form, but outside of the linear regime ($z \lesssim 20$), the Gaussian PDF clearly fails to capture fluctuations (especially underdensities) well, and better prescriptions are required; and
    \item \textbf{The variance of baryon fluctuations}. While the mean of the PDF is fixed to be zero by the fact that the average baryon density must be the homogeneous baryon density, the variance is not determined.\footnote{In this paper, we use only PDFs with functional forms that are fully defined by the mean and variance. Higher order statistics could play an important role in a full characterization of baryon fluctuations.} 
    The variance of baryon fluctuations will ultimately be determined by the power spectrum of matter or baryons as a function of redshift.
    Outside of the linear regime, one can no longer rely on Boltzmann codes to calculate these power spectra, and must instead make use of results informed by $N$-body simulations to obtain this information. 
    In all cases, the variance is ultraviolet-sensitive to the power spectrum at small scales, but this UV sensitivity is cut off by the Jeans scale, $k_\text{J}$, which we obtain from the CLASS linear baryon power spectrum. 
\end{enumerate}
We will devote Sec.~\ref{sec:PDFs} to examining more realistic alternative one-point PDFs to the Gaussian, and Sec.~\ref{sec:variance_of_fluctuations} to a discussion of how to obtain the variance of the PDF deep in the nonlinear regime. 

\section{One-Point Probability Density Functions}
\label{sec:PDFs}

\renewcommand*{\arraystretch}{1.5}

\afterpage{
    \begin{longtable*}{C{0.1 \textwidth}  C{0.3 \textwidth} L{0.24 \textwidth} L{0.31 \textwidth}}
    \toprule
    \textbf{PDF} & \textbf{Equation} & \textbf{Power spectrum} & \textbf{Remarks}\\
    \hline
    \hline
    Log-normal (fiducial) $\mathcal{P}_\text{LN}(\delta_\text{b};z)$ & $\frac{(1+\delta_\text{b})^{-1}}{\sqrt{2\pi \Sigma^2(z)}} \exp \left[- \frac{[\ln(1 + \delta_\text{b}) + \Sigma^2(z)/2]^2}{2 \Sigma^2(z)}\right]$ & Nonlinear baryon & $\Sigma^2(z) = \ln[1 + \sigma_\text{b}^2(z)]$. \\
    \hline 
    Analytic $\mathcal{P}_\text{an}(\delta_\text{b};z)$ & $\frac{\hat{C}(\delta_\text{b})}{\sqrt{2\pi \sigma_{R_\text{J}}^2(z)}} \exp \left[- \frac{F^2(\delta_\text{b})}{2 \sigma_{R_\text{J}}^2(z)}\right]$ & Linear matter, smoothed over baryon Jeans length $R_\text{J}$ & $\hat{C}(\delta_\text{b})$ and $F(\delta_\text{b})$ defined in App.~\ref{app:functions_analytic_pdf}.\\
    \hline 
    Log-normal with bias $\mathcal{P}_\text{LN}^b(\delta_\text{b};z)$ & $\frac{1}{b} \mathcal{P}_\text{LN} \left(\frac{\delta_\text{b}}{b}; z\right)$ & Nonlinear matter, with baryon Jeans scale cut-off & We adopt $b = 1.5$ following Ref.~\cite{Hurtado-Gil:2017dbm}. \\
    \hline 
    Voids $\mathcal{P}_\text{voids}(\delta_\text{b};z)$ & $\phi_\text{voids}(z) g_\text{voids}(1 + \delta_\text{b};z)$ & -- & $\phi_\text{voids}$ is the fractional volume of the simulation in a void, $g_\text{voids}$ is the PDF of the mean $1+\delta_\text{b}$ in voids~\cite{Adermann:2017izw,Adermann:2018jba}. Only used for underdensities. \\
    \botrule
    \\
    \caption{Summary of the baryon one-point PDFs used in this paper and in~\citetalias{Caputo:2020bdy}, with their defining equations and input power spectra used to determine the variance of these PDFs, where applicable.}
    \label{tab:PDFs}
    \end{longtable*}
}

\begin{figure*}[htbp]
    \centering
    \includegraphics[width=0.95\textwidth]{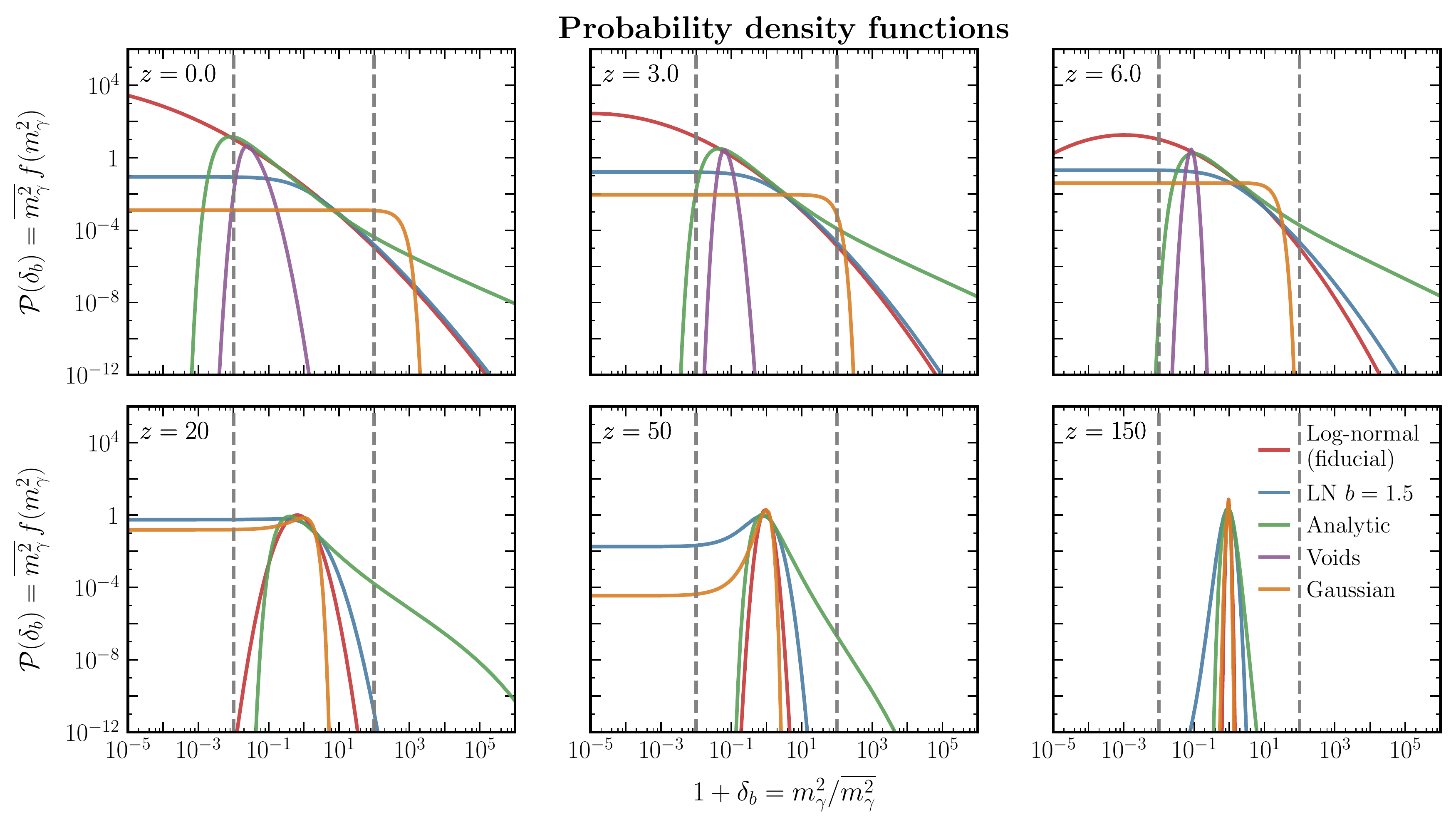}
    \caption{One point PDFs $\mathcal{P}(\delta_\text{b}; z)$ at six different redshifts. The fiducial log-normal $\mathcal{P}_\text{LN}$ (red), analytic $\mathcal{P}_\text{an}$ (green) PDFs, the log-normal PDF with bias $b = 1.5$, $\mathcal{P}_\text{LN}^{1.5}$ (blue), the PDF constructed from a model of voids $\mathcal{P}_\text{void}$ (purple)~\cite{Adermann:2018jba}, and the Gaussian PDF $\mathcal{P}_\text{G}$ (orange). Also shown are the fiducial $10^{-2} < 1+\delta < 10^2$ boundaries (dashed gray).~\nblink{14_multiple_pdf_plots}} 
    \label{fig:PDFs}
\end{figure*}

Table~\ref{tab:PDFs} shows a summary of all of the baryon one-point PDFs considered in this paper and in~\citetalias{Caputo:2020bdy}, and Fig.~\ref{fig:PDFs} shows a plot of these PDFs at a range of redshifts.  Beyond the linear regime, the log-normal PDF has been proposed as a phenomenological fit to the total matter distribution~\cite{1934ApJ....79....8H,Coles:1991if} for both observations~\cite{Clerkin:2016kyr,Gruen:2017xjj,Wild:2004me,Hurtado-Gil:2017dbm} and $N$-body simulations~\cite{Kofman:1993mx,Kayo:2001gu,Klypin:2017jjg}. 
The introduction of a bias parameter to the log-normal distribution has also been shown to produce good fits phenomenologically~\cite{Wild:2004me,Hurtado-Gil:2017dbm}.
There has also been a significant effort to calculate the matter one-point PDF from first principles~\cite{Bernardeau:1992zw,Bernardeau:2001qr} with the linear regime as a starting point, especially using a path-integral approach~\cite{Valageas:2001zr,Valageas:2001td,Matarrese:2000iz,Ivanov:2018lcg}. 
Finally, the study of cosmic voids has shed some light on the underdense tail of the one-point PDF~\cite{Zeldovich:1982zz,Plionis:2001gr,Einasto:2011eu,Jennings:2013nsa,Chan:2014qka,Adermann:2018jba}, and simulation results can be turned into a reasonable PDF at low densities.

To understand $\gamma \leftrightarrow A'$ oscillations, we need a PDF that is able to: \emph{(i)} capture baryonic effects, and not just the overall matter distribution; \emph{(ii)} capture the distribution of large overdensities and underdensities correctly, and \emph{(iii)} capture the behavior of baryonic fluctuations down to the Jeans scale of $k \sim 10^2$ -- \SI{e3}{\per\mega\parsec}. 
Existing studies of the one-point PDF cannot meet all three of these criteria simultaneously: first-principle, analytic results only apply to cold dark matter and do not account for baryonic effects, while the log-normal phenomenological fits have only been applied to simulations or data that have an effective smoothing scale much larger than the Jeans scale. 
Almost all results are validated with observations in a small range of density fluctuations ($10^{-1} \lesssim 1+\delta_\text{b} \lesssim 10$), or on one side of the PDF (\emph{e.g.},\ voids). These uncertainties surrounding the distribution of baryonic fluctuations make it a challenge to arrive at a rigorous conclusion regarding constraints on $\gamma \leftrightarrow A'$ oscillations.

Our approach is to adopt several independent models of the baryonic one-point PDF, in an attempt to capture the systematic uncertainties discussed here.
In our fiducial approach, we adopt a log-normal functional form for the one-point PDF, with the variance of this distribution determined by baryonic power spectra obtained from a combination of different hydrodynamic $N$-body simulation results, which we detail in Section~\ref{sec:power_spectra}. 
We truncate the PDF to the range $10^{-2} \leq 1 + \delta_\text{b} \leq 10^2$ to avoid the large uncertainties in the tails of the PDF. 
Our second approach relies on analytic results described in Ref.~\cite{Ivanov:2018lcg}, which takes as input the linear matter power spectrum and computes the one-point PDF for matter fluctuations as a function of redshift due to spherical collapse, which we then take to be equal to the baryon one-point PDF\@.
We find that at low redshifts, these two approaches lead to similar PDFs in the range $10^{-2} \lesssim 1 + \delta_\text{b} \lesssim 10^2$ at $z = 0$, as shown in Fig.~\ref{fig:PDFs}; this range decreases to $10^{-1} \lesssim 1 + \delta_\text{b} \lesssim 10$ at $z = 6$. 
Restricting the PDFs to the range $10^{-2} \lesssim 1 + \delta_\text{b} \lesssim 10^2$, the constraints on $\epsilon$ derived from $\gamma \leftrightarrow A'$ in~\citetalias{Caputo:2020bdy} differ by at most a factor of approximately three at $\mAp \sim \SI{e-12}{\eV}$ between our two prescriptions, suggesting that we have reasonable control over the uncertainties on the baryon PDF. 

In addition to the log-normal PDF and the analytically derived PDF, we also use two other PDFs as cross checks to our results. 
First, we use a log-normal distribution with a bias parameter $b$, with the variance of the distribution given by the nonlinear \textit{matter} power spectrum. 
This approach models the \textit{baryonic} fluctuations as simply a factor $b$ times the overall matter fluctuations, giving us an estimate of how reliant we are on baryonic physics modeled by the simulations we used to obtain the baryonic power spectrum for our fiducial log-normal PDF\@. 
Second, we use results from Ref.~\cite{Adermann:2018jba} for the probability distribution of finding voids of a certain volume with a certain underdensity in their simulations, and construct a PDF of underdensities to test the underdense tails of our PDFs. 
Both of these cross checks show that the constraints we derive in~\citetalias{Caputo:2020bdy} are likely to be robust to differences in systematics in the PDFs, and may improve if we can trust these PDF distributions to much larger underdense and overdense fluctuations. 

\subsection{Log-normal PDF}
\label{sec:PDFs_LN_PDF}

Our fiducial choice for the PDF in this paper is the log-normal PDF $\mathcal{P}_\text{LN}(\delta_\text{b};z)$, given by
\begin{multline}
    \mathcal{P}_\text{LN}(\delta_\text{b};z) = \frac{(1+\delta_\text{b})^{-1}}{\sqrt{2\pi \Sigma^2(z)}} \\
    \times \exp \left(- \frac{[\ln(1 + \delta_\text{b}) + \Sigma^2(z)/2]^2}{2 \Sigma^2(z)}\right)\,,
    \label{eq:log_normal_pdf}
\end{multline}
with $\Sigma^2(z) = \ln[1 + \sigma_\text{b}^2(z)]$ as defined in Eq.~\eqref{eq:sigma_sq_def}. 
The variable $\ln(1+\delta_\text{b})$ has a Gaussian distribution with mean $-\Sigma^2/2$ and time-dependent variance $\Sigma^2$. 
As an immediate consequence, unphysical fluctuations of $\delta_\text{b} < -1$ are forbidden, unlike the Gaussian PDF for $\delta_\text{b}$. With this choice of $\Sigma$, $\mathcal{P}_\text{LN}$ satisfies
\begin{alignat}{1}
    \int_{-1}^\infty \dd \delta_\text{b} \, \mathcal{P}_\text{LN}(\delta_\text{b};z) &= 1 \,,  \\
    \int_{-1}^\infty \dd \delta_\text{b} \, \delta_\text{b} \mathcal{P}_\text{LN}(\delta_\text{b};z) &= 0 \,,  \\
    \int_{-1}^\infty \dd \delta_\text{b} \, \delta_\text{b}^2 \mathcal{P}_\text{LN}(\delta_\text{b};z) &= \sigma_\text{b}^2(z) \,,
\end{alignat}
\emph{i.e.},\ $\mathcal{P}_\text{LN}$ is correctly normalized, with $\langle \delta_\text{b} \rangle = 0$ and $\langle \delta_\text{b}^2 \rangle = \sigma_\text{b}^2$, as required. 
These normalization conditions mean that as a function of $\ln(1+\delta_\text{b})$, the log-distribution is symmetric about $\ln(1+\delta_\text{b}) = -\Sigma^2/2$ and not zero. 
In the limit that $\sigma_\text{b}^2 \ll 1$ and $\delta_\text{b} \ll 1$, the log-normal PDF in Eq.~\eqref{eq:log_normal_pdf} reduces to the Gaussian PDF to $\mathcal{O}(\delta_\text{b})$ and $\mathcal{O}(\sigma_\text{b}^2)$; in the linear regime, with $\sigma_\text{b}^2 \ll 1$ and $\delta_\text{b}$ having an extremely low probability of approaching one, the fluctuations drawn from both the Gaussian and log-normal PDFs are virtually identical. The red band in Fig.~\ref{fig:m_Ap_single_LN} illustrates the middle-68\% containment of the inhomogeneous photon plasma mass assuming a log-normal PDF for the perturbations. Unlike in the case of a Gaussian PDF description (illustrated by the blue band), unphysically negative fluctuations are forbidden in this case. For our fiducial PDF, we limit the range of the PDF to $10^{-2} \leq 1 + \delta_\text{b} \leq 10^2$, removing the highly uncertain PDF tails.

\subsection{Analytic PDF}
\label{sec:PDFs_analytic_PDF}

Computing the PDF of matter fluctuations from first principles has been effectively studied in the language of path integrals, giving expressions that have been shown to be reliable in the nonlinear regime, even at large overdensities~\cite{Valageas:2001zr,Valageas:2001td,Matarrese:2000iz,Ivanov:2018lcg}. 
Here, we provide only a brief outline of the derivation of such an analytic PDF, and refer the reader to Ref.~\cite{Ivanov:2018lcg} for the details of the calculation.

Consider a spherical volume of radius $r_*$ at some redshift $z$ containing some density fluctuation $\delta_*$ obtained by integrating the spherical volume over a top-hat function.\footnote{We will only consider an averaging procedure using a top-hat windowing function, although more general arguments can be made for any arbitrary windowing function~\cite{Ivanov:2018lcg}.} 
This fluctuation was formed from some field configuration $\delta_\text{i}(\vec{x})$ deep in the linear regime undergoing gravitational collapse, where $\delta_\text{i}(\vec{x})$ can be described as a Gaussian random field. 
If the evolution of fluctuations is purely linear, then the size of linear fluctuations at the same redshift $z$ is $\delta_\text{L} = (1+z_\text{i}) \delta_\text{i} / (1 + z)$, since linear fluctuations grow in proportion to the scale factor of the Universe during matter domination.
The statistical properties of a Gaussian random field are governed entirely by the two-point correlation function $\xi(\vec{x} - \vec{y}) \equiv \langle \delta_\text{L}(\vec{x}) \delta_\text{L}(\vec{y}) \rangle$, which is related by the Fourier transform to the linear matter power spectrum $P_\text{mm,L}(k)$.\footnote{Translational and rotational invariance means that $\xi$ ultimately only depends on the magnitude $|\vec{x} - \vec{y}|$.} 
If the mapping between overdensities $\delta_*$ in a cell of size $r_*$ and field configurations in the linear regime $\delta_\text{L}$ is well-understood, then the PDF of finding such an overdensity can be mapped onto the statistical properties of the Gaussian random field. 

Concretely, let us define the functional $\overline{\delta}_W[\delta_\text{L}]$ which takes a given Gaussian field configuration $\delta_\text{L}$ expected by linear evolution to redshift $z$ of an initial (Gaussian) field configuration $\delta_\text{i}$, and maps it to the actual density contrast $\delta_*$ averaged over some spherical volume of radius $r_*$, produced by the actual gravitational evolution of $\delta_\text{i}$.
Then the PDF of $\delta_*$ is given by a path integral over all Gaussian field configurations $\delta_\text{L}$ with a Gaussian weight~\cite{Valageas:2001zr}: 
\begin{alignat}{1}
    \mathcal{P}(\delta_*) = \mathcal{N}^{-1} \int \mathcal{D} \delta_\text{L} \, e^{-S_\text{G}[\delta_\text{L}]} \delta_\text{D}(\delta_* - \overline{\delta}_W[\delta_\text{L}]) \,,
    \label{eq:pdf_path_integral}
\end{alignat}
where
\begin{alignat}{1}
    S_\text{G}[\delta_\text{L}] \equiv \frac{1}{2} \int \dd^3 \vec{x} \int \dd^3 \vec{y} \, \delta_\text{L}(\vec{x}) \xi^{-1}(\vec{x} - \vec{y}) \delta_\text{L}(\vec{y}) \,,
    \label{eq:gaussian_action}
\end{alignat}
with $\xi^{-1}$ defined as the functional inverse of $\xi$,
\begin{alignat}{1}
    \int \dd^3 \vec{z} \, \xi^{-1}(\vec{x} - \vec{z}) \xi(\vec{z} - \vec{y}) = \delta^{(3)}_\text{D}(\vec{x} - \vec{y}) \,.
\end{alignat}
The overall normalization factor is simply
\begin{alignat}{1}
    \mathcal{N} = \int \mathcal{D}\delta_\text{L}\, e^{-S_\text{G}[\delta_\text{L}]} \,.
\end{alignat}
Taking the Fourier transform of the integrand in Eq.~\eqref{eq:gaussian_action} gives~\cite{Valageas:2001zr}
\begin{alignat}{1}
    S_\text{G}[\delta_\text{L}] = \frac{1}{2} \int \frac{\dd^3 \vec{k}}{(2\pi)^3} \frac{|\tilde{\delta}_\text{L}(\vec{k})|^2}{P_\text{mm,L}(k,z)} \,,
\end{alignat}
where $\tilde{\delta}_\text{L}(\vec{k})$ is the Fourier transform of the field configuration $\delta_\text{L}$. 

Ref.~\cite{Ivanov:2018lcg} showed that Eq.~\eqref{eq:pdf_path_integral} can be integrated using the saddle point approximation, by showing that the saddle point configuration is spherically symmetric, and by making use of the fact that the spherical collapse model provides a mapping $F$ between $\delta_*$ and $\overline{\delta}_\text{L}(R_*)$, where $\overline{\delta}_\text{L}(R_*)$ is the mean density of the configuration $\delta_\text{L}$ smoothed over a radius $R_* \equiv r_*(1 + \delta_*)^{\nicefrac{1}{3}}$, with
\begin{alignat}{1}
    F(\delta_*) \equiv \overline{\delta}_\text{L}(R_*) \,.
\end{alignat}
With this, they were able to show that taking into account only spherically-symmetric fluctuations, the probability distribution function is
\begin{alignat}{1}
    \mathcal{P}(\delta_*;z) = \frac{\hat{C}(\delta_*)}{\sqrt{2\pi \sigma^2_{R_*}(z)}} \exp \left(- \frac{F^2(\delta_*)}{2 \sigma_{R_*}^2(z)}\right) \,,
    \label{eq:analytic_pdf_raw}
\end{alignat}
where $\sigma^2_{R_*}$ is the variance of \textit{linear} matter fluctuations smoothed with a top-hat of radius $R_*$, 
\begin{alignat}{1}
    \sigma^2_{R_*}(z) = \int \frac{d^3 \vec{k}}{(2\pi)^3} P_\text{mm,L}(k, z) \left| W_\text{th}(kR_*) \right|^2 \,,
\end{alignat}
with $W_\text{th}$ being the Fourier transform of the top-hat, $W_\text{th}(x) \equiv 3 j_1(x) / x$. 

The intuition behind this result is clear: a density fluctuation $\delta_*$ within a sphere of radius $r_*$ at redshift $z$ is formed through spherical collapse of some initial \textit{linear} density fluctuation, which under linear evolution corresponds to a linear density fluctuation of size $F(\delta_*)$ in a sphere of radius $R_*$ at the same redshift $z$. 
Since the linear density fluctuations follow a Gaussian distribution with variance $\sigma^2_{R_*}(z)$, $\mathcal{P}(\delta_*;z)$ is also Gaussian with respect to $F(\delta_*)$. 

Several further comments are in order before we are ready to use this PDF in our analysis:
\begin{enumerate}
\item Although Ref.~\cite{Ivanov:2018lcg} introduces an $\mathcal{O}(1)$ aspherical factor that includes the effects of aspherical fluctuations, this factor was not computed for the small scales of interest to this work.
Since we are mostly interested in understanding the systematics associated with the use of different PDFs, for simplicity, we neglect this aspherical factor throughout. 
In principle this prefactor can be computed from theory, allowing an improvement to the PDF\@. Nevertheless, this will be a small correction compared to the baryonic bias with respect to the matter fluctuations, which is not included in the analytic calculation at the moment.
We neglect all other baryonic effects that may cause a difference between $P_\text{bb,L}(k,z)$ and $P_\text{mm,L}(k,z)$, and take $\delta_* = \delta_\text{b}$. 

\item The PDF as defined in Eq.~\eqref{eq:analytic_pdf_raw} for $\delta_*$ is defined with respect to a sphere of size $r_*$. 
This is critical in light of the UV divergence exhibited by $P_\text{mm,L}(k,z)$, as discussed in Sec.~\ref{sec:linear_regime}, which leads to a divergence in $\sigma^2_{R_*}$ as $R_* \to 0$. As we argued in Sec.~\ref{sec:linear_regime}, baryons naturally have a cut-off length scale given by the Jeans length $R_\text{J}$, below which the power spectrum is suppressed. 
We therefore set the smoothing scale $R_* = R_\text{J}$ to approximately reproduce this suppression of power, and take the result to be the PDF for baryon density fluctuations. 

\end{enumerate}

In summary, the analytic PDF for baryon fluctuations that we adopt in this paper is
\begin{alignat}{1}
    \mathcal{P}_\text{an}(\delta_\text{b};z) \equiv \frac{\hat{C}(\delta_\text{b})}{\sqrt{2\pi \sigma_{R_\text{J}}^2(z)}} \exp \left[- \frac{F^2(\delta_\text{b})}{2 \sigma_{R_\text{J}}^2(z)}\right] \,.
\end{alignat}
We show the full expression for the terms $\hat{C}$ and $F$ in App.~\ref{app:functions_analytic_pdf}.

\subsection{Log-normal PDF with bias}
\label{sec:ln_bias}

The log-normal PDF can be generalized to include an additional parameter $b$, known as the bias~\cite{Dekel:1998eq}. 
This distribution is given by
\begin{alignat}{1}
    \mathcal{P}_\text{LN}^b(\delta_\text{b};z) \equiv \frac{1}{b} \mathcal{P}_\text{LN} \left(\frac{\delta_\text{b}}{b} ; z\right) \,,
\end{alignat}
where the choice of $b = 1$ gives us the log-normal PDF discussed in Sec.~\ref{sec:PDFs_LN_PDF}. For this distribution, however, we choose $\Sigma^2 = \ln[1 + \sigma_\text{m}^2(z)]$ where 
\begin{alignat}{1}
    \sigma_\text{m}^2(z) = \int \frac{\dd^3 \vec{k}}{(2\pi)^3} P_\text{mm}(k,z) 
\end{alignat}
is the variance of the \textit{matter} power spectrum. The bias parameter is a constant factor relating matter density fluctuations $\delta_\text{m}$ to baryonic density fluctuations $\delta_\text{b}$, \emph{i.e.},\ $\delta_\text{b} = b \delta_\text{m}$. 
With this in mind, the normalization conditions are now 
\begin{alignat}{1}
    \int_{-b}^\infty \dd \delta_\text{b} \, \mathcal{P}_\text{LN}^b(\delta_\text{b};z) &= 1 \,, \label{eq:LN_normalization} \\
    \int_{-b}^\infty \dd \delta_\text{b} \, \delta_\text{b} \mathcal{P}_\text{LN}^b(\delta_\text{b};z) &= 0 \,, \\
    \int_{-b}^\infty \dd \delta_\text{b}^2 \, \delta_\text{b}^2 \mathcal{P}_\text{LN}^b(\delta_\text{b}; z) &= b^2 \sigma_\text{m}^2 \label{eq:LN_variance} \,.
\end{alignat}
These normalization conditions follow naturally from having matter fluctuations $-1 \leq \delta_\text{m} < \infty$, and the fact that $\delta_\text{b} = b \delta_\text{m}$ implies $\sigma_\text{b}^2 = b^2 \sigma_\text{m}^2$. 
For $b > 1$, $\delta_\text{b}$ can have downward fluctuations of up to $-b$, which are clearly unphysical; however, $\mathcal{P}_\text{LN}^b$ has been shown to be a reasonable fit to data~\cite{Wild:2004me,Hurtado-Gil:2017dbm}, and we are once again using the PDF only as a way of capturing systematic uncertainties. In particular, $\mathcal{P}_\text{LN}^b$ relies on the distribution of \textit{matter} and not baryons, allowing us to arrive at a log-normal-like PDF without relying on $N$-body simulations with baryonic feedback included, using instead $P_\text{mm}$ from $N$-body simulations with cold dark matter only. 
We again use the Jeans scale as a UV cut-off for $P_\text{mm}$ to regulate the power spectrum. We will adopt the value of $b = 1.5$ below, consistent with Ref.~\cite{Hurtado-Gil:2017dbm}.  

\subsection{PDF from voids}

In Refs.~\cite{Adermann:2017izw,Adermann:2018jba}, a $\Lambda$CDM $N$-body simulation was performed in a box of volume $V_\text{sim} = 500^3 \, $\SI{}{\per\h\cubed \mega\parsec\cubed} over the redshift range $0 \leq z \leq 12$. 
The number of voids $N_\text{voids}(z)$, the PDF $f_\text{voids}(V;z)$ of the volume $V$ of voids, and the PDF $g_\text{voids}(\rho/\overline{\rho}; z)$ of the ratio of the mean matter density in voids to the mean cosmological matter density $\rho/\overline{\rho}$ are all reported. 
We can now construct a PDF for baryonic fluctuations by making the following simplifying assumptions: \emph{(i)} all underdensities are found in voids that are successfully detected by the simulation; \emph{(ii)} the density in the void is constant, and is given by the mean matter density in the void, and iii) no conversions happen outside of voids. First, we can work out the fractional volume of the simulation that is in a void, given by
\begin{alignat}{1}
    \phi_\text{voids}(z) = \frac{N_\text{voids}(z)}{V_\text{sim}} \int \dd V \, V f_\text{voids}(V;z) \,.
\end{alignat}
$\phi_\text{voids} \sim 0.1$ across the entire redshift range simulated. Under the simplifying assumptions outlined above, we can now write
\begin{alignat}{1}
    \mathcal{P}_\text{voids}(\delta_\text{b};z) \equiv \phi_\text{voids}(z) g_\text{voids}(1 + \delta_\text{b}; z) \,.
\end{alignat}
The normalization of $\mathcal{P}_\text{voids}$ is $\phi_\text{voids} < 1$; in obtaining the ensemble average in Eqs.~\eqref{eq:dP_dz} and~\eqref{eq:dE_dz_general}, this is equivalent to discarding all worldlines at redshift $z$ that are not in voids. 
This PDF therefore is, by construction, aimed at modeling only underdensities. 
The assumptions made here can certainly be improved: not all underdensities are found in voids, which necessarily must have a local minimum in density in 3D space, and the void density profile should also be taken into account. 
However, the main purpose of constructing this PDF is less about getting an accurate model for the density fluctuations and more to provide a sanity check on our modeling of underdensities using the log-normal or analytic PDFs. 

\section{Variance of Fluctuations}
\label{sec:variance_of_fluctuations}

A key input to calculating the photon-to-dark photon oscillation probability in the presence of inhomogeneities is a description of the spectrum of fluctuations of the photon plasma. A particular challenge at late times is posed by nonlinear effects, which can be quantified using input from $N$-body simulations. At early times post-recombination on the other hand, spatial fluctuations in the fraction of free electrons come into play and have to be accounted for. We now describe in turn the calculation of the variance of fluctuations and relevant inputs in each regime.

\subsection{Free electron fraction perturbations}
\label{sec:free_elec_frac_perturb}

\begin{figure}[tbp]
    \centering
    \includegraphics[width=0.45\textwidth]{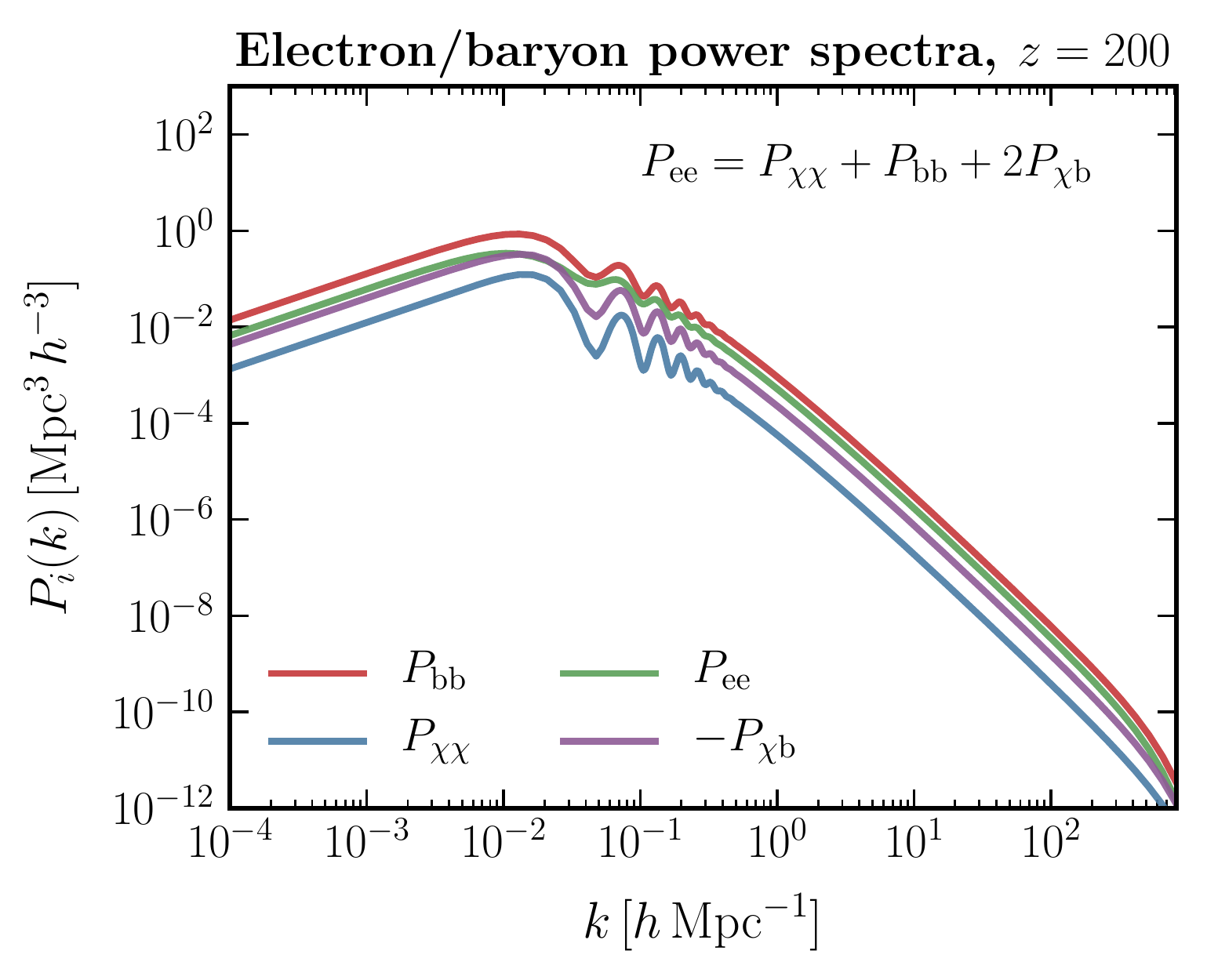}
    \caption{Baryon (red), electron (green), and free electron fraction (blue) power spectra; and negative of the baryon-free electron fraction cross-power spectrum (purple) at $z=200$. The electron fluctuations are reduced compared to the baryon ones due to the baryon and free electron fraction densities being anti-correlated.~\nblink{21_PS_plot_high_z}} 
    \label{fig:pspecs_high_z}
\end{figure}

Eq.~\eqref{eq:m_gamma_sq} shows that there are two sources of fluctuations for $m_\gamma^2(t)$: fluctuations in the baryon density, as well as fluctuations in the free electron fraction, which we define as $x_\text{e} \equiv n_\text{e}/n_\text{H}$, where $n_\text{H}$ is the number density of both neutral and ionized hydrogen atoms. 
So far, we have neglected fluctuations in $x_\text{e}$; we will now show how fluctuations in $m_\gamma^2$ are related to fluctuations in both baryon density and $x_\text{e}$, and discuss the conditions under which $x_\text{e}$ can be neglected.

Consider a point $t$ along a worldline of a photon with some HI density $n_\text{HI}(t)$ and free electron density $n_\text{e}(t)$, each with a fluctuation from the mean values $\overline{n}_\text{HI}$ and $\overline{n}_\text{e}$ given by $\delta_\text{HI}$ and $\delta_\text{e}$ respectively, so that 
\begin{alignat}{1}
    n_\text{HI} = (1 + \delta_\text{HI}) \overline{n}_\text{HI} \,, \qquad n_\text{e} = (1 + \delta_\text{e}) \overline{n}_\text{e} \,. 
\end{alignat}
We can further rewrite $\delta_\text{e}$ in terms of baryon density fluctuations $\delta_\text{b}$ and free electron density fluctuations
\begin{alignat}{1}
    \delta_\chi \equiv \frac{x_\text{e}}{\overline{x}_\text{e}} - 1 \,.
\end{alignat}
Writing $\overline{n}_\text{e}(1 + \delta_\text{e}) = \overline{x}_e(1 + \delta_\chi) \overline{n}_\text{H}(1 + \delta_\text{b})$,
\begin{alignat}{1}
    \delta_\text{e} = \delta_\text{b} + \delta_\chi + \delta_\chi \delta_\text{b} \,.
\end{alignat}
We can see that as long as $\delta_\chi \ll \delta_\text{b}$ and $\delta_\chi \ll 1$, we have $\delta_\text{e} = \delta_\text{b}$ to leading order, \emph{i.e.}, perturbations in the free electron density are given entirely by fluctuations in the baryon density when free electron fraction perturbations are small, even in the nonlinear regime. On the other hand, if $\delta_\chi \sim \delta_\text{b} \ll 1$, then
\begin{alignat}{1}
    \delta_\text{e} = \delta_\chi + \delta_\text{b} \,.
    \label{eq:delta_e_linear}
\end{alignat}

With this new notation, we can rewrite the plasma mass fluctuation $\delta_{m_\gamma^2}$ as
\begin{alignat}{1}
    \delta_{m_\gamma^2} \overline{m_\gamma^2} \equiv m_\gamma^2 - \overline{m_\gamma^2} = A \delta_\text{e} \overline{n}_\text{e} - B \omega^2 \delta_\text{HI} \overline{n}_\text{HI} \,,
    \label{eq:delta_m_gamma_sq}
\end{alignat}
where we have defined for convenience the constants
\begin{alignat}{1}
    A \equiv \SI{1.4e-21}{\eV\squared\centi\meter\cubed} \,, \quad B \equiv \SI{8.4e-24}{\centi\meter\cubed} \,.
\end{alignat}

In the linear regime, with $\delta_\text{e}$ and $\delta_\text{HI}$ being small and Gaussian, $m_\gamma^2$ is also Gaussian:
\begin{alignat}{1}
    f(m_\gamma^2;z) = \frac{1}{\sqrt{2\pi \sigma_{m_\gamma^2}^2}} \exp \left[-\frac{(1 - m_\gamma^2/\overline{m_\gamma^2})^2}{2 \sigma_{m_\gamma^2}^2}\right] \,,
\end{alignat}
where
\begin{alignat}{1}
    \sigma^2_{m_\gamma^2} \equiv \langle \delta_{m_\gamma^2} \delta_{m_\gamma^2} \rangle \,.
\end{alignat}
We can now make use of Eq.~\eqref{eq:delta_m_gamma_sq} to obtain an expression for this variance. For simplicity, we consider the redshift range $20 \lesssim z \lesssim 1600$, during which helium was almost completely neutral, so that we can write $n_\text{HI} = (1 - x_\text{e})n_\text{H}$.\footnote{Outside of this range, one must take into account that $\overline{x}_\text{e}$ can exceed one, which would require a simple modification to the results shown here; we omit these modifications since fluctuations in $x_\text{e}$ are not important outside the specified redshift range.} We find
\begin{multline}
    \overline{m_\gamma^2}^2 \sigma^2_{m_\gamma^2} = (A + B\omega^2)^2 \overline{n}_\text{e}^2 \langle \delta_\text{e} \delta_\text{e} \rangle + B^2 \omega^4 \overline{n}_\text{H}^2 \langle \delta_\text{b} \delta_\text{b} \rangle \\
    - 2 (A + B \omega^2) B \omega^2 \overline{n}_\text{e} \overline{n}_\text{H} \langle \delta_\text{e} \delta_\text{b} \rangle \,,
    \label{eq:sigma_m_gamma_sq_delta_e}
\end{multline}
where
\begin{alignat}{1}
    \langle \delta_i \delta_j \rangle = \int \frac{d^3 \vec{k}}{(2\pi)^3} P_{ij,\text{L}} (k) \,,
\end{alignat}
where $P_{ij,\text{L}}$ is the linear (auto) power spectrum of $i$ for $i=j$, and the cross power spectrum for $i$ and $j$ for $i \neq j$, with $i,j = \text{b}$ or $\text{e}$. 

A more mathematically transparent form of Eq.~\eqref{eq:sigma_m_gamma_sq_delta_e} is obtained by rewriting $\delta_\text{e}$ in terms of $\delta_\chi$ and $\delta_\text{b}$, which in the linear regime is simply given by Eq.~\eqref{eq:delta_e_linear}. 
This immediately leads to the following relation between auto- and cross-power spectra: 
\begin{alignat}{1}
    P_\text{ee} &= P_{\chi \chi} + P_\text{bb} + 2 P_{\chi \text{b}} \,, \\
    P_\text{eb} &= P_{\chi \text{b}} + P_\text{bb} \,.
\end{alignat}

Putting together these results, we find
\begin{multline}
    \overline{m_\gamma^2}^2 \sigma_{m_\gamma^2}^2 = \overline{m_\gamma^2}^2\langle \delta_\text{b} \delta_\text{b} \rangle + \left(A + B \omega^2\right)^2 \overline{n}_\text{e}^2 \langle \delta_\chi \delta_\chi \rangle \\ 
     + 2 \overline{n}_\text{e} \overline{m_\gamma^2} \left(A + B \omega^2 \right)\langle  \delta_\chi \delta_\text{b} \rangle \,.
    \label{eq:var_of_fluctuations_and_chi_PS}
\end{multline}

The power spectra that enter into Eq.~\eqref{eq:var_of_fluctuations_and_chi_PS} are all calculable in the linear regime after photons decouple from baryons at $z \sim 1089$ using the theory of perturbed recombination~\cite{Lewis:2007zh}.  

We can also see immediately that neglecting perturbations in $x_\text{e}$ leads to the previous result, $\sigma^2_{m_\gamma^2} = \sigma_\text{b}^2$. 
The coefficients for the terms on the right-hand side of Eq.~\eqref{eq:var_of_fluctuations_and_chi_PS}, however, are of comparable size, and hence the simplification of taking $\delta_\chi \to 0$ is only a good approximation if $\delta_\chi \ll \delta_\text{b}$. To get a sense of how important these terms are, we plot the power spectra required to compute the two-point correlations shown in Eq.~\eqref{eq:var_of_fluctuations_and_chi_PS} in Fig.~\ref{fig:pspecs_high_z} at $z=200$. Since the baryon $\delta_\text{b}$ and free electron $\delta_\chi$ fluctuations are anti-correlated,\footnote{The anticorrelation is due to the fact that recombination is more efficient when there are more hydrogen atoms present~\cite{Lewis:2007zh}.} the presence of free-electron fluctuations causes a reduced variance in electron fluctuations $\langle \delta_\text{e} \delta_\text{e} \rangle$ at higher redshifts. We see that at $z \sim 200$, we have $P_{\chi \chi} < |P_{\chi \text{b}}| < P_\text{bb}$, with the spectra becoming more comparable in magnitude for $z > 200$, and less so at $z < 200$. We use a slightly modified version of CLASS\footnote{Available at \url{https://github.com/smsharma/class_public}.} to extract the transfer functions associated with perturbations in the free electron fraction.

With this, we can now discuss the importance of $\delta_\chi$ on our results at the following redshifts: 
\begin{enumerate}
    \item $\mathbf{\boldsymbol{z} \gtrsim 1089}$. The Universe is completely ionized prior to recombination, and there are no significant perturbations in $x_\text{e}$. We may neglect $\delta_\chi$; 
    \item $\mathbf{200 \lesssim \boldsymbol{z} \lesssim 1089}$. At this time, $\delta_\chi \sim \delta_\text{b}$, both perturbations are small, and aside from differences in the functional form of $\dd \langle P_{\gamma \to A'} \rangle/\dd z$, this redshift range is well approximated by the homogeneous limit; 
    \item $\mathbf{20 \lesssim \boldsymbol{z} \lesssim 200}$. During this period, $\delta_\chi \ll \delta_\text{b}$, and we may once again neglect $\delta_\chi$ to a good approximation; 
    \item $\mathbf{6 \lesssim \boldsymbol{z} \lesssim 20}$. This is the period of reionization, an increasingly nonlinear regime where the behavior of $\delta_\chi$ depends on the details of reionization, and can have potentially large effects on the PDF of plasma mass fluctuations.
    In principle, $\delta_\chi$ can be calculated from reionization codes like 21cmFAST~\cite{Mesinger:2010ne, Munoz:2019rhi}, but to avoid this complication, we neglect any $\gamma \leftrightarrow A'$ transitions in this epoch throughout our work; and
    \item $\mathbf{\boldsymbol{z} \lesssim 6}$. Reionization is complete, and once again there are no significant perturbations in $x_\text{e}$. We may once again neglect $\delta_\chi$, even though baryon density fluctuations are highly nonlinear. 
\end{enumerate}
In summary, we avoid the redshift regime during which reliably predicting the effect of $x_\text{e}$ perturbations is nontrivial, staying in regimes where the effect is either absent, or has a minimal and calculable effect on the total conversion probability.
This latter regime, $200 \lesssim z \lesssim 1089$, is well-characterized by small Gaussian fluctuations, justifying our linear treatment above.
The effect on the conversion probability width or the redshift dependence of the conversion probability during the dark ages will be quantified in Sec.~\ref{sec:results}.

\subsection{Low-redshift power spectra}
\label{sec:power_spectra}

\begin{figure}[tbp]
    \centering
    \includegraphics[width=0.45\textwidth]{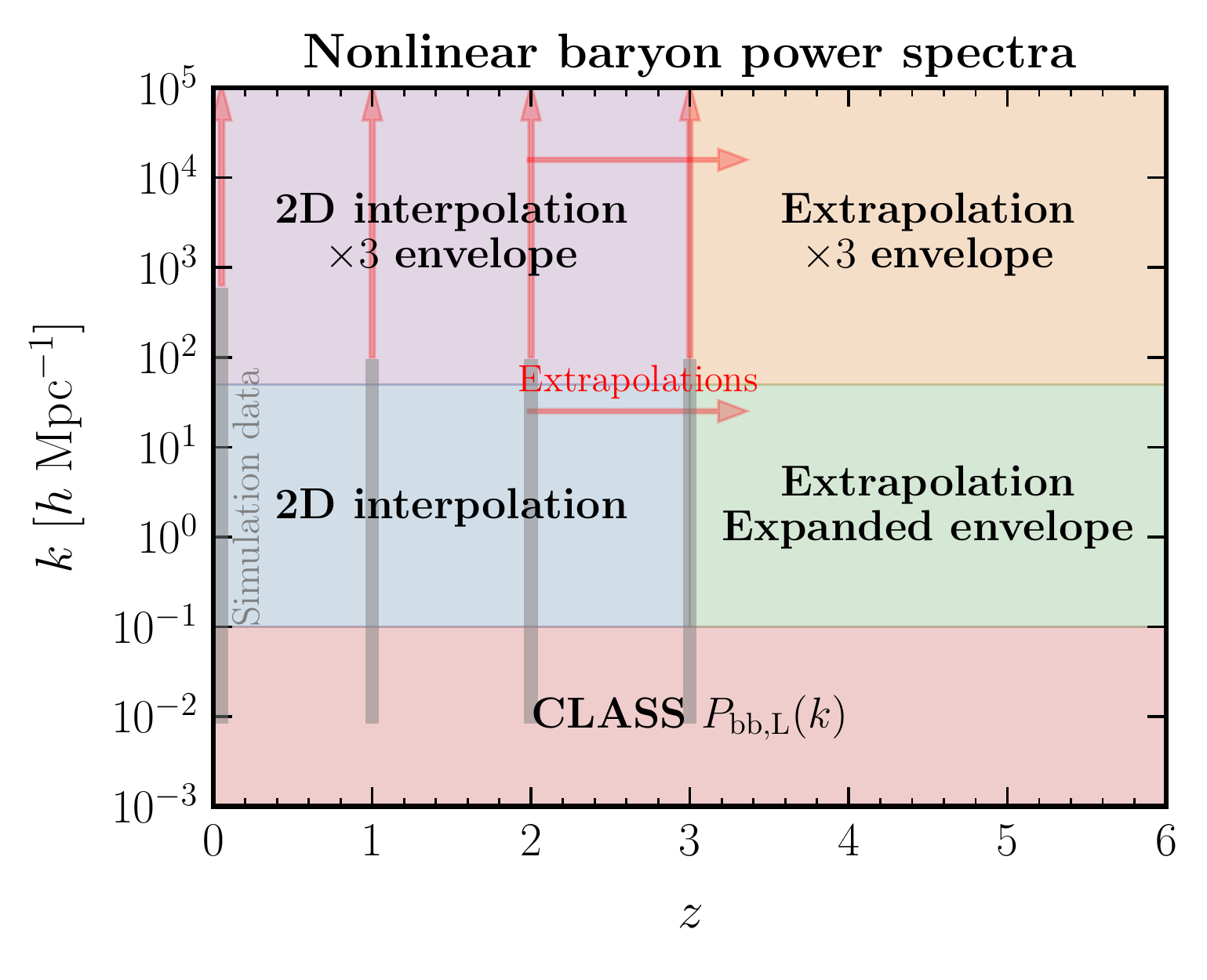}
    \caption{Illustration of the scheme used to construct an envelope of the nonlinear baryon power spectra at low redshifts, $0 \lesssim z \lesssim 6$, in different redshift $z$ and scale $k$ regimes. We use as input the CLASS linear baryon power spectrum $P_{\text{bb,L}}$ as well as the envelope of simulation data from Refs.~\cite{vanDaalen:2019pst,Foreman:2019ahr}, and linearly extrapolate the bias $P_\text{bb}/P_\text{mm}$ into regions without data (red arrows). For $k \leq \SI{0.1}{\h\per\mega\parsec}$, we use the CLASS linear baryon power spectrum (red). In the range $\SI{0.1}{\h\per\mega\parsec} < k < \SI{80}{\h\per\mega\parsec}$ and $0 \leq z \leq 3$, a 2D interpolation over available data is performed (blue). We then extrapolate into the region $3 < z \leq 6$, multiplying the resulting envelope by a factor of 3 (green). For $k > \SI{80}{\h\per\mega\parsec}$, we extrapolate the power spectra using the CLASS linear baryon power spectrum as a guide. We then perform a 2D interpolation in the range $0 \leq z \leq 3$, taking as an envelope a factor of 3 above and below the central value of the interpolated bias (purple), and then extrapolate this into $3 < z \leq 6$ (orange). See the text for more details.~\nblink{13_power_spectrum_scheme}} 
    \label{fig:franken_grid}
\end{figure}
\begin{figure*}[htbp]
    \centering
    \includegraphics[width=0.95\textwidth]{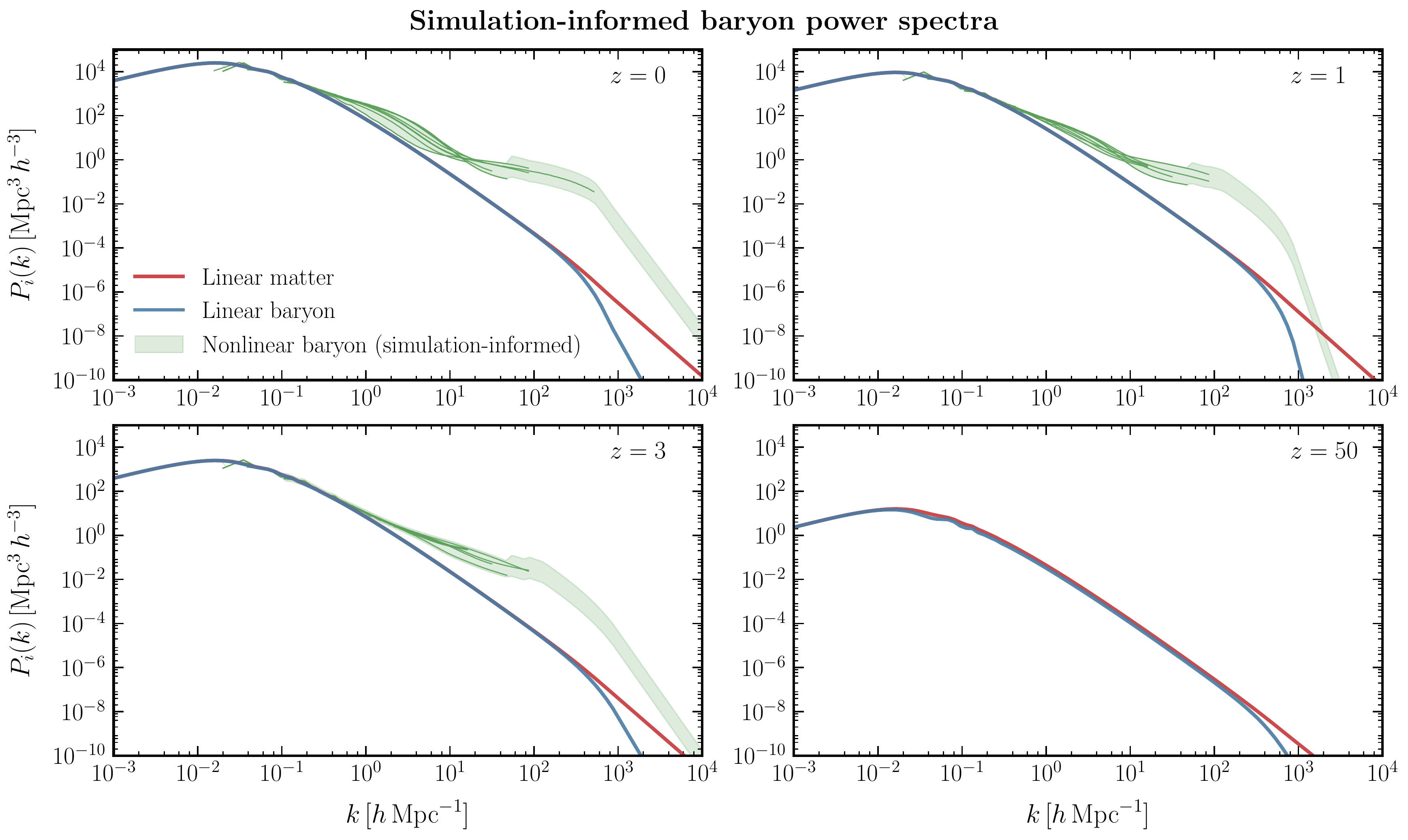}
    \caption{Simulation-informed baryon power spectra at low redshifts, bracketed with the green band and obtained using the method outlined in Sec.~\ref{sec:power_spectra}, shown at redshifts $z = 0, 1$, and 3. Solid green lines correspond to baryon power spectra from individual hydrodynamic simulations as obtained in Ref.~\cite{Foreman:2019ahr}. Also shown for comparison are the linear matter and baryon power spectra as the solid red and blue lines, respectively, also at $z=50$. Suppression due to the baryonic Jeans scale can clearly be seen.~\nblink{20_PS_plot_low_z}} 
    \label{fig:pspecs}
\end{figure*}

As described in the last section, at late times $z \lesssim 6$ after reionization is complete, fluctuations in the electron plasma mass track fluctuations in the number density of baryons, which is characterized by the baryonic power spectrum. Description of baryon density fluctuations at these late times is challenging, however, due to the highly nonlinear evolution of perturbations. Furthermore, even though nonlinear \emph{matter} fluctuations have been extensively studied in the literature, the distinction between baryonic and total matter fluctuations must be taken into account as the two components (baryons and dark matter) evolve separately and baryonic effects become increasingly important at late times, especially at the smaller scales of interest here. In this subsection, we describe our approach for constructing the nonlinear baryonic power spectra at low redshifts $z < 6$ using input from hydrodynamic simulations as well as the Boltzmann code CLASS. 

Ref.~\cite{Foreman:2019ahr} provides baryonic power spectra from different configurations of the hydrodynamic simulation suites IllustrisTNG~\cite{Nelson:2018uso}, Illustris~\cite{Genel:2014lma}, EAGLE~\cite{McAlpine:2015tma}, and BAHAMAS~\cite{McCarthy:2016mry} up to $k \sim \SI{80}{\h\per\mega\parsec}$ at the discrete redshifts $z=0,1,2$, and $3$, with Ref.~\cite{vanDaalen:2019pst} further providing baryonic spectra from the BAHAMAS simulation at redshift $z=0$ up to $k = \SI{500}{\h\per\mega\parsec}$. We use the following algorithmic procedure for constructing the nonlinear baryonic power spectra from these. We first construct lower and upper envelopes encoding the uncertainty on the power spectra extracted from simulations. Where fewer than three simulations are available, we obtain the median spectra over the available simulations and multiply and divide these by a factor of 3 to obtain upper and lower uncertainty envelopes, respectively, motivated by the magnitude of the typical spread in the regime where the full suite of simulations is available. Where three or more simulations are available, we use the extremal values over those simulations to construct the envelopes. At large scales $\lesssim \SI{0.1}{\h\per\mega\parsec}$ where simulations are not available, we use the well-constrained linear power spectrum from CLASS\@. At smaller scales and redshifts $0 < z < 6$ where simulations are not available, we linearly interpolate the nonlinear baryon bias (defined as the ratio of the nonlinear baryon power spectrum to the nonlinear matter spectrum), further applying a suppression due to the baryonic Jeans scale at small scales (see Sec.~\ref{sec:jeans_scale}). Above $z > 3$, we linearly extrapolate the nonlinear baryonic bias, multiplying and dividing the resulting power spectra by a factor of 3 to obtain the uncertainty envelope. In the regime above $z > 20$, we simply use the linear baryonic power spectrum from CLASS.

An illustration of this algorithmic procedure is provided in Fig.~\ref{fig:franken_grid}, showing how the nonlinear baryon power spectra are estimated at different redshifts $z$ and scales $k$. The resulting baryon power spectra at several different redshifts obtained using this procedure are shown in Fig.~\ref{fig:pspecs} (green envelopes), with the power spectra from individual simulations shown as green lines for reference. 

\begin{figure}[tbp]
    \centering
    \includegraphics[width=0.45\textwidth]{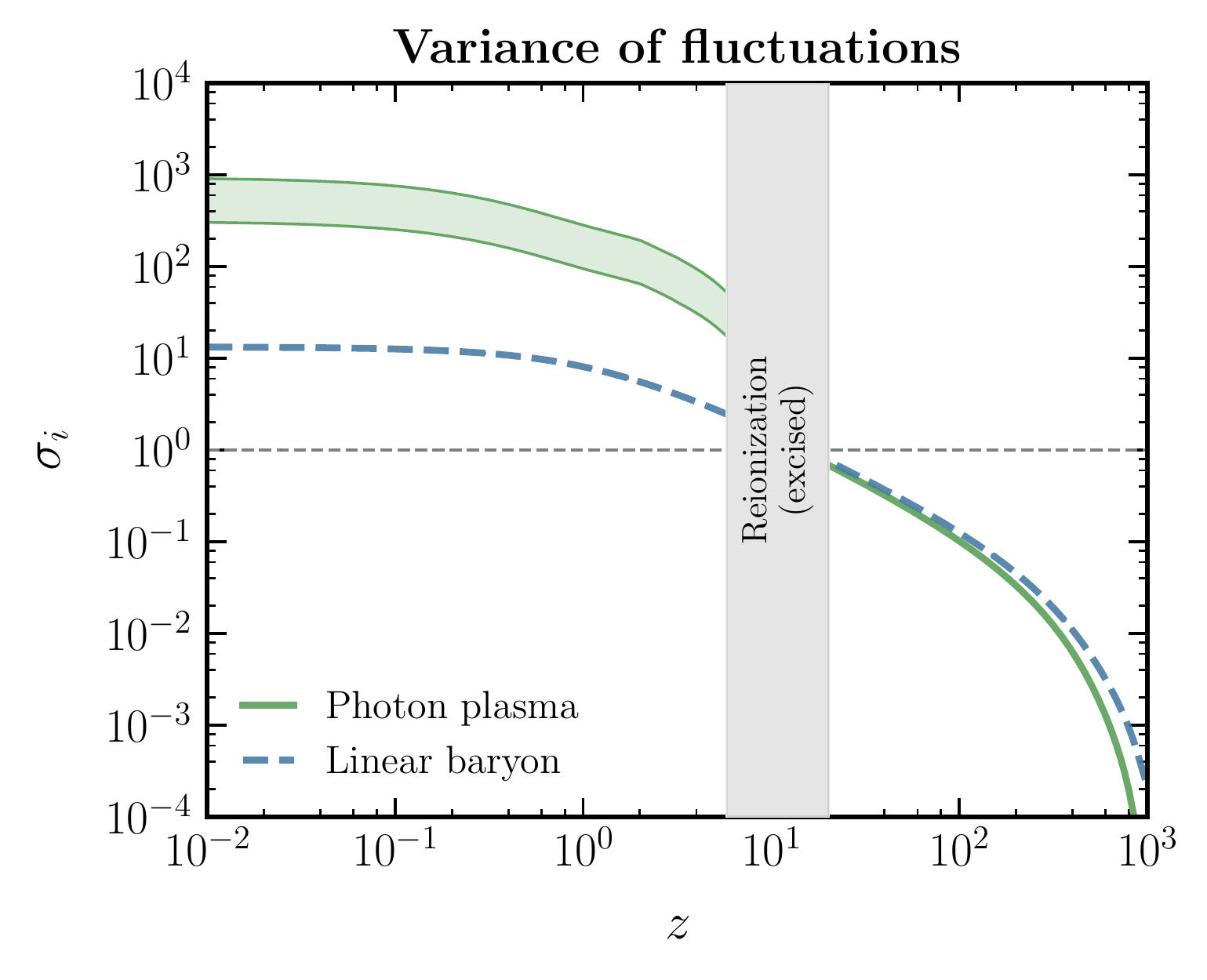}
    \caption{Variance of fluctuations as a function of redshift for the various power spectra configurations considered in this work. The photon plasma mass variance is informed by the nonlinear baryon power spectrum from simulations at late times $z < 6$ and is shown bracketed by the green band, while at late times $z > 20$ it is informed by the linear baryon and free electron fraction perturbation spectra. The variance of linear baryon fluctuations is shown as the dashed blue line, for comparison.~\nblink{16_formalism_prob_plots_high_z}} 
    \label{fig:sigma_all}
\end{figure}

The inferred variance of fluctuations as a function of redshift is shown in Fig.~\ref{fig:sigma_all}. At late times $z < 6$, the variance is informed by the nonlinear baryon power spectrum extracted from hydrodynamic simulations and is shown bracketed by the green band. The variance from the linear baryon power spectrum in this regime is shown as the blue line for comparison. Pre-reionization, the variance of photon plasma mass fluctuations is given by Eq.~\eqref{eq:var_of_fluctuations_and_chi_PS} and involves the (linear) baryon and free electron perturbations, shown as the red line. The variance due to just baryon perturbations, ignoring the effects of free electron perturbations, is shown as the dashed blue line for comparison.

\section{Simulation studies}
\label{sec:simulations}

\begin{figure*}[htbp]
    \centering
    \includegraphics[width=0.95\textwidth]{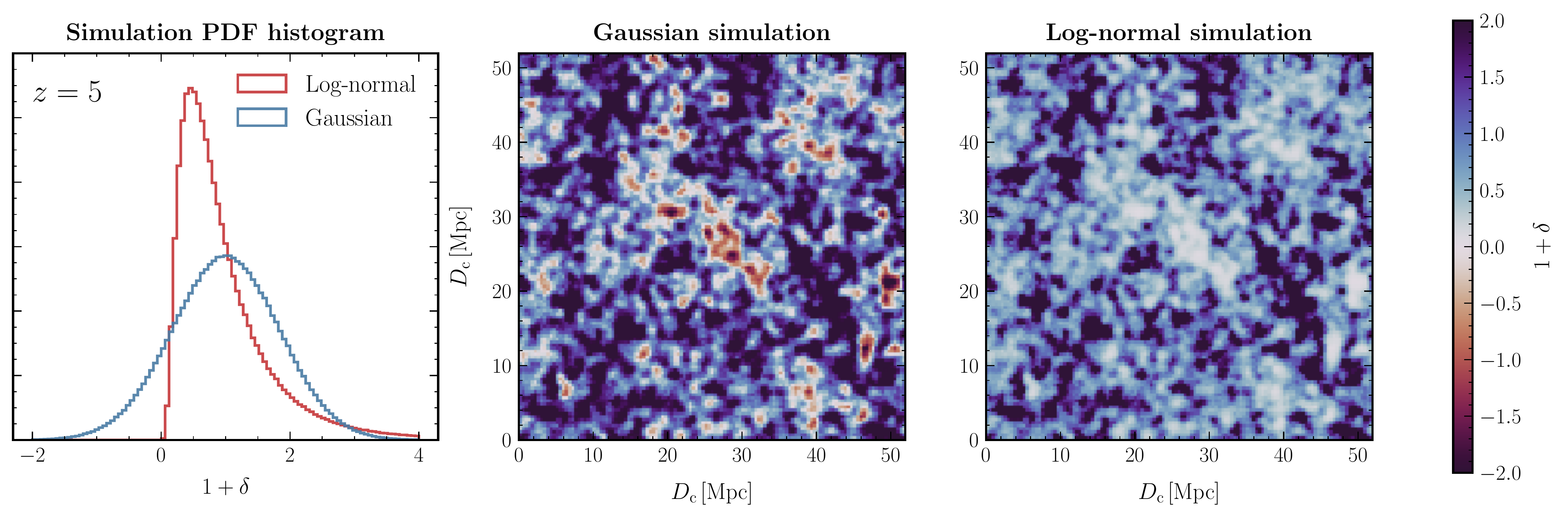}
    \caption{2D section through a Gaussian random field simulated at $z\sim 4$ (middle panel) and the corresponding log-normal-transformed field section (right panel). The left panel shows the histogrammed pixel count for both boxes, illustrating the skewed distribution of overdensities in the log-normal case restricted to positive values.~\nblink{19_lognormal_simulations}} 
    \label{fig:ln_sims}
\end{figure*}

We use Gaussian and log-normal simulations, which are relatively cheap to generate, to validate key aspects of the analytic approach presented in this paper. In particular, we verify that:
\begin{enumerate}
    \item The width of the oscillation probability is described by Eq.~\eqref{eq:dP_dz_simplified}, even when the fluctuations in the plasma are non-Gaussian, and
    \item Averaged over a large number of photon paths, the differential transition probability depends \emph{only} on the one-point PDF of the underlying plasma density field, and not higher-order moments such as two-point correlations.
\end{enumerate}

We note that the simulations we generate in this section are fundamentally different from the $N$-body simulations used to inform the baryon power spectra in the previous section---these simulations simply produce a Gaussian or log-normal random field with statistics consistent with a given input power spectrum.

We create realizations of the perturbed plasma mass by creating instances of baryon density fluctuations $1 + \delta_\mathrm{b}$, described either as a Gaussian or log-normal field, and then obtaining the perturbed plasma mass as $m_\gamma^2 = \overline{m_\gamma^2}(1 + \delta_\mathrm{b})$, where $\overline{m_\gamma^2}$ is the homogeneous plasma mass. Gaussian random fields consistent with the baryon power spectrum described in the last section are generated using \texttt{nbodykit}~\cite{Hand:2017pqn}, and log-normal fields as described in Sec.~\ref{sec:PDFs_LN_PDF} are generated by rescaling these as
\begin{equation}
    \ln(1 + \delta_\mathrm{b}^{\mathrm{LN}}) = -\frac{\Sigma^2}{2} + \frac{\delta_\mathrm{b}}{\sigma}\times \Sigma
\end{equation}
where $\delta_\mathrm{b}$ are the Gaussian overdensities and $\delta_\mathrm{b}^{\mathrm{LN}}$ the corresponding log-normal overdensities. This transformation ensures that the resulting log-normal field has the same mean and variance as the initial Gaussian field, as in Eqs.~\eqref{eq:LN_normalization} and~\eqref{eq:LN_variance}.

We choose a benchmark dark photon mass of $\mAp = 10^{-13}$\,eV, which would correspond to a broad resonance around $z \sim 5$ in the regime where the underlying fields are highly non-Gaussian. We generate boxes of Gaussian random field realizations between $4 < z < 6$, going up to scales of $k_\mathrm{max} = \SI{20}{\h\per\mega\parsec}$ and up to $\texttt{n\_points}=100$ points in each of the simulated boxes. Several boxes are created within the specified redshift range for computational efficiency and also to capture the redshift dependence of the power spectrum of fluctuations. While this does not capture the full spectrum of fluctuations relevant to oscillations (since $k_\mathrm{max} < k_\mathrm{J}$), the realized fields have large enough fluctuations ($\delta < -1$) so as to not be physically describable as Gaussian. We additionally impose a top-hat filter of 4 times the grid size in order to mitigate against the effects of finite gridding at the smallest simulated scales. 

An example 2D section through a Gaussian random field box generated with this procedure is shown in the middle panel of Fig.~\ref{fig:ln_sims}, with the corresponding section through a log-normally-transformed field in the right panel. Blue and red patches correspond to positive and negative (unphysical) values of the resulting field. The left panel shows the PDF of fluctuations in both boxes. The Gaussian random field description leads to frequent unphysical, negative fluctuations in this case.

\begin{figure*}[htbp]
    \centering
    \includegraphics[width=0.45\textwidth]{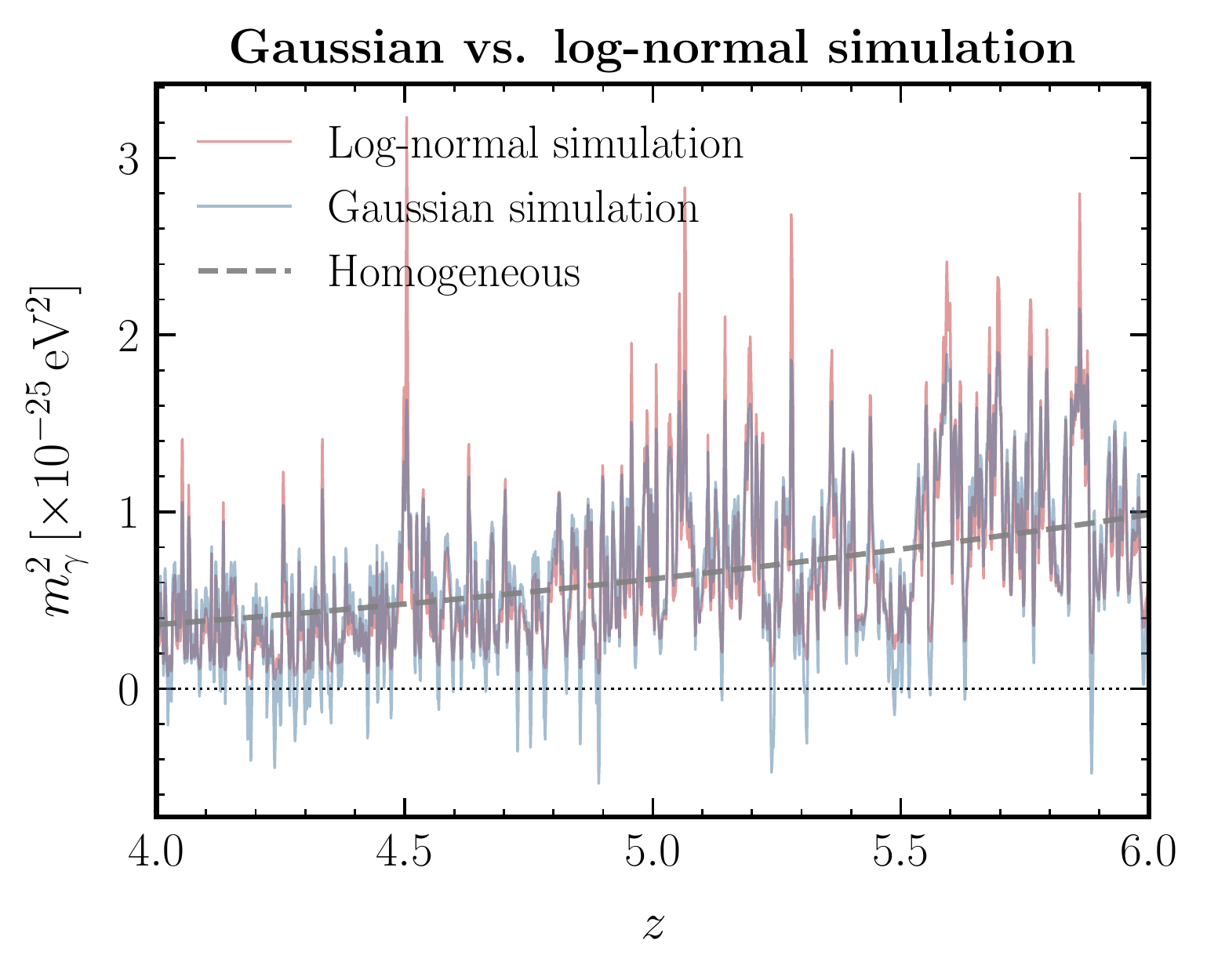}
    \includegraphics[width=0.45\textwidth]{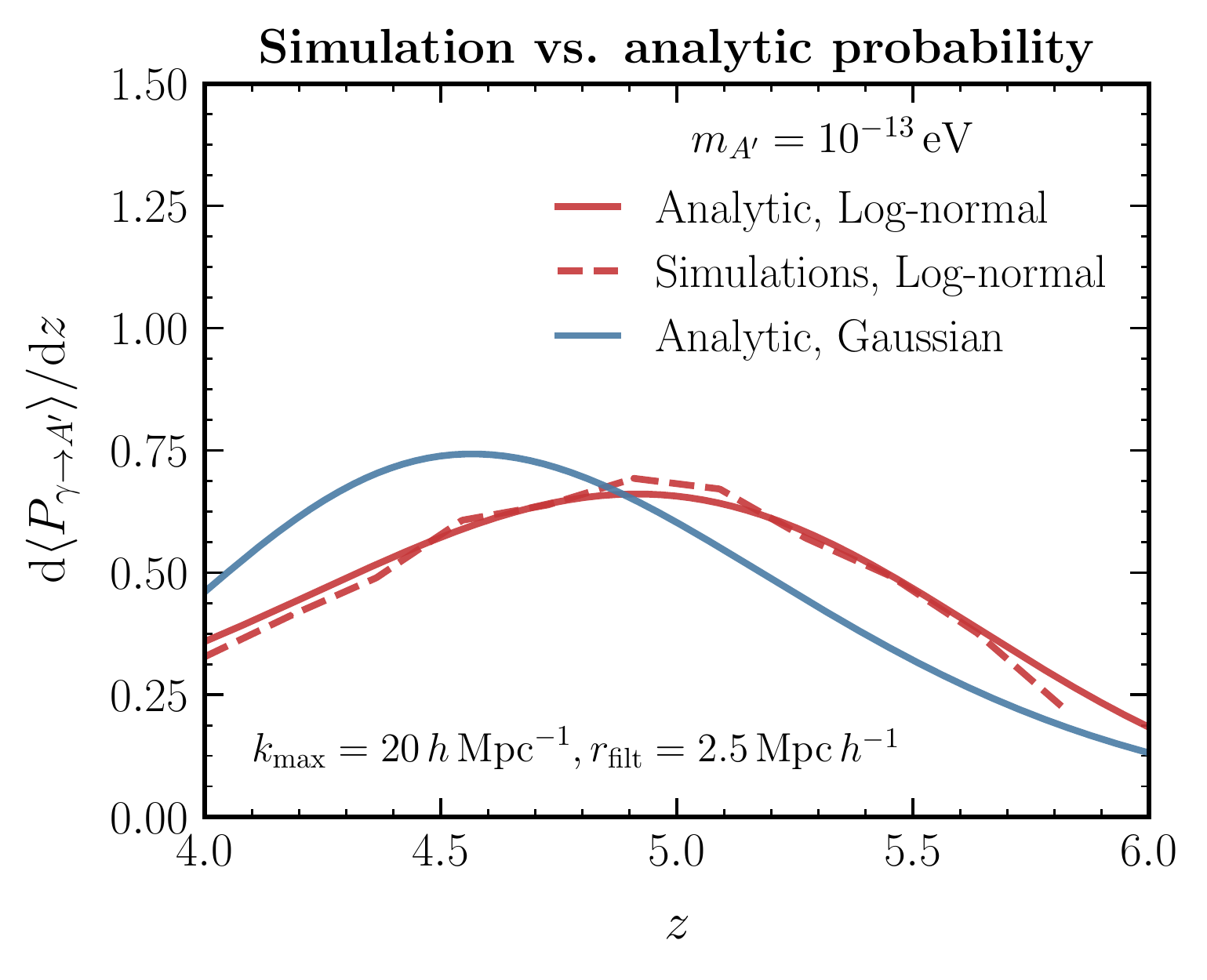}
    \caption{\emph{(Left)} 1D sections through realizations of Gaussian (blue) and log-normal (red) perturbations in the squared plasma mass. The homogeneous plasma mass is shown in dashed gray. \emph{(Right)} The log-normal differential oscillation probability averaged over a large number of photon paths drawn through simulations (dashed red) and derived analytically (solid red), with good agreement between the two. The analytic Gaussian description in shown in solid blue.~\nblink{19_lognormal_simulations}} 
    \label{fig:plasma_sims}
\end{figure*}

The perturbed squared plasma mass over the considered redshift range for one particular sequence of boxes is shown in the left panel of Fig.~\ref{fig:plasma_sims}, for the Gaussian (blue) and log-normal (red) descriptions. The homogeneous plasma mass is shown as the dashed black line. Again, frequent unphysically negative values of the squared plasma mass can be seen in the Gaussian description. We obtain the averaged conversion probability by creating a large number of such simulations, drawing photon paths separated by at least twice the size of the top-hat filter (to ensure they are sufficiently uncorrelated) through them, and numerically calculating transition probabilities at each crossing using Eq.~\eqref{eq:prob_gamma_to_Ap}. Probabilities over a large number of photon paths are then histogrammed to obtain the numerical estimates for $\dd \langle P_{\gamma \to A'} \rangle/\dd z$, shown in the right panel of Fig.~\ref{fig:plasma_sims} as the dashed red line for the log-normal case. The analytically-computed differential conversion probability for this configuration is shown in solid red, and provides a good match to the numerical results. The analytic Gaussian description, shown in blue, does not accurately described the conversion probability in this regime. 

\begin{figure}[htbp]
    \centering
    \includegraphics[width=0.45\textwidth]{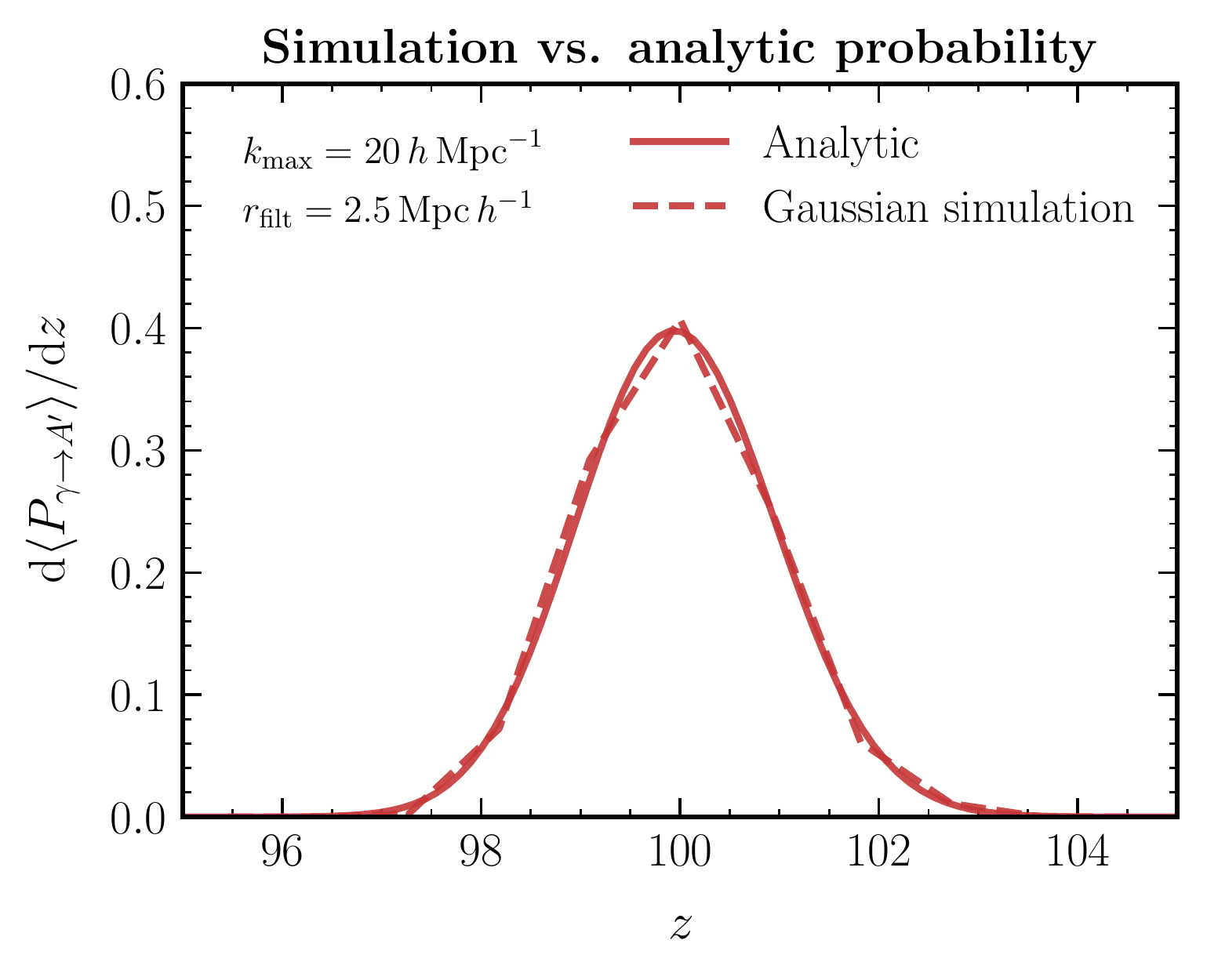}
    \caption{Differential conversion probability obtained by drawing photon paths through Gaussian random field simulations (dashed red) and computed analytically (solid red), for a resonance around $z=100$. Good agreement between simulations and the analytic description can be seen.~\nblink{19_lognormal_simulations}} 
    \label{fig:dPdz_GRF_sim}
\end{figure}

At higher redshifts and in the linear regime, on the other hand, the Gaussian PDF is an excellent description of the plasma mass fluctuations. Fig.~\ref{fig:dPdz_GRF_sim} shows a comparison of the analytically-computed differential conversion and the probability derived by considering photon paths through Gaussian random field-simulations of the plasma mass, showing once again good agreement between the two. 

\begin{figure*}[htbp]
    \centering
    \includegraphics[width=0.98\textwidth]{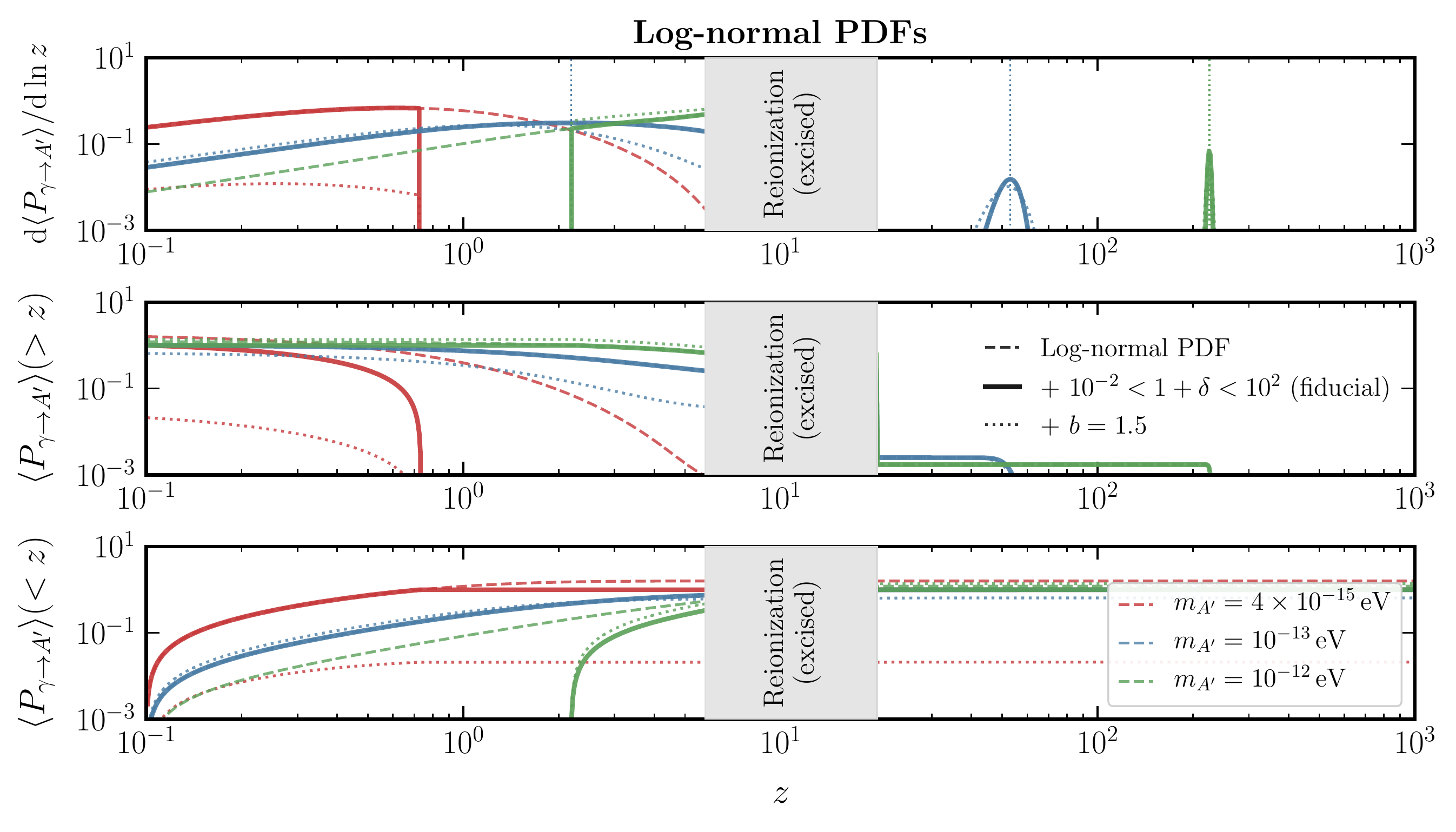}
\caption{The differential conversion probability $\dd \langle P_{\gamma \to A'} \rangle/\dd\ln z$ (top row), cumulative conversion probability above a given redshift $z$ (middle row), and cumulative conversion probability below a given redshift $z$, shown for a log-normal PDF (dashed lines), our fiducial log-normal PDF with $10^{-2} \lesssim 1 + \delta_\text{b} \lesssim 10^2$ (solid lines), and additionally with a bias $b=1.5$ (dotted lines). Masses $\mAp = 4\times10^{-15}$\,eV (red), $10^{-13}$\,eV (blue), and $10^{-12}$\,eV (green) are shown. Lines are normalized such that the cumulative probabilities for the $10^{-2} \lesssim 1 + \delta_\text{b} \lesssim 10^2$-bounded log-normal PDF cases are unity.~\nblink{17_formalism_prob_plots_nonlinear}} 
    \label{fig:dP_dz_ln}
\end{figure*}
\begin{figure*}[htbp]
    \centering
    \includegraphics[width=0.98\textwidth]{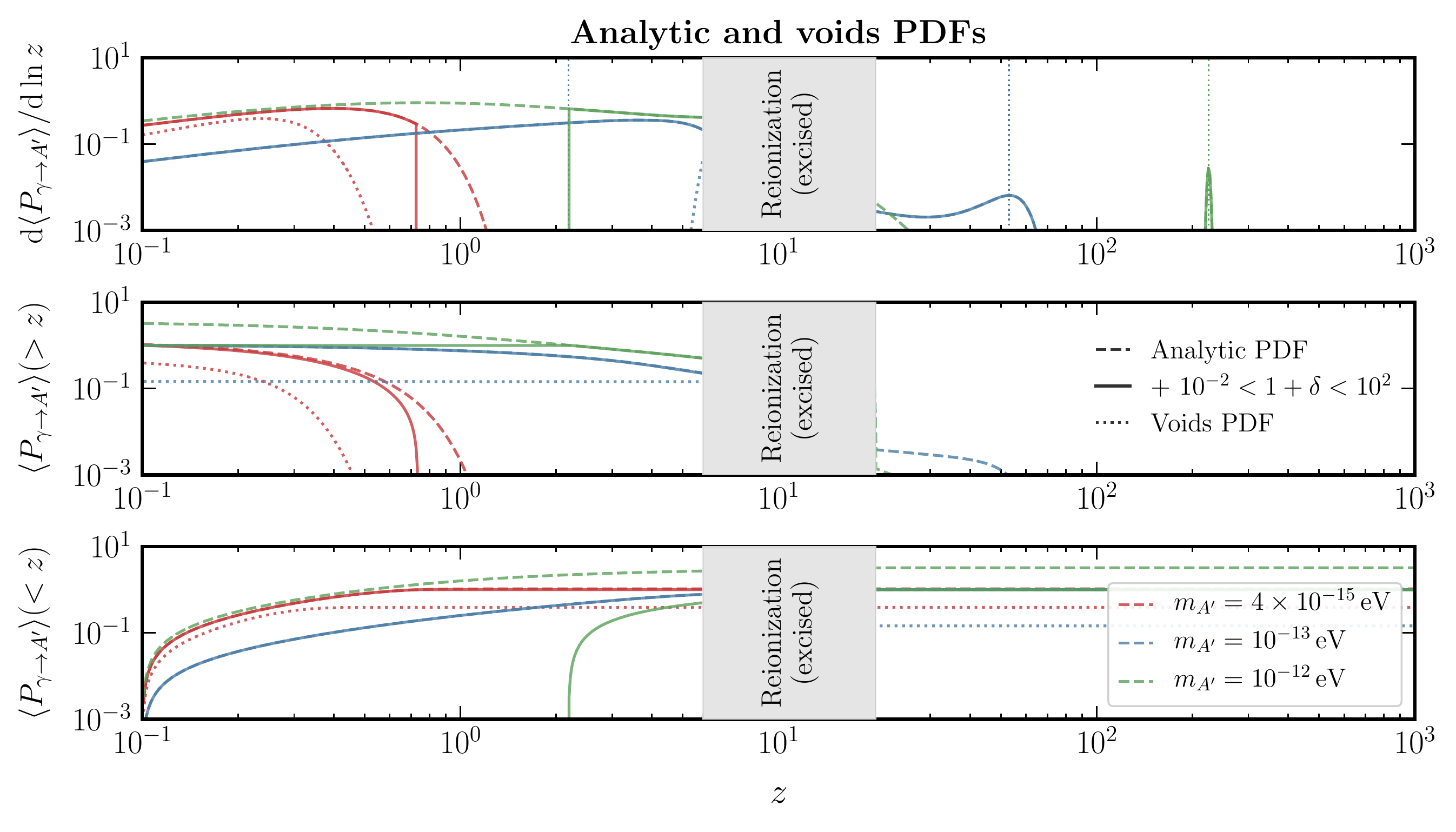}
    \caption{The same as Fig.~\ref{fig:dP_dz_ln}, shown for the analytic PDF (dashed lines), additionally imposing $10^{-2} \lesssim 1 + \delta_\text{b} \lesssim 10^2$ (solid lines), and the voids PDF (dotted lines).~\nblink{17_formalism_prob_plots_nonlinear}} 
    \label{fig:dP_dz_an}
\end{figure*}

\section{Systematics of conversion probability and energy injection}
\label{sec:results}

Given a PDF of density fluctuations and a description of the fluctuations through the power spectra, the differential conversion probability $\dd \langle P_{\gamma \to A'} \rangle / \dd \ln z$ at a given redshift, for a given dark photon mass, can be computed. This is the main deliverable of this paper, and is plotted in the top rows of Fig.~\ref{fig:dP_dz_ln} and Fig.~\ref{fig:dP_dz_an} for various PDF descriptions and benchmark masses $\mAp = 4\times10^{-15}$\,eV (red), $10^{-13}$\,eV (blue), and $10^{-12}$\,eV (green). The cumulative probabilities above (below) a given redshift are plotted in the middle(bottom) panels of these figures. Fig.~\ref{fig:dP_dz_ln} shows various log-normal PDFs---including all overdensities and underdensities (dashed lines), imposing $10^{-2} \lesssim 1 + \delta_\text{b} \lesssim 10^2$ (solid lines), and additionally with a bias $b=1.5$ (dotted lines) as described in Sec.~\ref{sec:ln_bias}. Fig.~\ref{fig:dP_dz_an} shows these for the analytic PDF (dashed lines), additionally imposing $10^{-2} \lesssim 1 + \delta_\text{b} \lesssim 10^2$ (solid lines), and the voids PDF (dotted lines). For ease of comparison, these are normalized such that the cumulative probabilities for the fiducial $10^{-2} \lesssim 1 + \delta_\text{b} \lesssim 10^2$-bounded log-normal PDF cases are unity. The primary focus here is on dark photons of masses $\lesssim 10^{-12}$\,eV, where the conversion probability is dominated by a broad efficiency of conversions at late times $z \lesssim 6$. The lower uncertainty envelope of the simulation-informed power spectrum described in Sec.~\ref{sec:power_spectra} was used to inform the variance for the PDFs in these plots; using the power spectrum corresponding to the upper uncertainty envelope produces qualitatively similar results. 

\begin{figure*}[htbp]
    \centering
    \includegraphics[width=0.45\textwidth]{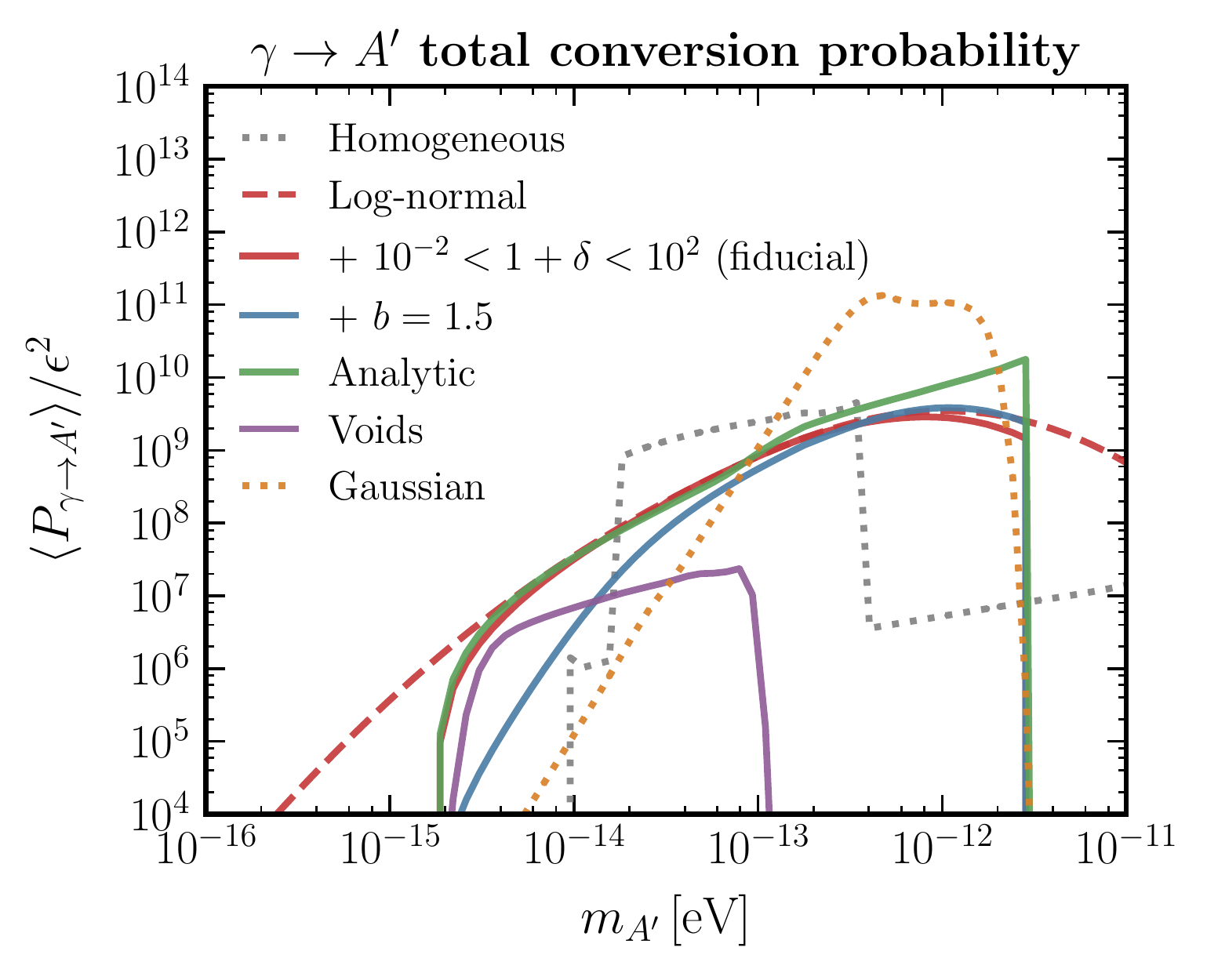}
    \includegraphics[width=0.45\textwidth]{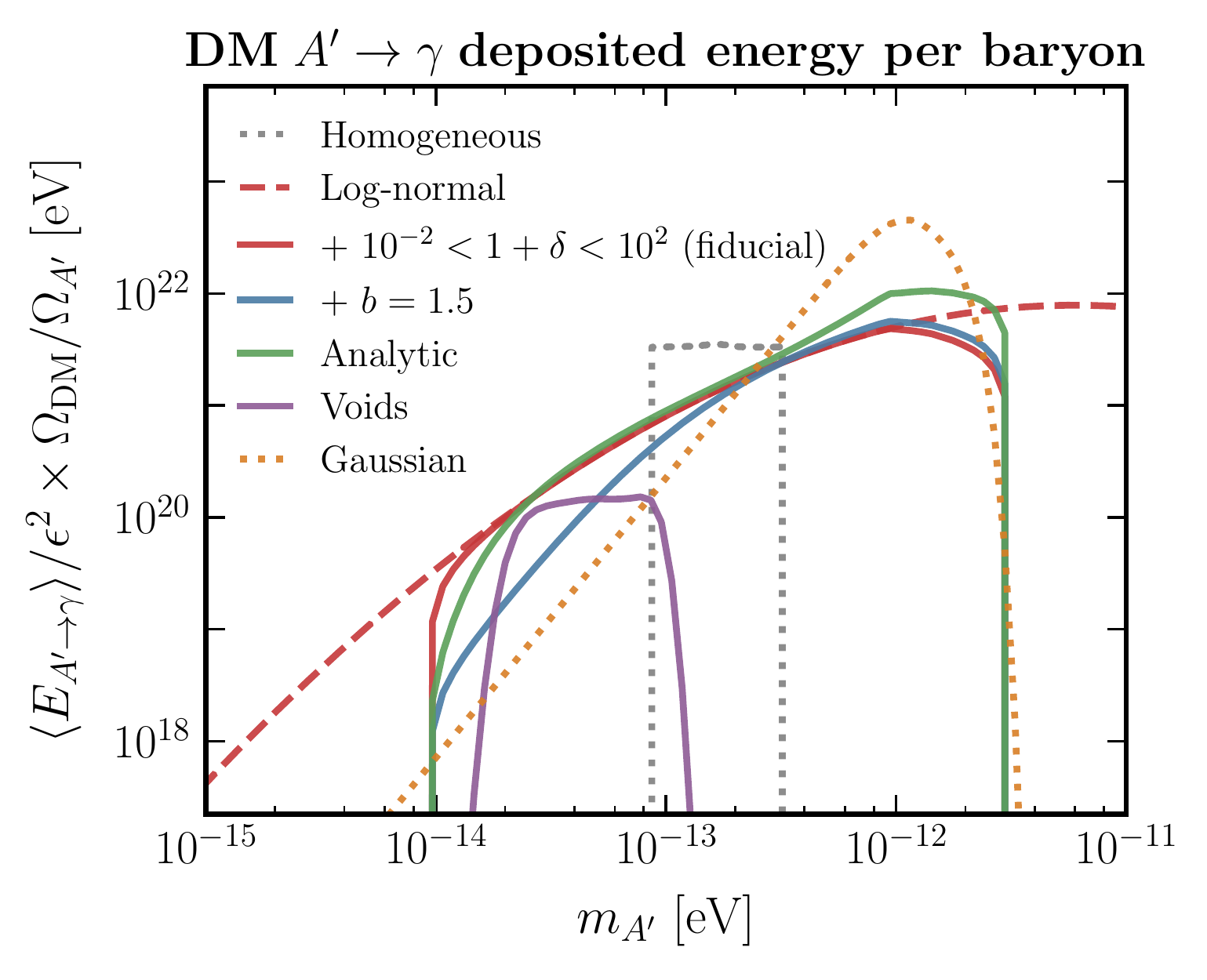}
    \caption{\emph{(Left)} The total $\gamma\leftrightarrow A'$ conversion probability as a function of dark photon mass, and \emph{(Right)} The $A' \to \gamma$ dark photon dark matter energy deposited per baryon as a function of dark photon mass, shown for different choices of PDFs explored in this work: log-normal (red dashed), the fiducial log-normal with $10^{-2} \lesssim 1 + \delta_\text{b} \lesssim 10^2$ (red solid), log-normal with a bias $b=1.5$ (blue solid), analytic (green solid), voids (purple solid), and Gaussian  (orange dotted).~\nblink{17_formalism_prob_plots_nonlinear}~\nblink{15_dp_dm_He_energy_per_baryon}} 
    \label{fig:P_tot_m_Ap}
\end{figure*}

In order to illustrate how the total $\gamma \rightarrow A'$ conversion probability is affected by various PDFs for different dark photon masses, the total conversion probability per squared kinetic mixing parameter $\epsilon$ is shown in the left panel of Fig.~\ref{fig:P_tot_m_Ap} for the different PDFs we have considered. Log-normal (dashed red), log-normal imposing $10^{-2} \lesssim 1 + \delta_\text{b} \lesssim 10^2$ (solid red), log-normal with bias $b=1.5$ (blue), analytic (green), voids (purple), and Gaussian (orange dotted) PDFs are illustrated. Similarly, the total energy deposited per baryon when a non-zero ambient density of dark photons is present (\emph{e.g.}, in the case of dark photon dark matter) is shown in the right panel of Fig.~\ref{fig:P_tot_m_Ap}. In each case, the corresponding quantities under the assumption of a homogeneous photon plasma are shown in dotted gray. It can be seen that inhomogeneities have a significant effect on the nature of photon-to-dark photon oscillations, either underestimating or overestimating the total conversion probability and energy deposition depending on the dark photon mass point considered. Variation is also observed across the different PDFs considered; however, after restricting to fluctuations of size $10^{-2} \lesssim 1 + \delta_\text{b} \lesssim 10^2$, the log-normal and analytic PDFs show quantitatively similar behavior, with the log-normal PDF being somewhat more conservative. For this reason, henceforth in this paper and in~\citetalias{Caputo:2020bdy}, we use the log-normal PDF with variance informed by hydrodynamic simulations as the benchmark for computing the effects of $\gamma\leftrightarrow A'$ conversions. In the absence of dedicated PDFs capturing baryonic effects and their uncertainties to the smallest relevant scales, we advocate for its use in applications beyond those considered in these papers where the effects of inhomogeneities in the nonlinear regime on $\gamma\leftrightarrow A'$ conversions may be important.

\begin{figure*}[htbp]
    \centering
    \includegraphics[width=0.45\textwidth]{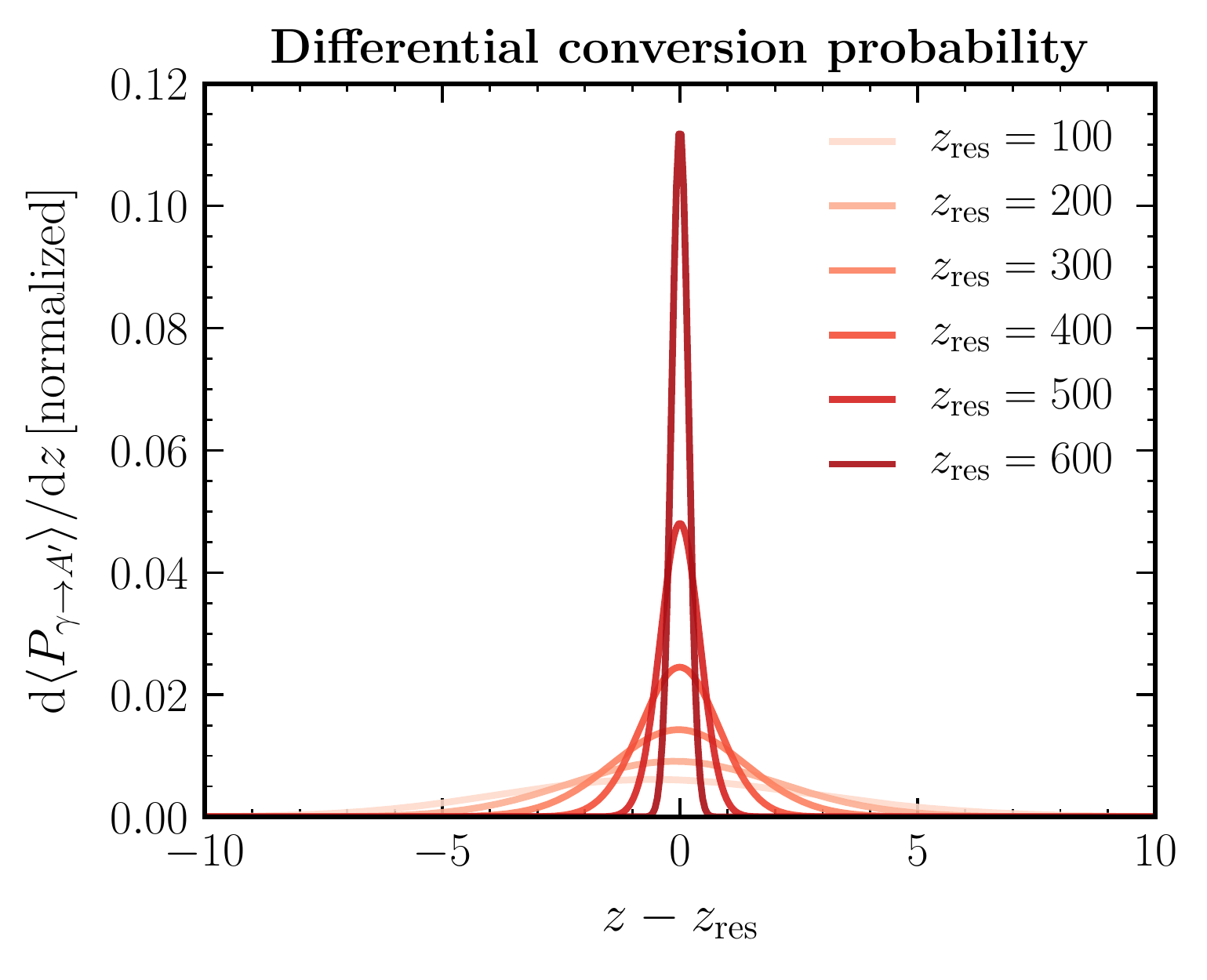}
    \includegraphics[width=0.45\textwidth]{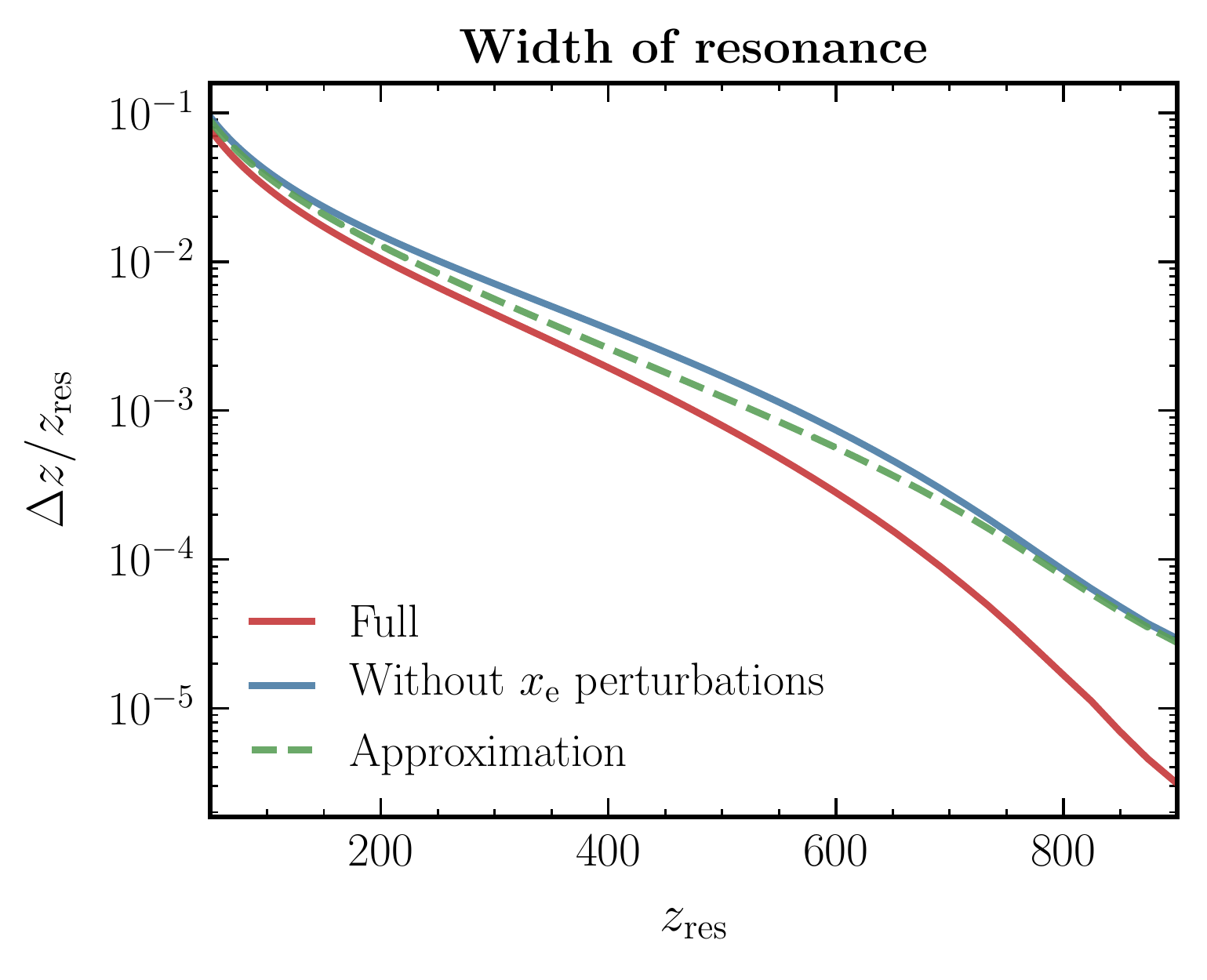}
    \caption{\emph{(Left)} The differential conversion probability $\dd \langle P_{\gamma\to A'}\rangle / \dd z$ for resonant conversion at higher redshifts $z_\mathrm{res} = 100$ to $600$, shown centered on the resonant redshift $z_\mathrm{res}$. \emph{(Right)} Relative width of the resonance as a function of resonance redshift $z_\mathrm{res}$. Shown with (red) and without (blue) accounting for perturbations in the electron ionization fraction $x_\mathrm{e}$. The dotted green line shows the approximate width as given by Eq.~\eqref{eq:z_range_of_conversion_less_approx}, showing good agreement with the numerical estimate without accounting for $x_\text{e}$ perturbations.~\nblink{16_formalism_prob_plots_high_z}} 
    \label{fig:dP_dz_highz}
\end{figure*}

Conversions at earlier times $z\gtrsim 100$ can be well-described by a Gaussian in redshift with a weakly redshift-dependent variance, described in Eq.~\eqref{eq:dP_dz_gaussian}. Example differential conversion probabilities are shown in the left panel of Fig.~\ref{fig:dP_dz_highz} for resonance redshifts spanning $100 \leq z_\mathrm{res} \leq 600$, centered on the resonance redshift and normalized to unity. The approximate relative width of the resonance is shown in the right panel of Fig.~\ref{fig:dP_dz_highz}, with (without) accounting for perturbations in the free electron fraction in red (blue). The width is numerically computed as the interval $\Delta z$ between redshifts where the squared plasma mass is $\pm \sigma_{m_\gamma^2}/2$ of its central value, approximately corresponding to a middle 1-$\sigma$ containment interval. The presence of spatial perturbations in the free electron fraction becomes increasingly important closer to the redshift of recombination, although the relative width of the conversion feature is already less than one part in $10^{-3}$ by $z_\mathrm{res} = 600$.

\begin{figure}[htbp]
    \centering
    \includegraphics[width=0.45\textwidth]{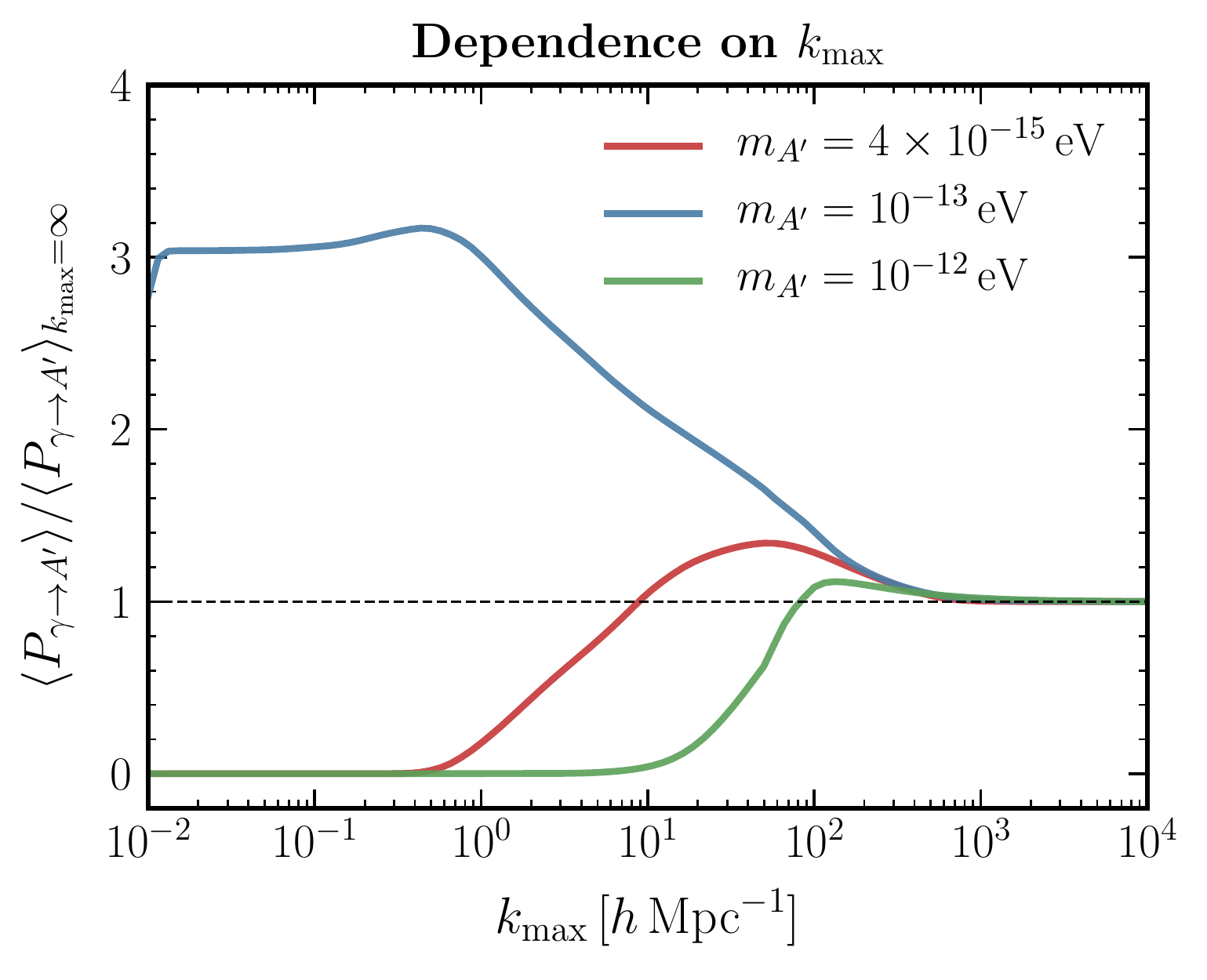}
    \caption{For the fiducial log-normal PDF, the ratio of the total conversion probability using fluctuations only up to a given scale $k_\mathrm{max}$ and its asymptotic value, shown for masses $\mAp = 4\times10^{-15}$\,eV (red), $10^{-13}$\,eV (blue), and $10^{-12}$\,eV (green).~\nblink{17_formalism_prob_plots_nonlinear}} 
    \label{fig:P_k_max}
\end{figure}

Due to the sensitive dependence of the conversion probability on small-scale physics as discussed in Sec.~\ref{sec:linear_regime}, it is illustrative to see how the total conversion probability depends on the maximum scale $k_\mathrm{max}$ considered.  This is illustrated in Fig.~\ref{fig:P_k_max} for our benchmark masses, shown as the ratio of the total probability considering scales up to $k_\mathrm{max}$ to the asymptotic probability. We see that the total probability approaches the asymptotic value around the characteristic baryon Jeans scale at late times, $k_\mathrm{J} \sim 500\,h$\,Mpc$^{-1}$. Note that neglecting the effect of small scales is not necessarily conservative and may significantly underestimate or overestimate the conversion probability.

\begin{figure*}[htbp]
    \centering
    \includegraphics[width=0.45\textwidth]{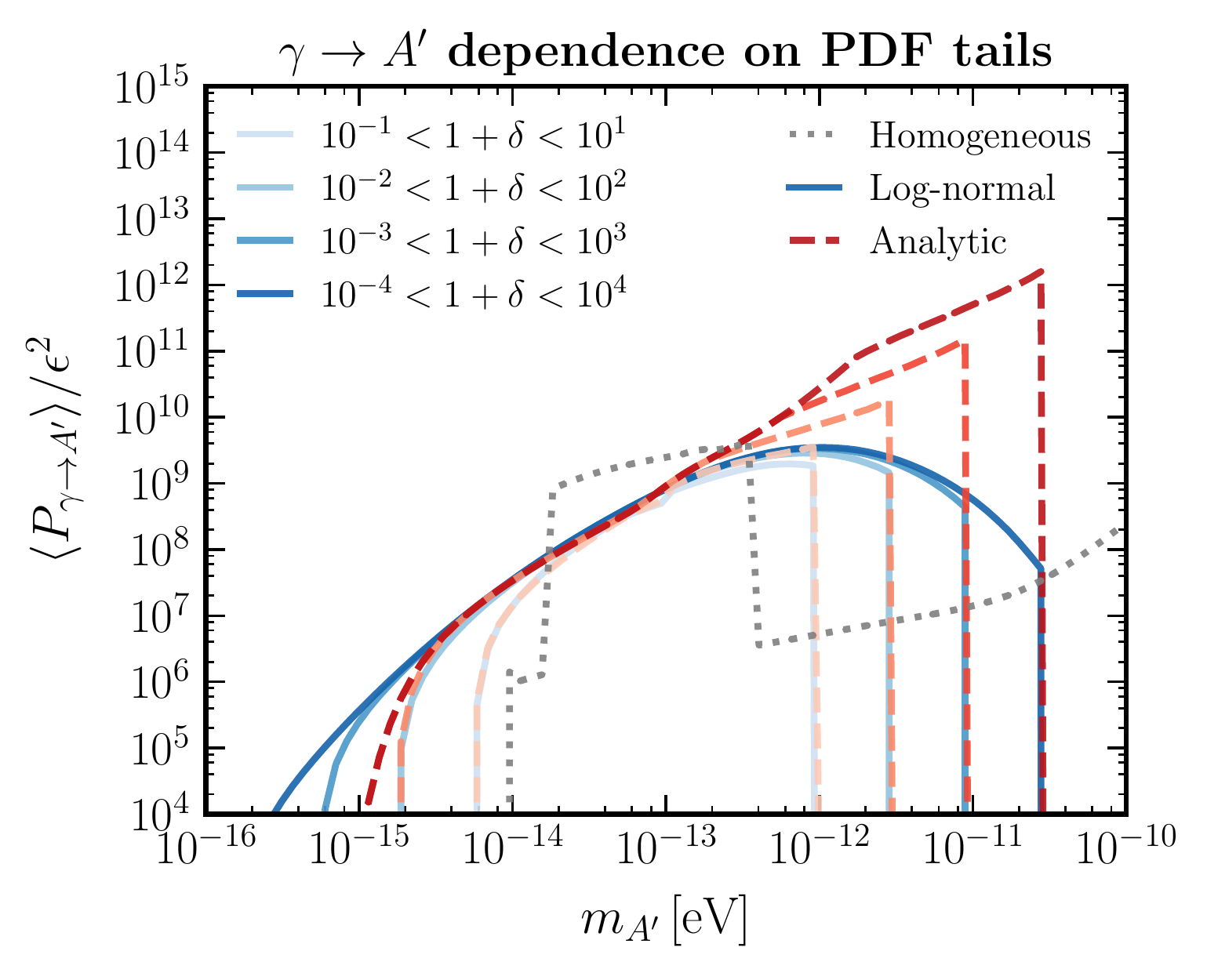}
    \includegraphics[width=0.45\textwidth]{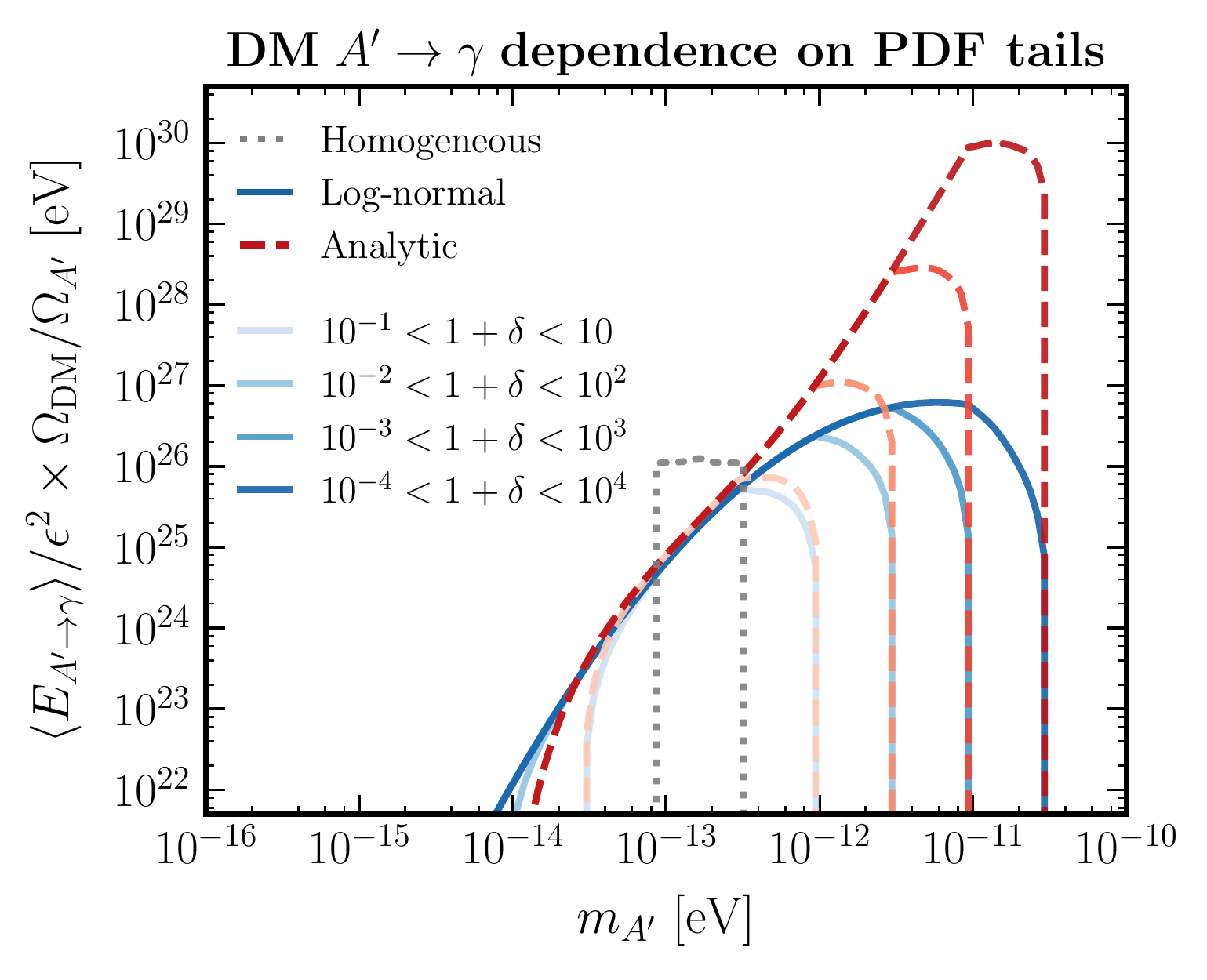}
    \caption{Dependence of \emph{(Left)} the total $\gamma\leftrightarrow A'$ conversion probability and \emph{(Right)} the energy injected by $A' \to \gamma$ dark-photon dark matter on the tails of the plasma mass PDF, shown for the log-normal (blue) and analytic (dashed red) PDFs. The total homogeneous probability is shown as dotted grey, for comparison.~\nblink{17_formalism_prob_plots_nonlinear}~\nblink{15_dp_dm_He_energy_per_baryon}} 
    \label{fig:P_tot_m_Ap_tails}
\end{figure*}

Finally, although we advocate restricting to fluctuations in the range $10^{-2} \lesssim 1 + \delta_\text{b} \lesssim 10^2$ where the different PDF descriptions considered show qualitative agreement, it is instructive to ask how expanding this range and including larger underdensities and overdensities in the tails of the PDFs can affect the oscillation physics. In Fig.~\ref{fig:P_tot_m_Ap_tails}, we show the total conversion probability as a function of dark photon mass varying the range of fluctuations from  $10^{-1} \lesssim 1 + \delta_\text{b} \lesssim 10$ to $10^{-4} \lesssim 1 + \delta_\text{b} \lesssim 10^4$ for the log-normal (solid red lines) and analytic (dashed blue lines) PDFs. Although the two descriptions disagree for fluctuations beyond $10^{-2} \lesssim 1 + \delta_\text{b} \lesssim 10^2$, in either case larger conversion probabilities over a much wider range of dark photon masses can be seen when including conversions from fluctuations deeper in the tails of the PDFs. This motivates the need for a better understanding of the nonlinear baryon PDF at late times. A similar conclusion can be drawn for $A' \to \gamma$ dark-photon dark matter conversions, also shown in Fig.~\ref{fig:P_tot_m_Ap_tails}. 


\section{Conclusions}
\label{sec:conclusion}

In this paper, we have studied photon-dark photon oscillations in the early Universe, deriving a formalism for computing the averaged probability of conversions in both directions, taking into account the effect of inhomogeneities in the photon plasma. 
We found that the average probability of $\gamma \leftrightarrow A'$ and the average energy injected per baryon for $A' \to \gamma$ for dark photon dark matter are completely specified given the standard $\Lambda$CDM parameters as well as three inputs: \emph{(i)} a description of the one-point PDF of baryon fluctuations, \emph{(ii)} the baryon power spectrum which, to a good approximation in the low-redshift Universe, provides the variance of plasma mass fluctuations, and \emph{(iii)} fluctuations in the free electron fraction, which contributes to the variance of plasma mass fluctuations at high redshift. To understand the systematic uncertainties associated with the PDF and the variance of fluctuations, we studied several independent choices of the one-point PDF. 
We also constructed a nonlinear baryon power spectrum that is informed by high-resolution hydrodynamic $N$-body simulations, allowing us to characterize the behavior of baryons at small scales. 
Finally, we also performed a series of Gaussian and log-normal random field simulations in order to validate our analytic results, finding agreement between theory and simulations. 

In our companion work~\citetalias{Caputo:2020bdy}, we have applied this formalism in order to derive constraints on the dark photon kinetic mixing parameter $\epsilon$ by through the effect of $\gamma \to A'$ conversions on the CMB spectrum as measured by COBE/FIRAS in the general case, as well as dedicated constraints for the case of dark photon dark matter obtained by computing the amount of IGM heating due to $A' \to \gamma$ conversions. 
We found that previous constraints assuming a homogeneous plasma were not conservative, and were able to expand the mass range over which resonant oscillations are possible due to conversions in plasma underdensities and overdensities. 
We also found good agreement between constraints obtained using different PDFs and power spectra, showing that we have a sufficiently good understanding of baryon fluctuations to set reliable constraints. 

The formalism that we have developed across both papers has additional applications. For example, perturbations in the photon plasma mass will also modify resonant oscillations of photons into axion-like-particles, which can occur in the presence of primordial magnetic fields~\cite{Mirizzi:2009nq}. Moreover, relativistic dark photons can also resonantly inject photons, which can be tested by 21-cm observations~\cite{Choi:2019jwx,Pospelov:2018kdh,Moroi:2018vci}. Photon-to-dark photon oscillations in an inhomogeneous background will also imprint anisotropies in the CMB that may be testable by \emph{Planck}~\cite{Aghanim:2019ame} or future CMB probes~\cite{Abazajian:2016yjj}, as also explored in Ref.~\cite{Garcia:2020qrp}. 

A comparison with our results and methodology with those presented in related recent studies, auxillary information about the analytic PDF employed in this work, and a complementary derivation of the Landau-Zener formula for resonant conversions in the language of thermal field theory is provided in the appendices. The code used to obtain the results in both papers is available at \url{https://github.com/smsharma/dark-photons-perturbations}.

\vspace{.3cm}

\begin{acknowledgments}

We thank Yacine Ali-Ha\"{i}moud, Masha Baryakhtar, Asher Berlin, Julien Lesgourgues, Sam McDermott, Alessandro Mirizzi, Julian Mu\~{n}oz, Stephen Parke, Maxim Pospelov, Josef Pradler, Javier Redondo, Roman Scoccimarro, Anastasia Sokolenko, Alfredo Urbano, Edoardo Vitagliano, Sam Witte, and Chih-Liang Wu for helpful conversations. We thank Marcel van Daalen for providing baryonic power spectra from high-resolution BAHAMAS simulations. We are especially grateful to Misha Ivanov for many enlightening discussions regarding the analytic PDF of density fluctuations utilized in this work. AC acknowledges support from the ``Generalitat Valencian'' (Spain) through the ``plan GenT'' program (CIDEGENT/2018/019), as well as national grants FPA2014-57816-P, FPA2017-85985-P, and the European projects H2020-MSCA-ITN-2015//674896-ELUSIVES. HL is supported by the DOE under contract DESC0007968. SM and JTR are supported by the NSF CAREER grant PHY-1554858 and NSF grant PHY-1915409. SM is additionally supported by NSF grant PHY-1620727 and the Simons Foundation. 
JTR acknowledges hospitality from  the Aspen Center for Physics, which is supported by the NSF grant PHY-1607611.
This work made use of the NYU IT High Performance Computing resources, services, and staff expertise. The authors are pleased to acknowledge that the work reported on in this paper was substantially performed using the Princeton Research Computing resources at Princeton University which is a consortium of groups including the Princeton Institute for Computational Science and Engineering and the Princeton University Office of Information Technology's Research Computing department. This research has made use of NASA's Astrophysics Data System. We acknowledge the use of the Legacy Archive for Microwave Background Data Analysis (LAMBDA), part of the High Energy Astrophysics Science Archive Center (HEASARC). HEASARC/LAMBDA is a service of the Astrophysics Science Division at the NASA Goddard Space Flight Center. This research made use of the \texttt{astropy}~\cite{Price-Whelan:2018hus,Robitaille:2013mpa}, CAMB~\cite{Lewis:1999bs,Lewis:2002ah}, CLASS~\cite{Blas:2011rf}, \texttt{HyRec}~\cite{AliHaimoud:2010dx}, \texttt{IPython}~\cite{PER-GRA:2007}, Jupyter~\cite{Kluyver2016JupyterN}, \texttt{matplotlib}~\cite{Hunter:2007}, \texttt{nbodykit}~\cite{Hand:2017pqn}, \texttt{NumPy}~\cite{numpy:2011}, \texttt{seaborn}~\cite{seaborn}, \texttt{pandas}~\cite{pandas:2010}, \texttt{SciPy}~\cite{2020SciPy-NMeth}, and \texttt{tqdm}~\cite{da2019tqdm}  software packages. 
\end{acknowledgments}

\appendix

\section{Comparison with other work}
\label{app:comparison_with_previous_work}

In this section we present a comparison of the formalism and results described in this work and in~\citetalias{Caputo:2020bdy} with those presented in several recent studies which also attempt to model inhomogeneous $\gamma \leftrightarrow A'$ oscillations and their observational consequences.

In Refs.~\cite{Bondarenko:2020moh,Garcia:2020qrp}, the conversion probability as photons pass through inhomogeneities was determined through the use of the EAGLE simulation~\cite{McAlpine:2015tma} with baryons.
Lines were drawn at random for each redshift snapshot in the simulation, and one hundred continuous lines-of-sight in the range $0 < z < 6$ were constructed. 
These lines-of-sight are then used to compute the probability of $\gamma \to A'$ conversion with the inhomogeneities encountered in the simulation, and used to set limits on the kinetic mixing parameter $\epsilon$. 
Ref.~\cite{Garcia:2020qrp} found good agreement between their results and those presented in~\citetalias{Caputo:2020bdy}. 
They also use a similar approach to obtain CMB power spectrum constraints by comparing the fluctuation in conversion probability between line-of-sights, finding a weaker limit than that obtained from the COBE/FIRAS energy spectrum measurement. 

We note that while we also use input from the same EAGLE simulation~\cite{McAlpine:2015tma}, we only rely on the baryon power spectrum from this and other simulations, rather than the full spatial information. This significantly simplifies the process of understanding $\gamma \leftrightarrow A'$ oscillations, and allows us to do two things: \emph{(i)} avoid the need to smooth the simulation excessively, and \emph{(ii)} capture the uncertainty associated with different choices of the one-point PDF. We will now discuss each point in turn:

\begin{enumerate}

\item \textbf{Smoothing.} $N$-body simulations have a finite resolution, and it is often the case that some smoothing of the data needs to be done prior to analysis. 
Finite resolution effects and smoothing ultimately introduce an effective cut-off $k_\text{res}$ in the power spectrum of fluctuations.
For values of $k_\text{res} \lesssim \SI{e2}{\kilo\parsec}$, Fig.~\ref{fig:P_k_max} shows that the calculated conversion probability can deviate significantly from the asymptotic value we infer using the procedure described in Sec.~\ref{sec:power_spectra}. 
In Ref.~\cite{Bondarenko:2020moh}, the lines-of-sight are smoothed over a comoving pixel size of $\SI{20}{\kilo\parsec} \times \SI{20}{\kilo\parsec} \times \SI{250}{\kilo\parsec}$, while in Ref.~\cite{Garcia:2020qrp}, this is reduced to $\SI{20}{\kilo\parsec} \times \SI{20}{\kilo\parsec} \times \SI{25}{\kilo\parsec}$, with the authors of Ref.~\cite{Garcia:2020qrp} finding no difference in the conversion probability between the two smoothing scales. 
We have checked that performing this anisotropic smoothing over comoving $\SI{20}{\kilo\parsec} \times \SI{20}{\kilo\parsec} \times \SI{250}{\kilo\parsec}$ pixels produces a variance of fluctuations $\sigma_\text{b}$ that is similar to having $k_\text{res} \sim \SI{170}{\h\per\mega\parsec}$ in the redshift range $0 < z < 6$.
This should therefore lead to similar results for the conversion probability, as shown in Fig.~\ref{fig:P_k_max}. 
This also explains why Ref.~\cite{Garcia:2020qrp} observes no difference in results between the two pixel sizes. 
In general, however, smoothing must be used with caution due to the ultraviolet divergence of the variance of fluctuations, as described in Secs.~\ref{sec:jeans_scale} and~\ref{sec:results}. 
Too large of a smoothing scale, either due to the finite resolution of a simulation or post-processing of the results, may lead to very different and incorrect (not necessarily conservative) outcomes. 
It is important to use high resolution results and smooth as little as possible.

\item \textbf{Capturing uncertainties.} As we showed in Sec.~\ref{sec:simulations}, the full simulation data is not necessary to determine the $\gamma \to A'$ conversion probability in the presence of inhomogeneities; knowledge of the one-point PDF alone is sufficient for that. 
Our work therefore represents a significant simplification compared to constructing lines-of-sight through simulation results.
In particular, we do not need to rely on the outcome of a single simulation to extract our results, as was done in Refs.~\cite{Bondarenko:2020moh,Garcia:2020qrp}; we have shown how our results change depending on our choice of one-point PDFs and baryon power spectra, allowing us to study the uncertainty associated with these inputs based a large array of theoretical and simulation results. 
This is particularly important for conversions in large under- and overdensities, where the PDFs are highly uncertain. 

\end{enumerate}

The authors of Ref.~\cite{Witte:2020rvb} on the other hand reconsidered the bounds on dark photon dark matter $A' \rightarrow \gamma$ conversions, obtained from Ly-$\alpha$ observations of the IGM temperature, in the presence of inhomogeneities. 
Their overall approach to the problem is similar to ours, although they do not generalize their results to treat $\gamma \to A'$ as we do in our work, where a CMB photon passes through multiple level crossings along its path at which $m_\gamma^2 = m_{A'}^2$. 
Our results, however, differ from Ref.~\cite{Witte:2020rvb} for the following reasons:

\begin{enumerate}
\item \textbf{Value of the Jeans scale.} The authors of Ref.~\cite{Witte:2020rvb} adopt a value of the Jeans scale close to $R_\text{J} \sim \SI{1}{\mega\parsec}$ after reionization, which derives from Eq.~\eqref{eq:jeans_length_numeric} with a baryon temperature of approximately $T_\text{b} \sim \SI{e4}{\kelvin}$. 
However, as we discussed in Sec.~\ref{sec:jeans_scale}, a suppression at these scales is not seen in any of the $N$-body simulations (with baryonic physics included) we used to infer our cut-off scale $k_\text{J}$, which is then smaller by roughly two orders of magnitude. 
This is due to the increasingly nonlinear behavior of baryons at late times, which makes difficult to analytically predict the scale at which structure formation is suppressed. 
Their choice of the Jeans scale is therefore an underestimate, leading to overly narrow $\dd P/\dd z$ as a function of redshift. This can have a large effect on the derived constraints, as we show in Fig.~\ref{fig:P_k_max}. 

\item \textbf{Ly-$\boldsymbol{\alpha}$ observations sensitivity.} The authors of Ref.~\cite{Witte:2020rvb} note that IGM temperature measurements from Ly-$\alpha$ observations are not sensitive to large under- or overdensities, which is of particular importance if the energy injection is deposited locally (see the following point). 
Too large values of $\delta_\text{b}$ lead to a large optical depth of the IGM medium, leading to near-total absorption of Ly-$\alpha$ photons, preventing us from learning anything about optically thick regions; on the other hand, too low $\delta_\text{b}$ would mean no absorption lines at all, which is required to deduce the IGM temperature~\cite{Becker:2010cu}. 
Ref.~\cite{Witte:2020rvb} proposed two heuristic ways of correcting for this; their fiducial method, for example, rescales the energy deposited by a factor proportional to the derivative of the Ly-$\alpha$ absorption probability, while their alternative method simply assumes that no temperature measurements are possible outside of some optical depth range.
Both prescriptions adopted in Ref.~\cite{Witte:2020rvb} are reasonable, but nevertheless only heuristic, and have many caveats. 
They depend, for example, on the IGM temperature-density relation, assumed to be $T \propto (1+\delta_\text{b})^{\gamma-1}$ where $\gamma \sim 1.5$; it is unclear if this power-law relation is valid at low densities~\cite{Bolton:2007xi, Rorai:2016jfm}.

\item \textbf{Energy injection.} We worked under the assumption that energy injection is a global phenomenon, \emph{i.e.},\ energy injected from $A' \to \gamma$ conversions is shared evenly among all baryons. The authors of Ref.~\cite{Witte:2020rvb}, on the other hand, assume local energy injection, where the energy is deposited only into baryons at the point where conversions occur. 
For completeness, we have also derived the energy deposition per baryon under the local assumptions, shown in Eq.~\eqref{eq:dE_dz_local}. This expressions agrees with the expression derived in Ref.~\cite{Witte:2020rvb}, although we show that it reduces to a much simpler form shown in Eq.~\eqref{eq:dE_dz_local_simple} within our framework, as compared to the results shown in Ref.~\cite{Witte:2020rvb}. 
The authors of Ref.~\cite{Witte:2020rvb} justify the local assumption by noting that the electrons that absorb this energy are nonrelativistic, and so the energy transport timescale has to be much longer than the age of the Universe. 

We expect the transport of energy from $A' \to \gamma$ conversions to lie somewhere in between both regimes.
The argument in Ref.~\cite{Witte:2020rvb} about nonrelativistic electrons applied to reionization, for example, would seem to preclude the possibility of complete reionization across the entire Universe. 
Instead, as in the process of reionization, we expect photons with energy above the ionization threshold of HI to play a large role in energy transport. 
During HeII reionization, the epoch in which we derive our constraints in~\citetalias{Caputo:2020bdy}, the IGM is already at $T_\text{b} \sim \SI{e4}{\kelvin} $ and will be heated beyond that due to the $A'$ conversion. 
The blackbody spectrum of the IGM contains ionizing photons, which have a long interaction path length, potentially comparable to the size of the Universe at redshifts $2 \lesssim z \lesssim 6$. This may allow for energy transport over large distances. 

\end{enumerate}

Whether or not the energy injection is local is a nontrivial problem which requires a more involved treatment of the complete transport equations describing the system under consideration; we defer such an effort to future work. 
To account for general uncertainties regarding large under- and overdensities, especially with regard to uncertainties in the tails of the baryon one-point PDFs, we presented our limits on $\epsilon$ as a function of the expected range in $\delta_\text{b}$ in~\citetalias{Caputo:2020bdy}.
In addition, we show the $A' \to \gamma$ dark photon dark matter constraints derived from Ly-$\alpha$ temperature measurements of the IGM during HeII reionization in~\citetalias{Caputo:2020bdy}, neglecting densities which lead to an optical depth for Ly-$\alpha$ photons that satisfy $\exp(-\tau) < 0.05$ or $\exp(-\tau) > 0.95$, the `alternate' method adopted by Ref.~\cite{Witte:2020rvb}. 
Our results broadly agree with those obtained in Ref.~\cite{Witte:2020rvb}. 

\section{Functions for the analytic PDF}
\label{app:functions_analytic_pdf}

Following Ref.~\cite{Ivanov:2018lcg}, the function $F(\delta_*)$ is defined as the composition of two functions
\begin{alignat}{1}
    F \equiv \mathcal{G} \circ \mathcal{F}^{-1} \,,
\end{alignat}
where
\begin{alignat}{1}
    \mathcal{G}(\theta) \equiv \frac{3}{20} \left( 6[\theta - s(\theta)] \right)^{\nicefrac{2}{3}} \,,
\end{alignat}
and
\begin{alignat}{1}
    \mathcal{F}(\theta) \equiv \frac{9[s(\theta) - \theta]^2}{2[c(\theta) - 1]^3} - 1 \,,
\end{alignat}
with
\begin{alignat}{1}
    s(\theta) \equiv \begin{cases}
        \sin \theta\,, & \delta_* > 0 \,, \\
        \sinh \theta \,, & \delta_* \leq 0 \,,
    \end{cases} \quad
    c(\theta) \equiv \begin{cases}
        \cos \theta\,, & \delta_* > 0 \,, \\
        \cosh \theta \,, & \delta_* \leq 0 \,.
    \end{cases}
\end{alignat}
$\hat{C}(\delta_*)$ is then defined as
 \begin{alignat}{1}
     \hat{C}(\delta_*) \equiv F'(\delta_*) + \frac{F(\delta_*)}{1 + \delta_*} \left(1 - \frac{\xi_{R_*}}{\sigma^2_{R_*}}\right) \,,
 \end{alignat}
 and
 \begin{alignat}{1}
     \xi_{R_*} \equiv \frac{1}{2\pi^2} \int \dd k \, k^2 \frac{\sin(kR_*)}{kR_*} W_\text{th}(kR_*) P_\text{m,L}(k) \,,
 \end{alignat}
 where $W_\text{th}$ is the Fourier transform of the top-hat function defined in Sec.~\ref{sec:PDFs_analytic_PDF}, and $P_\text{m,L}$ is the linear matter power spectrum. 

\section{Thermal Field Theory derivation of Landau-Zener probability}
\label{app:thermal}

Here we give a brief derivation of the Landau-Zener formula using thermal field theory techniques. Indeed the conversion of CMB photons to dark photons can be seen as the production of dark photons from a thermal bath of photons following a blackbody spectrum. Following Refs.~\cite{Arias:2012az, Redondo:2013lna,Hardy:2016kme}, we can write the production rate of dark photons as 
\begin{alignat}{1}
\Gamma_{\mathrm{prod}} &= \left(\frac{1}{e^{\omega / T} - 1} \right) \frac{\epsilon^2 m_{A'}^4 \Gamma}{\omega^2 \Gamma^2 + (m_{\gamma}^2 - m_{A'}^2)^2} \, \nonumber \\ 
&\equiv f_\gamma(\omega,T)  \frac{\epsilon^2 m_{A'}^4 \Gamma}{\omega^2 \Gamma^2 + (m_{\gamma}^2 - m_{A'}^2)^2} \,,
\end{alignat}
where $\Gamma$ is the damping rate of the plasmon quanta, $m_{\gamma}$ is the plasma mass acquired by the photons in the plasma. The first factor $f_\gamma(\omega,T)$ is the photon occupation number, with $T$ being the CMB temperature. The second factor is the probability of conversion per unit time. 
In the limit of the narrow width approximation, assuming that the plasmons are weakly damped, the probability of conversion reduces to 

\begin{align}
\frac{\Gamma_{\mathrm{prod}}}{f_\gamma(\omega,T)} \rightarrow \, \frac{\epsilon^2 m_{A'}^4}{\omega^2} \delta_\text{D}\Big(\frac{m_{\gamma}^2 - m_{A'}^2}{\omega}   \Big) \,, 
\end{align}

where we used the definition of the Dirac $\delta_\text{D}$-function 
\begin{equation}
\lim_{\alpha \rightarrow 0} \frac{\alpha}{\alpha^2 +x^2} = \delta_\text{D}(x) \,.
\end{equation}

We can then integrate it over time along the photon path to find
\begin{align}
P_{\gamma \rightarrow A'} &= \int \dd t \, \frac{\epsilon^2 m_{A'}^4}{\omega^2} \delta_\text{D}\Big(\frac{m_{\gamma}^2 - m_{A'}^2}{\omega}   \Big) \nonumber \\ 
&= \sum_i  \frac{\epsilon^2 m_{A'}^2}{\omega(t_i)} \left| \frac{\dd\ln m_{\gamma}^2}{\dd t} \right|^{-1}_{t = t_i} \,,
\end{align}
which is indeed in agreement with Eq.~\eqref{eq:prob_gamma_to_Ap}.

\bibliographystyle{apsrev4-1}
\bibliography{perturbations-formalism}

\end{document}